\documentclass[fleqn,usenatbib]{mnras}

\usepackage{newtxtext,newtxmath}

\usepackage[T1]{fontenc}
\usepackage{color}
\usepackage{xcolor}

\usepackage{natbib}
\DeclareRobustCommand{\VAN}[3]{#2}
\let\VANthebibliography\thebibliography
\def\thebibliography{\DeclareRobustCommand{\VAN}[3]{##3}\VANthebibliography}

\usepackage{graphicx}	
\usepackage{amsmath}	

\title[A comprehensive separation of dark matter and baryons in galaxy clusters I]{A comprehensive separation of dark matter and baryonic mass components in galaxy clusters I: Mass constraints from Abell S1063}

\author[Beauchesne et al.]{
Benjamin Beauchesne,$^{1,2}$\thanks{E-mail: benjamin.e.beauchesne@durham.ac.uk}
Benjamin Cl\'ement,$^{3}$
Marceau Limousin,$^{4}$
Belén Alcalde Pampliega,$^{5,6,7}$
\newauthor Mathilde Jauzac,$^{1,2,8,9}$
Anna Niemiec,$^{10}$
Johan Richard,$^{11}$
Guillaume Mahler,$^{12}$
Jose M. Diego,$^{13}$
\newauthor Pascale Hibon,$^{5}$
Anton M. Koekemoer,$^{14}$
Thomas Connor,$^{15}$
Jean-Paul Kneib,$^{3}$
Andreas L. Faisst$^{16}$
\\
$^{1}$Centre for Extragalactic Astronomy, Department of Physics, Durham University, South Road, Durham DH1 3LE, UK\\
$^{2}$Institute for Computational Cosmology, Department of Physics, Durham University, South Road, Durham DH1 3LE, UK\\
$^{3}$Institute of Physics, Laboratory of Astrophysics, Ecole Polytechnique Fédérale de Lausanne (EPFL), Observatoire de Sauverny, 1290 Versoix, Switzerland\\
$^{4}$Aix Marseille Univ, CNRS, CNES, LAM, Marseille, France\\
$^{5}$ESO Vitacura, Alonso de Córdova 3107,Vitacura, Casilla 19001, Santiago de Chile, Chile\\
$^{6}$Instituto de Estudios Astrofísicos, Facultad de Ingeniería y 455 Ciencias, Universidad Diego Portales, Av. Ejército Libertador 441, Santiago, Chile\\
$^{7}$SKA Observatory, Jodrell Bank, SK11 9FT, UK\\
$^{8}$Astrophysics Research Centre, University of KwaZulu-Natal, Westville Campus, Durban 4041, South Africa\\
$^{9}$School of Mathematics, Statistics \& Computer Science, University of KwaZulu-Natal, Westville Campus, Durban 4041, South Africa\\
$^{10}$Univ. Grenoble Alpes, CNRS, Grenoble INP, LPSC IN2P3, 53, Avenue des Martyrs, 38000 Grenoble, France \\
$^{11}$Univ Lyon, Univ Lyon1, Ens de Lyon, CNRS, Centre de Recherche Astrophysique de Lyon (CRAL) UMR5574, F-69230 Saint-Genis-Laval, France\\
$^{12}$STAR Institute, Quartier Agora - All\'ee du six Ao\^ut, 19c B-4000 Li\`ege, Belgium\\
$^{13}$Instituto de F\'isica de Cantabria (CSIC-UC). Avda. Los Castros s/n. 39005 Santander, Spain \\
$^{14}$Space Telescope Science Institute, 3700 San Martin Dr., Baltimore, MD 21218, USA\\
$^{15}$Center for Astrophysics $\vert$\ Harvard\ \&\ Smithsonian, 60 Garden St., Cambridge, MA 02138, USA\\
$^{16}$Caltech/IPAC, 1200 E. California Blvd. Pasadena, CA 91125, USA
}

\date{Accepted XXX. Received YYY; in original form ZZZ}

\pubyear{2025}

\begin{document}
\label{firstpage}
\pagerange{\pageref{firstpage}--\pageref{lastpage}}
\maketitle

\begin{abstract}
In this two-part series, we present a multi-probe mass modelling method for massive galaxy clusters, designed to disentangle the contributions of individual mass components (Dark matter, intra-cluster gas, stellar masses). In this first paper, we focus on recovering the mass constraint datasets required for the modelling approach introduced in the second paper. Specifically, we measure the light distribution, stellar mass, and kinematics of the cluster members, the brightest cluster galaxy (BCG), and the intra-cluster light (ICL) in Abell S1063. To that end, we developed a new method to extract the light profiles of the cluster members, BCG, and ICL, while accounting for contamination from nearby foreground and background galaxies in \textsc{Hubble Space Telescope} (HST) imaging. We obtained light profiles for $289$ cluster members using a dual Pseudo-Isothermal Elliptical (dPIE) model based on the HST F160W filter, while the BCG \& ICL is modelled as a single component using a multi-Gaussian expansion. To estimate stellar masses and velocity dispersions, we rely on multi-band HST photometry and \textsc{VLT/MUSE} integral field spectroscopy, respectively. Stellar masses are derived using three different spectral energy distribution (SED) models. We measure the line-of-sight velocity dispersions of the cluster members at their half-light radii, as determined from their light profiles, while for the BCG \& ICL components, we use elliptical annular apertures. Thanks to these measurements, we will be able to constrain the cluster stellar mass content, which is detailed in the second paper of the series. We publicly release these measurements with intermediary data products.
\end{abstract}



\section{Introduction}
In the context of the standard $\Lambda$ cold dark matter ($\Lambda$CDM) cosmological model, galaxy clusters are the largest gravitationally bound structures and they are largely dominated by dark matter (DM). In that regard, they are among the best natural laboratories to study DM properties \citep{Clowe2004,Bradac2008,Natarajan2017} and its (possible) deviation from the CDM paradigm \citep{Harvey2015,Meneghetti2020,Limousin2022,Andrade2022,Eckert2022b}. Statistical analyses of their properties allow us to put constraints on DM parameters such as an upper bound on the DM interaction cross-section \citep{Sagunski2020,Andrade2022,Eckert2022b}. The study of outliers in the cluster population, such as the bullet cluster, has provided evidence in favour of the DM paradigm \citep{Clowe2006} while challenging modified theories of gravity \citep{Brownstein2007,Dai2008}.

As DM manifests itself through gravitational interactions, its observational studies in clusters rely on mass modelling methods. Different probes can be used, such as the heating of the intra-cluster gas, allowing for a total mass estimation if the cluster follows the hydrostatic equilibrium \citep{Ettori2013}, or the dynamics of the cluster members, which are floating in the overall cluster gravitational potential well \citep{Mamon2013,Mamon2022}. Among these probes, gravitational lensing is a frontrunner, as it only relies on projection hypotheses instead of cluster relaxation states or symmetry. This phenomenon refers to the bending of light near a massive object, in our case, a galaxy cluster, which permits us to probe the projected gravitational potential. At the core of the cluster, we are in the strong gravitational lensing regime where the background object appears highly distorted to observers, forming giant arcs and/or appearing multiple times in the cluster field. This last phenomenon is referred to as the multiple image phenomenon, which forms the main constraints used by strong lensing modelling methods (see \citet{Kneib2012,Natarajan2024} for reviews on this topic). 

Each of these mass probes can help derive relevant constraints on DM properties by mapping the total mass distribution. Upper limits on the self-interacting dark matter \citep[SIDM]{Spergel2000} cross-section have been derived from hydrostatic analyses of galaxy clusters using X-ray observations and large-scale simulations \citep{Eckert2022b}. \citet{Andrade2022} obtained similar constraints by fitting the lensing profile with a phenomenological SIDM model \citep{Robertson2021}. However, none of these methods accurately measure the DM profile as baryons in cluster members or intra-cluster gas are usually not modelled and accounted for in the total mass budget. This baryonic fraction can reach up to $20$ per cent of the total mass, primarily due to the gas contributions \citep{Bonamigo2018,Beauchesne2024}. Accounting for baryonic contributions is important when comparing observational results with large-scale simulations, as both can converge on the total mass profile while still exhibiting discrepancies in the DM distribution.

To obtain an accurate census of DM, each cluster baryonic component has to be accounted for and thus constrained. Several analyses provided methods to account for the intra-cluster gas in the context of lensing analyses via different optimisation approaches \citep{Sereno2010,Bonamigo2017,Bonamigo2018,Beauchesne2024}. In particular, \citet[][B24 hereafter]{Beauchesne2024} provided a self-consistent framework to combine lensing and X-ray observations in a single process, where the mass model is constrained by a likelihood combining these two sets of constraints. Such a framework can be extended to include other types of constraints, such as the stellar kinematics of the brightest cluster galaxy (BCG). This galaxy sits at the centre of the cluster mass distribution, which lacks lensing features, but can offer valuable constraints on the SIDM cross-section \citep{Sagunski2020}. \citet{Newman2013b} and \citet{Cerny2025} provided a decomposition of DM and baryons in the BCG based on lensing and stellar kinematic constraints, allowing them to report a deviation from the CDM paradigm with a flatter inner DM slope. Combining such an approach with the method used in B24 would allow us to constrain the innermost part of the cluster where the BCG sits, while providing an exhaustive mapping of baryons via the inclusion of the intra-cluster gas. Additionally, intra-cluster stars, from which the intra-cluster light \citep[ICL; see][for a review]{Montes2022} originates, can be used to extend such constraints. 
Similarly to the BCG stellar kinematics, probing the BCG total mass distribution, the same measurement from ICL stars can constrain the cluster total mass profile. Both components are known to present different patterns as the velocity dispersion of the stellar distribution is known to increase from the BCG to the ICL \citep{Longobardi18}. Adding a mass component representing the ICL in a mass model similar to B24, would allow us to disentangle all main mass components at the cluster scale: DM, intra-cluster gas and intra-cluster stars.

In this series of papers, we aim to present a modelling method that accounts for each main mass component of the cluster to fully separate DM from baryonic masses. As state-of-the-art cosmological simulations are now hydrodynamic and thus include baryonic physics, cluster mass models have to be more detailed to properly compare observed DM distributions as well as the interplay with baryons. The main focus of this series is to combine the B24 approach, which includes the intra-cluster gas with \citet{Newman2013b} method extended to account for the ICL, allowing the resulting mass models to include all cluster main mass components. As DM will be disentangled at the cluster scale, we will also detail a new modelling of cluster members, based on a baryonic/DM decomposition, to fully disentangle DM from baryons down to the galaxy scale. We organise this series in two papers, in the first and current one, we present the mass constraints and, in particular, how we recover the supplementary constraints required to map the stellar mass distribution. In the second paper \citep{Beauchesne2025b}, which we refer to as B25b, we focus on the modelling method and the reliability of the stellar mass estimate.

We test this new method on a well-studied and observationally constrained cluster, Abell\,S1063 \citep{Abell1989}. The mass distribution of Abell\,S1063 has been studied multiple times using strong gravitational lensing \citep[][, B24]{Monna2014,Johnson2014,Richard2014,Zitrin2015,Caminha2016,Diego2016,Bonamigo2018,Bergamini2019,Granata2022,Limousin2022,Andrade2022}. Due to its morphology simplicity and wealth of observational data as one of the Hubble Frontier Field clusters \citep{lotz2017}, this cluster has been used as a test case for modelling improvement with the work of \citet{Bonamigo2018,Bergamini2019,Granata2022} and B24. It is also the main motivation behind the present work, in addition to reusing the same modelling of the gas mass as in B24.

In this first paper, we present the test object along with the observational datasets in Sect.~\ref{sect:obs}. In Sect.~\ref{sect:mass-cons}, we present the pre-existing mass constraints, i.e. strong lensing and X-ray observations. In Sect.~\ref{sect:cluster-member-cons}, we detail how we extract cluster member constraints using their light distribution (Sect.~\ref{sect:galaxy-light}), stellar kinematics (Sect.~\ref{sect:Cluster-member-LOSVD}) and stellar masses (Sect.~\ref{sect:cluster-member-stellar-masses}). We then describe the measurements of the same quantities for the BCG and the ICL from Sect.~\ref{sect:BCG-ICL-cons} to Sect.~\ref{sect:ICL-light}, \ref{sect:BCG-Vrms} and \ref{sect:BCG_ICL_M_*}.

Throughout this work, we adopt a flat $\Lambda$CDM cosmology with $\Omega_\Lambda=0.7$, $\Omega_m=0.3$ and $H_0=70$ km~s$^{-1}$Mpc$^{-1}$. Magnitudes are quoted in the AB system. Regarding the statistical treatment of all analyses, we compute uncertainties using median-centred credible intervals ($\rm CI$) based on the posterior distribution of the considered random variable. We refer to the interval containing $100\times \rm erf\left(\frac{n}{\sqrt{2}}\right)$ per cent of the posterior with $n$ an integer by $n\sigma\, \rm CI$. If we use $\sigma$ to denote a standard deviation, we explicitly notify it in the text.

\section{Test object: Abell\,S1063}
\label{sect:obs}

\begin{figure*}
    \centering
    \includegraphics[width=\linewidth]{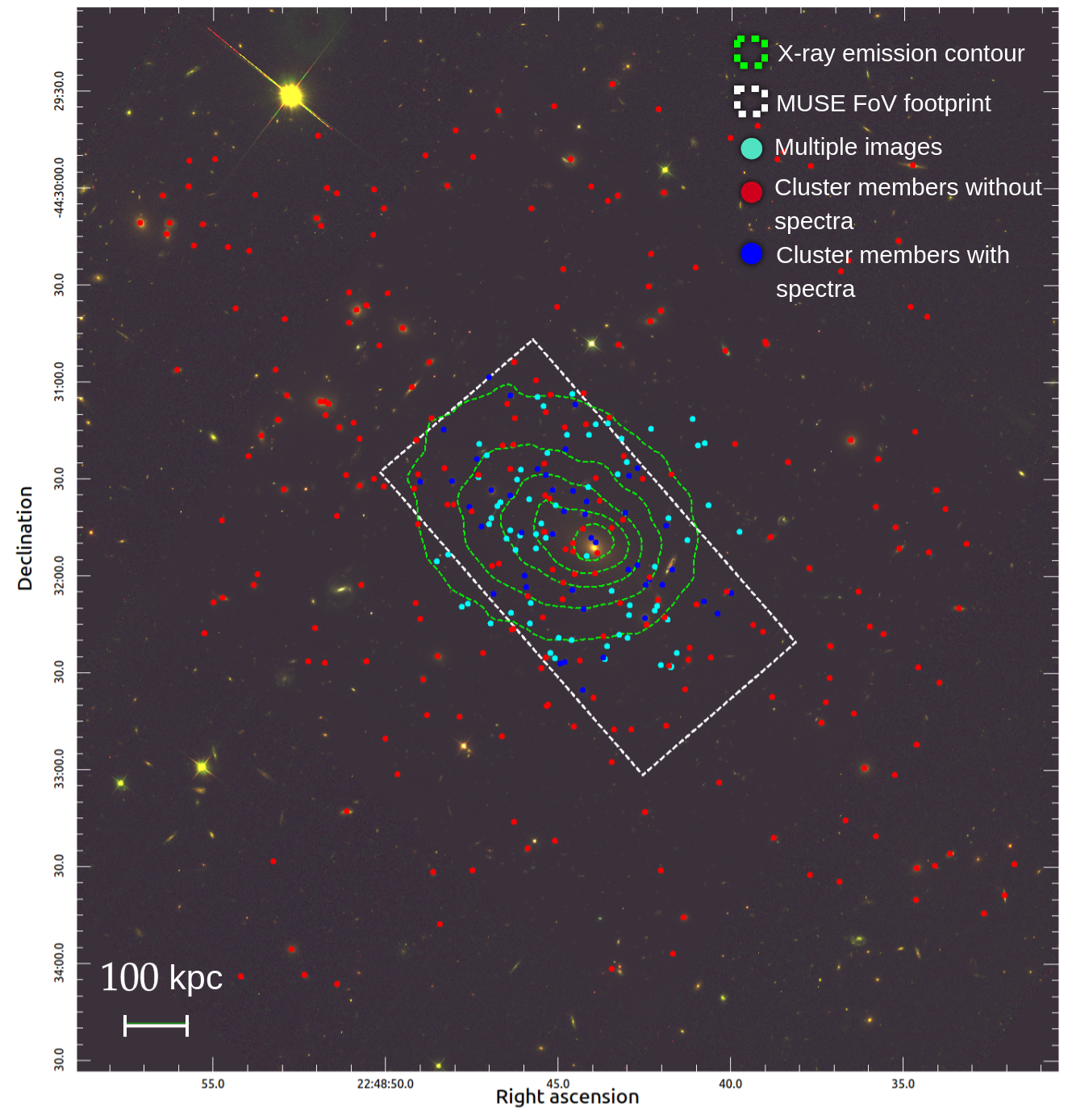}
    \caption{BUFFALO composite colour image of AS1063 with the following \emph{HST} filters: ${\rm F435W}$ (blue), ${\rm F606W}$ (green) and ${\rm F814W}$ (red). The footprint of the combined MUSE observations is highlighted by the white dashed rectangle. The X-ray surface brightness from the \textsc{Chandra X-ray Observatory} is shown by the green dashed contours. The set of multiple images used in this work is highlighted by the cyan dots. Red and blue dots represent spectroscopically confirmed cluster members, the latter colour indicates the availability of velocity dispersions measured from the MUSE datacube.}
    \label{fig:summary_S1063}
\end{figure*}

Galaxy cluster Abell\,S1063 (AS1063; also known as RXC J2248.7-4431 and SPT-CL J2248-4431), is a massive galaxy cluster at $z=0.3475$, first identified by \citet{Abell1989}. It is an ideal laboratory for this work as it has a fairly simple morphology for the amount of archival observations available. The reasoning behind this choice follows that of B24, to which we refer the reader, as the methodology developed in this work builds upon and extends their framework. To set-up the general observational context, we show in Fig.~\ref{fig:summary_S1063} a composite colour image of the cluster made with the \textit{Hubble Space Telescope} (\textit{HST}) broad band filters: ${\rm F435W}$ (blue), ${\rm F606W}$ (green) and ${\rm F814W}$ (red). We present in Fig.~\ref{fig:summary_S1063} the X-ray surface brightness and the footprint of the spectroscopic observations from the Very Large Telescope (VLT)/Multi Unit Spectroscopic Explorer (MUSE) \citep{Bacon2010}. We introduce each observational dataset in the following sections (X-rays in Sect.~\ref{sect:X-ray-observations}, and optical imaging and spectroscopy in Sect.~\ref{sect:photo-cat} and \ref{sect:spectra-cat}).

\subsection{X-ray observations}
\label{sect:X-ray-observations}

We use archival observations from the \textit{Chandra X-ray observatory} of AS1063 as our X-ray datasets. These observations are contained in the \emph{Chandra} Data Collection (CDC) 389~\href{https://doi.org/10.25574/cdc.389}{doi:10.25574/cdc.389}. Their combined observing times account for $123$~ks. Regarding the reduction and analysis of those datasets, we use the same reduction and analysis as B24, and we refer the reader to that reference for the detailed procedure.

\subsection{Photometric data}
\label{sect:photo-cat}

AS1063 has been extensively observed with the \textsc{Hubble Space Telescope} (\textsc{HST}) through multiple programs, including The Cluster Lensing And Supernova survey with Hubble \citep[CLASH;][]{Postman2012}; the Hubble Frontier Fields \citep[HFF;][]{lotz2017}, and the Beyond Ultra-deep Frontier Fields And Legacy Observations \citep[BUFFALO;][]{steinhardt2020}. The bulk of the observations is formed by the HFF dataset that has been spatially extended with the BUFFALO program.

For the following analyses, we use the mosaic images reprocessed by the BUFFALO collaboration, combining all available \textsc{HST} observations in the filters obtained by the HFF program. All the HST exposures from these programs had been recalibrated using the latest reference files, aligned to Gaia-DR3\footnote{\url{https://www.cosmos.esa.int/web/gaia/dr3}}, and combined into mosaics at a pixel scale of 0$\farcs$03, following the techniques first described by \citet{Koekemoer2011}, updated for these new programs. For the complete description of the datasets, we refer the reader to \citet{lotz2017} and \citet{steinhardt2020} for the HFF and BUFFALO programs, respectively. The BUFFALO collaboration has produced weak lensing catalogues, although we are not making use of them in this work. They are presented in the work of \citep{Niemiec2026}.

\subsection{Spectroscopic data}
\label{sect:spectra-cat}

AS1063 was observed by MUSE with two pointings covering the South-West and North-East regions of the cluster, with the following proposal IDs:
60.A-9345(A) (PI: Caputi \& Grillo) and 095.A-0653(A) (PI: Caputi)

We use a new reduction of the MUSE datacube, performed for the analysis of B24. To extend the redshift measurements, we use the complete redshift catalogue from the CLASH-VLT program (ESO ID: 186.A-0798; PI: P. Rosati) produced by \citet{Mercurio2021} within the BUFFALO footprint.

\section{Existing mass constraints}
\label{sect:mass-cons}

As we aim to disentangle DM from baryons, we need to gather several sets of constraints to probe the total mass as well as each baryonic component. 
In this section, we present the datasets adopted in this analysis, originally obtained in B24, along with a brief description of their derivation.

\subsection{Lensing constraint}
\label{sect:lens-cons}
We constrain the total mass using strong gravitational lensing, in particular, multiple images. We adopt the same constraints as in B24 and refer the reader to that work for a detailed description. This set comprises the positions of $67$ multiple images distributed over $25$ systems. All systems have spectroscopic confirmation, ensuring high-quality lensing constraints. The position of each image is presented in Fig.~\ref{fig:summary_S1063}.

\subsection{X-ray analysis}
\label{sect:X-ray-cons}
The X-ray properties and associated ICM maps adopted in this analysis are drawn from B24; we refer the reader to that work for full methodological details, which we briefly outline below.

To model the ICM mass, we rely on its X-ray emission. We use a combination of surface brightness and spectral data. The former is accessed via maps of photon counts, while the latter provides the conversion factor from surface brightness to gas mass. We binned the cluster field with the Python package \textsc{VorBin} \citep{Cappellari2003} in the $[4,7]$ keV band, with a signal-to-noise threshold of $10$. For each of these cells, we extract the spectra. Those spectra are fitted with the Astrophysical Plasma Emission Code (APEC)\footnote{\url{http://atomdb.org/}} model combined with a photoelectric-absorption (PHABS)\footnote{\url{https://heasarc.gsfc.nasa.gov/docs/xanadu/xspec/manual/XSmodelPhabs.html}} model to account for the foreground galactic gas. Hence, we recovered the temperature and metallicity of the ICM, which allowed us to compute the conversion factor from photon counts to gas mass.

Our surface brightness estimation is based on the count maps in the soft, medium and hard \textit{Chandra} science energy bands \citep{Evans2010} that are combined with their associated exposure maps. To estimate the background emission, we recover the \textit{blank-sky} observations associated with each observation.

\section{Supplementary mass constraint 1: cluster members}
\label{sect:cluster-member-cons}
To model the cluster member mass distribution and, in particular, separate its baryonic and DM counterparts, we rely on photometric and spectroscopic measurements. To model the stellar mass distribution, we fit the light profiles with a method that we tailored towards crowded fields such as galaxy clusters, presented in Sect.~\ref{sect:galaxy-light}. Some of the parameters of these light profiles are also used in the modelling of their DM distribution. As the light profile offers only a 2D representation, a scaling factor must be introduced to connect it to the underlying DM and baryonic mass distributions. We address this in Sect.~\ref{sect:Cluster-member-LOSVD} where we measure the line-of-sight velocity dispersions (LOSVD) to probe the total mass, and in Sect.~\ref{sect:cluster-member-stellar-masses}, where we recover stellar masses from spectral energy density (SED) fits. 

All these measurements form the basis of our cluster member modelling, which is detailed in B25b (sections~2.2 and 3.3). In particular, section~2.2 presents the modelling assumptions used to define the cluster member DM and stellar mass distribution models. In section~3.3, we detail how we specifically constrain the total mass of cluster members using their stellar kinematics. We publicly release each of these measurements and some intermediary data products as presented in Sect.~\ref{sect:data-availability}. For the cluster member, we provide light profile parameters, LOSVD, stellar mass estimates, the empirical Point Spread Function (PSF) and galaxy masks used in Sect.~\ref{sect:galaxy-light}.

\subsection{Light profiles}
\label{sect:galaxy-light}

We aim to improve the modelling of the cluster members by accounting for their baryonic and DM masses using a single mass profile for each, which requires extracting sufficient information to characterise the baryonic contribution. The simplest approximation of baryonic mass is to assume that it is proportional to the cluster members light distributions. Hence, in this section, we focus on extracting the light profile from imaging data.

In the following, we define cluster members as all objects which have a spectroscopic redshift in the $[0.30905,0.3716]$ range, as measured by \citet{Mercurio2021} from the complete CLASH-VLT data on AS1063. We made the selection based on spectroscopic catalogues presented in Sect.~\ref{sect:spectra-cat}. We do not consider photometrically selected cluster members, as the CLASH-VLT data cover the full FoV of the BUFFALO mosaic.

\begin{figure*}
    \centering
    \includegraphics[width=0.8\linewidth]{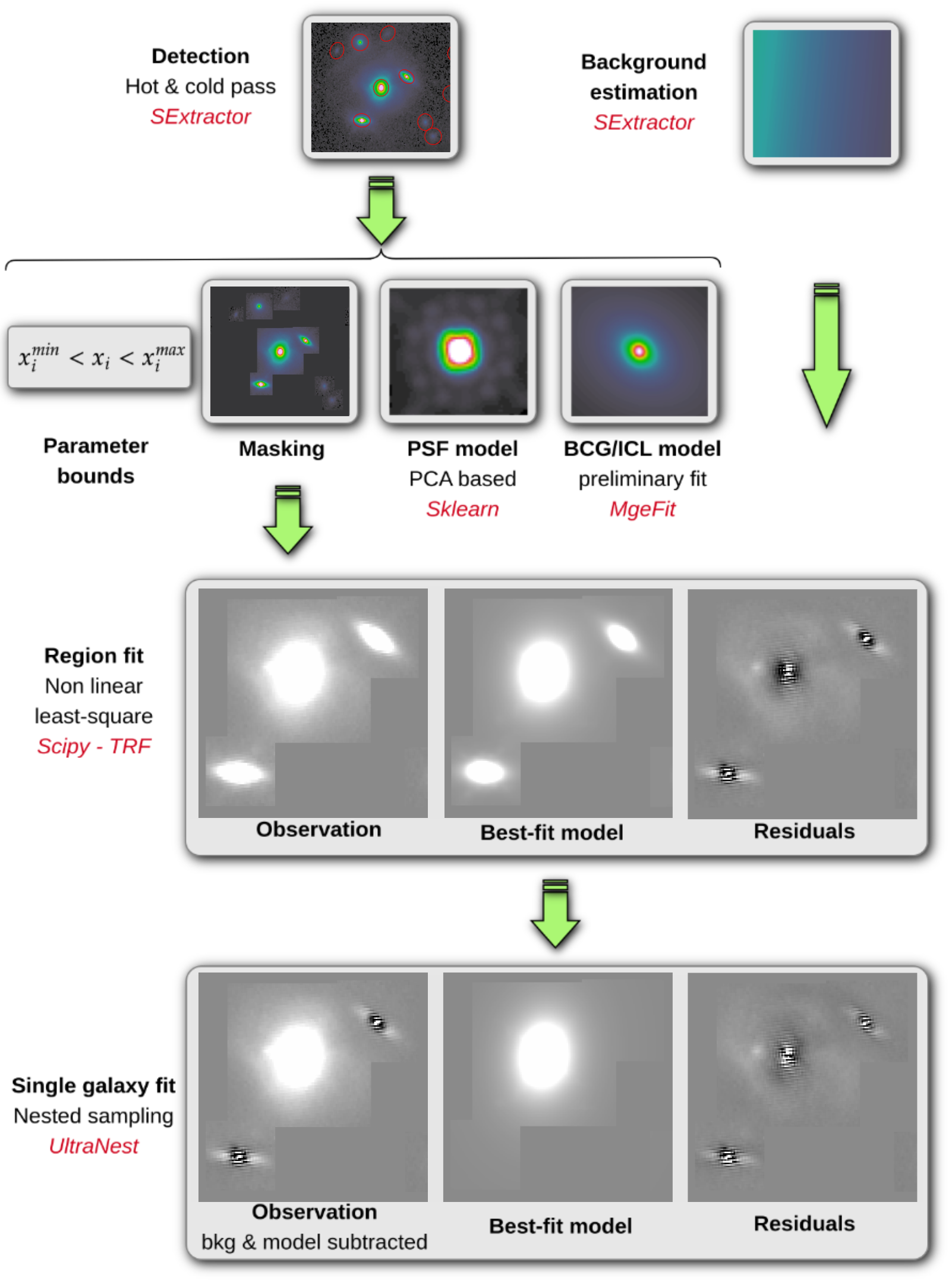}
    \caption{Diagram of the workflow to fit the cluster member light distribution. The first step is split between estimating the background and building the galaxy catalogue. We extract the necessary quantities from this step to proceed to the first fitting step: PSF estimation, masking, parameter bounds, and preliminary ICL estimation. The second step is based on fitting regions that include multiple galaxies. Regions are defined from the galaxy catalogue estimation of galaxy sizes. The final step consists of a Bayesian inference for each galaxy, where the influence of neighbouring objects is accounted for with the model computed in the previous step.}
    \label{fig:galaxy-fitting-workflow}
\end{figure*}

As cluster fields are particularly crowded, we need a method to extract light profiles while taking into account other galaxies in the field. \citet{tortorelli2018} showed that light profile measurements done independently for each object have an increasing bias toward the cluster centre as the field becomes increasingly crowded. They relied on the \textsc{Morphofit} \citep{tortorelli2023} method to perform 2D profile fitting in a cluster field. However, this method is not satisfactory for our goals as it does not provide profiles that can easily be included in lensing mass models, and it requires prohibitive computing costs. Hence, we developed our own method, which minimises the computing cost of such analysis and takes benefits from modern hardware accelerators such as GPU through the \textsc{JAX} python package \citep{JAX}. A scheme of the method workflow is presented in Fig.~\ref{fig:galaxy-fitting-workflow} and highlights the different steps of the process that we detail here: object detection (Sect.~\ref{sect:galaxy-light:detect}), region fit (Sect.~\ref{sect:galaxy-light:region-fit}) and individual Bayesian optimisation (Sect.~\ref{sect:galaxy-light:bayes}). In particular, this method aims at measuring the profiles of cluster members only. However, we use the light model, including cluster and non-cluster galaxies, to extract the ICL light distribution through a separate analysis presented in Sect.~\ref{sect:ICL-light}.

\subsubsection{Object detection and fitting setup}
\label{sect:galaxy-light:detect}

To set up our fitting process, use a similar approach as \textsc{Galapagos} or \textsc{Morphofit} to detect objects by using \textsc{SExtractor} \citep{Bertin1996}. Following a standard procedure, we start by performing a \textit{hot} and \textit{cold} pass to detect objects \citep{Rix2004}. We provide in Appendix~\ref{app:sextractor_param} the \textsc{SExtractor} parameters that we use for this work. We concatenate the two catalogues by excluding hot components for which the centre is within the \textsc{KRON\_RADIUS} of the objects detected by the cold pass. In our particular case, we aim at measuring the profile of galaxies in different bands; thus, we use the double image mode and choose as our base detection image the \textit{HST} filter ${\rm F105W}$. This filter provides a good balance between the depth and FoV of the BUFFALO data, as well as the number of detectable objects.

We are also relying on \textsc{SExtractor} to estimate the background emission. We here use the same parameters as one of the previous pass with the exception of \textsc{BACK\_SIZE} parameters that we fix to $3096$ pixels to smooth out the effect of the ICL. We cannot use the background produced by the initial passes as a significant part of the ICL is included in it. With such a large \textsc{BACK\_SIZE}, we limit the leaking of the ICL light distribution in the background estimation, but that set of parameters is not well-suited to detect objects in the image. Hence, we run \textsc{SExtractor} three times in total, a cold and a hot pass with the parameter presented in Appendix~\ref{app:sextractor_param}, and we run a single pass with the large \textsc{BACK\_SIZE} where we only aim at estimating the background. The background estimation being independent of the source detection in \textsc{SExtractor}, that last pass can be a cold or hot one, as they produce identical backgrounds. 

To estimate a PSF for each band of interest, we use a Principal Component Analysis (PCA) based method \citep{tortorelli2023,Amara2012} similar to the one implemented in \textsc{Morphofit}. The more stars are included, the more accurate and complex such a PSF will be \citep{Herbel2018}. As an empirical method, it naturally accounts for effects due to the reduction of the raw data to the final mosaic, such as drizzling. In preliminary analysis, this method was superior to a stack of unsaturated stars once we used more than $\sim 10$ stars, as it presented fewer artefacts. To compute such PSF, we extract postage stamp images of unsaturated stars of size $4.2\times4.2$ arcsec, representing $70$ pixels at the resolution of our drizzled images. These postage stamp images are then background-subtracted. We estimate the background using a sigma-clipped median of the postage stamp with a $5\sigma$ limit. We finish the preparation of the postage stamp images by normalising them. We perform a PCA decomposition as implemented in the \textsc{Scikit-learn} package \citep{scikit-learn} on the array of the flattened postage stamp images. We select the number of PCA components by taking the minimal number of components, which explains $95$ per cent of the data variance. We finish our estimation by subtracting the sigma-clipped median for the individual stamps and normalising the final PSF. We use a single PSF for the whole field. We manually select the unsaturated stars in each band in the field with the \textsc{FLUX\_RADIUS}-\textsc{MAG\_AUTO} plane, where stars distinguish themselves as they present similar radii with varying magnitudes.

\subsubsection{Region fit}
\label{sect:galaxy-light:region-fit}

To deal with the crowded field of the cluster, we start with a crude model of the BCG and the ICL using a Multi-Gaussian Expansion \citep[MGE;][]{Cappellari2002} fit with the python package \textit{MgeFit}. To capture the light distribution, Gaussians only share the same centre and are fitted on twisted isophotes. We restrict the fit to an elliptical aperture following the position angle of the BCG as fitted by \textsc{SExtractor} and a major and minor axis of $150$~kpc and $130$~kpc, respectively. We restrict our analysis to the brightest part of ICL, as our procedure includes only a simple estimation of the background. Hence, we preferred to focus on its brighter part, where the light profile estimation is the most robust. We mask all galaxies in this aperture. We find that the fit robustness is improved by removing the BCG and the ICL before fitting the other galaxies rather than accounting for them with elliptical profiles as in \textsc{Morphofit}.

We continue our fitting procedure by defining the fitting regions. All galaxies are fitted (cluster members and other field galaxies) in a square based on a radius of four times their half-light radii, as estimated by \textsc{FLUX\_RADIUS} measurements. We rescale \textsc{FLUX\_RADIUS} values by $(1-\epsilon)^{-1}$, with $\epsilon$ being the ellipticity of the considered object. This is the equivalent of rescaling that radius to the major axis of the associated ellipse. It is comparable but slightly smaller than the value used by \textsc{Morphofit} as it considers a radius of five half-light radii instead \citep{tortorelli2023}. We define a region for each contiguous area of overlapping postage stamps associated with individual galaxies. In addition to these regions, we manually mask objects such as stars and their flares as well as galaxies with morphologies too complex for a singular elliptical profile, such as lensing arcs. However, we retain all cluster members as their light profiles are required by the mass model.

For each object except the BCG, we either fit a dual Pseudo Isothermal Elliptical model \citep[dPIE;][]{Limousin2005} for cluster members or a S\'ersic \citep{sersic1963} profile for field galaxies combined with a convolution with the empirical PSF defined in sect.~\ref{sect:galaxy-light}. That way, the dPIE light profiles can be added to the mass model easily, as a dPIE has an analytical expression for all its lensing quantities \citep{Kassiola1993}. Another parametrisation of the dPIE profile is known as the Chameleon profile, which differs in how the core and cut radii are defined. This dPIE variant has been tested against S\'ersic by \citet{Dutton2011}. Depending on the S\'ersic parameters, dPIE can struggle to reproduce the core or the outskirts of a galaxy. Hence, we expect the dPIE residuals to be larger at the core than with a traditional S\'ersic fit. We re-parameterise these potentials to ease the numerical optimisation. The profile expressions with the re-parametrisation are detailed in Appendix~\ref{app:light_model}. We detail the specific treatment of the BCG and ICL in Sect.~\ref{sect:ICL-light} which are fitted by a MGE.

We split the fitting process into two steps. First, we fit each region independently for $250\times\lfloor {\rm Nb}_{\rm pot}/2\rfloor$ iterations, where ${\rm Nb}_{\rm pot}$ is the number of profiles fitted simultaneously in the considered postage stamp. These fits can be performed in parallel and aim to establish a satisfactory model in each region before accounting for neighbouring regions.

In the second step, we fit each region sequentially for $50\times\lfloor {\rm Nb}_{\rm pot}/2\rfloor$ iterations and repeat until convergence. Here, we subtract the global model from neighbouring regions before optimizing each region. This two-step approach avoids poor models in neighbouring regions from negatively impacting the current fit. Ideally, we would perform only one iteration per region to fully account for fit evolution across all regions, but this generates significant computational overhead, so we increase the iteration count for efficiency.

To order the sequential fits, we compute a weight for each region from the luminosity of its objects. We construct a Gaussian kernel density estimator using galaxy positions, weighted by their magnitude, then integrate this density over each region to obtain the final weight. The regions are then ordered from the highest weight to the lowest, which is equivalent to processing the regions from the one containing the brightest to the faintest objects.

Preliminary tests showed that applying the least square fit to the regular weighted residual $\frac{f_{\rm obs,i}-f_{\rm model,i}}{\sigma_i}$ is not the most robust function to minimise. We improve the stability of our fit by transforming the data and model with the $\sinh^{-1}$ function, and propagating the error with the transformation derivative. It leads to the following formula for the $\chi^2$ in transformed space:
\begin{equation}
    \chi^2_{\rm transformed}=\left(\frac{\left(\sinh^{-1}(f_{\rm obs,i})-\sinh^{-1}(f_{\rm model,i})\right)\sqrt{1+f_{\rm obs,i}^2}}{\sigma_i}\right)^2
\end{equation}
As the $\sinh^{-1}$ function evolve as $x$ close to $x=0$ and $\ln{2x}$ as $x$ become larger than $1$. We obtain the $\chi^2$ in the normal space for $f_{\rm obs,i},f_{\rm model,i}<<1$. If $f_{\rm obs,i},f_{\rm model,i}>>1$, it creates a log compression of the flux dynamic range and a different pixel weighting scheme as shown by its expression:
\begin{equation}
    \chi^2_{\rm transformed}\approx\left(\frac{\left(\ln(f_{\rm obs,i})-\ln(f_{\rm model,i})\right)|f_{\rm obs,i}|}{\sigma_i}\right)^2
\end{equation}
During the optimisation as $f_{\rm obs,i}$ is fixed, it compresses the dynamic range of $f_{\rm model,i}$ and smooth the likelihood landscape for bad models. Specifically for the least-square algorithm, the gradient became normalised when $f_{\rm model,i}$ becomes large, which prevents the optimiser from moving rapidly in one direction. A drawback of that transformation is that it creates an asymmetry of $\chi^2_{\rm transformed}$ with respect to $f_{\rm obs,i}$ and $f_{\rm model,i}$, which biases parameter estimates toward higher flux values. We present robustness tests in Appendix~\ref{app:Likelihood_checks} to show the benefit of this likelihood against the regular one. The uncertainty of each pixel $\sigma_i$ is estimated in a similar way to \textsc{Galfit} and \textsc{Morphofit} (see equation~$11$ from \citet{tortorelli2023}).

The $\sinh^{-1}$ function transition from a linear to a logarithmic behaviour happens around $1$, which fits well in our case as most of the pixel have a flux inferior to $1\,\rm{e^{-1}/s}$, thus only affecting the bright pixels. Other datasets may need a rescaling of the flux to adapt the transformation transition to their flux distribution.

We oversample the model to obtain a more accurate estimate of the flux within each observed pixel and to reduce numerical artefacts. We define a square region centred on each fitted galaxy, extending $\pm15$ pixels from the galaxy centre in both directions (i.e. a region 30 pixels on a side), where we use an oversampling factor of 3. We define this region around each galaxy centre as measured by \textsc{SExtractor}. Thus, these regions remain fixed throughout the optimisation. We use a large oversampling area to ensure the profile centre remains within it during optimisation.

We use the bounded version of the trust region reflective algorithm implemented in \textsc{Scipy} as our non-linear least square optimiser\footnote{\url{https://docs.scipy.org/doc/scipy/reference/generated/scipy.optimise.least_squares.html}}. We translate the original code to the \textsc{JAX} framework to allow the code to run on hardware accelerators. We define the bounds and the starting points of each parameter based on \textsc{SExtractor} measurements. We present the details of this procedure in Appendix~\ref{app:init_param_LM}.

\subsubsection{Independent Bayesian fit}
\label{sect:galaxy-light:bayes}

\begin{figure*}
    \vspace{-0.5cm}
    \centering
    \includegraphics[width=0.68\linewidth]{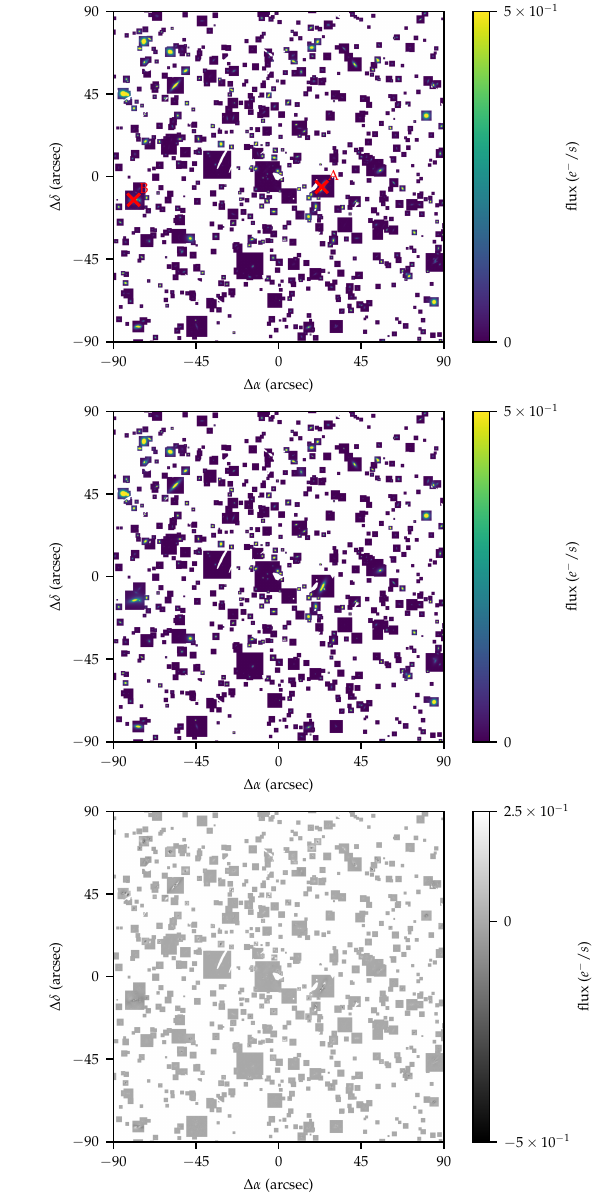}
    \vspace{-0.4cm}
    \caption{\textit{Region of $180^{"}\times180^{"}$ around the centre of AS1063:} Original (\textit{top}), model (\textit{middle}) and residual (\textit{bottom}) from the galaxy fitting procedure for the ${\rm F160W}$ filter.}
    \label{fig:galaxy-fitting-ICL-res}
\end{figure*}

To have a final estimation of the galaxy model, we use the Monte Carlo nested sampling algorithm \textsc{MLFriends} \citep{Buchner2014,Buchner2019} from the \textsc{UltraNest}\footnote{\url{https://johannesbuchner.github.io/UltraNest/}} package \citep{Buchner2021}. By performing this optimisation galaxy per galaxy, we effectively neglect the correlation between parameters from different objects. However, for most of them, we expect these covariances to be lower than those between parameters within the same profile. That way, we can estimate our model posterior distribution while avoiding the curse of dimensionality. 

We define the fitting regions with the same individual squares for each galaxy as in the previous step, but we do not merge them into larger postage stamps. We subtract the previously estimated global model and background sky in each region, which accounts for neighbouring objects. We use the parameter estimation from the first step to resize the large bounds for computational efficiency in the second step. These resized bounds then serve as intervals for uniform priors, but only for parameters associated with profile normalisation or specific radii (see Appendix~\ref{app:init_param_Bayes}). For other parameters, we found that using the best-fitting estimation does not yield efficient priors.

\begin{figure}
    \centering
    \includegraphics[width=\linewidth]{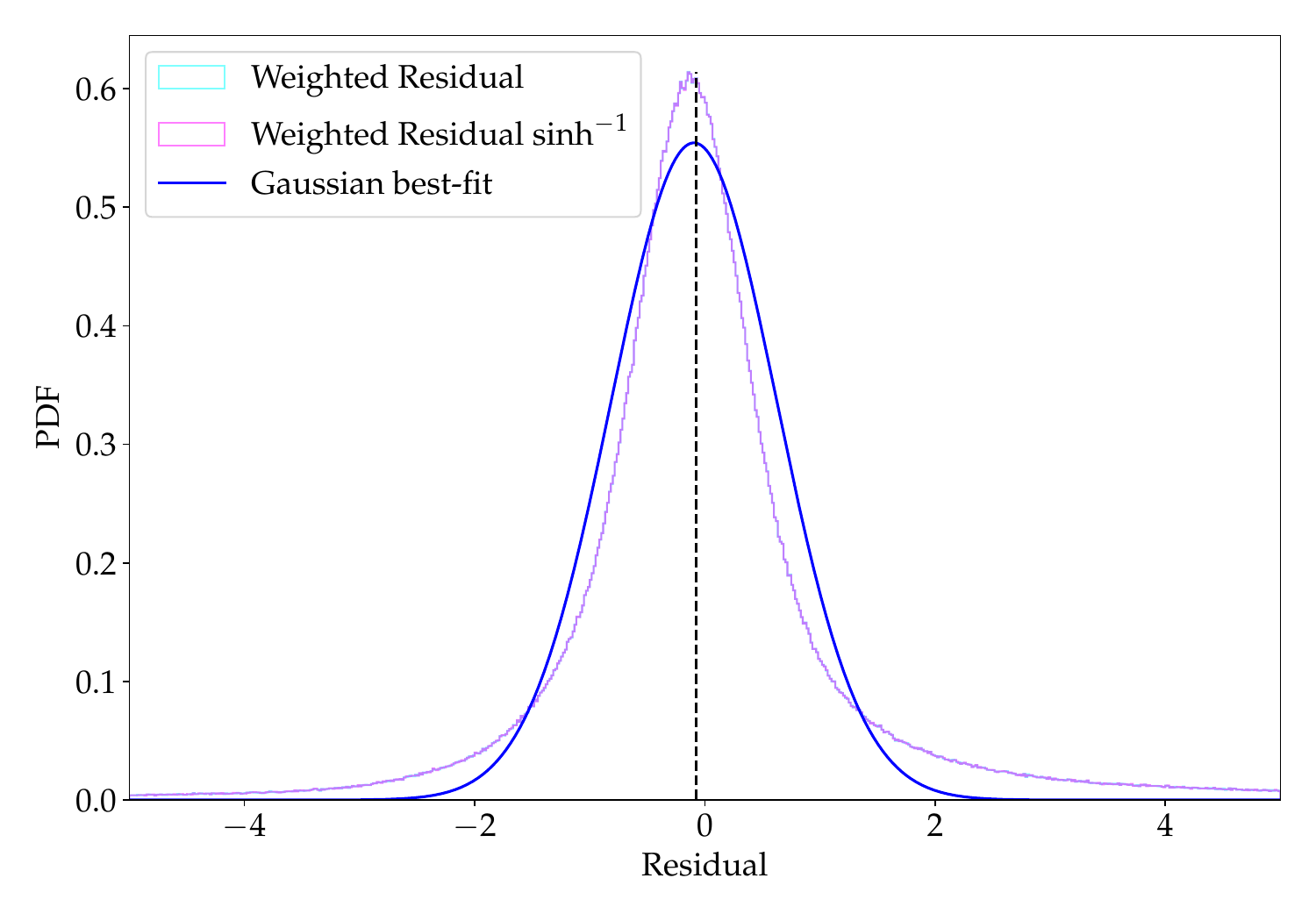}
    \caption{Distribution of the weighted residuals with and without $\sinh^{-1}$ transform of the galaxy fit procedure for the filter ${\rm F160W}$. This distribution is extracted from the masked area where galaxy fits are performed. The residuals without the transform are presented with the cyan plain line while the ones with the $\sinh^{-1}$ transform are in magenta. Both distributions are almost indistinguishable in the range presented in the figure. Plain blue line represent the best-fitting results of a Gaussian distribution. That fit has been performed with the weighted residual in the $[-2,2]$ range. The dashed black line shows the median of the weighted residuals.}
    \label{fig:galaxy-fitting-res-laplace}
\end{figure}

Fig.~\ref{fig:galaxy-fitting-ICL-res} presents the results of our fitting procedure in the \textit{HST} image considering the ${\rm F160W}$ filter. The top panel presents the \textit{HST} image where the non-fitted areas are masked, the middle and bottom panels show the output model and residuals, respectively. Our procedure is able to capture globally the shapes of the fitted galaxies in the field of AS1063, as shown in the output model. However, residuals highlight the limitation of our approach and of the chosen galaxy model. Of the $1566$ fitted galaxies, we rejected the results of $63$ fits that did not reproduce the data well, all concerning field galaxies and not cluster members. Besides galaxies which significantly deviate from elliptical symmetry, it is usually possible to obtain a good fit by manually tuning priors, model subtraction and masking from neighbouring objects, which we only did for the cluster member.
We notice an overestimation of the flux around some galaxies in the residual maps, like the two objects highlighted by letters 'A' and 'B'. These galaxies exhibit a complex morphology with spiral features that cannot be accounted for by S\'ersic or dPIE profiles. Many galaxies also exhibit an over/underestimation pattern in their central part or outskirts, though this pattern is weaker than observed in spiral galaxies. These galaxies may be more accurately fitted with two profiles accounting for a bulge-disk decomposition.

Regarding the residuals, we present their distribution weighted by the inverse of the data error in Fig.~\ref{fig:galaxy-fitting-res-laplace} with and without the $\sinh^{-1}$ transform in the $[-5,5]$ interval. As one can see, both residual distributions are indistinguishable from each other. The differences between the two likelihood schemes appear in the tail of the distributions. This comparison is presented in Appendix~\ref{app:Likelihood_checks} with the likelihood robustness test. The median of the weighted residuals is $-0.078$ as shown by the black dashed line. As this median is computed on the residuals divided by the observational uncertainty, we can deduce that this overestimation is small compared to that error, representing only $7.8$ per cent. The width of the distribution is measured with the median absolute deviation (MAD), scaled to be equivalent to a Gaussian standard deviation. We obtain a value of $0.769$. Hence, according to the MAD, our models capture galaxy profiles within the observational errors with a slight over-fit, as in the ideal case we expect a value of $\sim1$.

In the ideal case of a Gaussian likelihood, we would expect the residuals to follow a normal distribution. However, the residuals distribution differs as one can see in Fig.~\ref{fig:galaxy-fitting-res-laplace}. We overlaid the best-fitting results of residuals by a Gaussian distribution fitted on the $[-2,2]$ range. Our distribution presents a narrower width but larger tails than a Gaussian distribution, which indicates that our galaxy light profiles are not able to fully capture the complexity of observed profiles. It is expected as we only use a single elliptical light profile for each galaxy and do not correct the residual by adding an intrinsic error to the fit.

\subsubsection{Caveats}

The goal of our procedure is to obtain an estimation of the best dPIE parameters for the light profiles of each cluster member. We aim at including them in our mass model presented in B25b to trace their baryonic distributions. We only include the best-fit parameter in the model. dPIE profiles struggle to reproduce a wide range of S\'ersic profiles \citep{Dutton2011}, particularly in the core, where a dPIE tends to be flatter. Hence, we do not expect to have light profile estimation as accurate as a standard analysis with S\'ersic. The likelihood transform implies a bias towards an overestimation of the bright pixels flux, which will be more present in the uncertainty than the best-fit model. This bias would probably be more pronounced if we only fitted the S\'ersic profile as it can yield a much steeper slope in the core.

The postage stamps model and residual of every dPIE fitted are presented in the supplementary materials. They do not perfectly fit the cluster member light profiles with the most discrepancies in the core estimation, as we could expect. In our use case, it is sufficient as the exact shape of those profiles does not have much effect on the final mass models.

\subsection{Line-of-sight velocity dispersions}
\label{sect:Cluster-member-LOSVD}
To probe the cluster member total mass, we use the line-of-sight velocity dispersions (LOSVD), which can be expressed through their light and mass distributions (see B25b, section~3.3). For this task, we use the same method as in B24 with the MUSE datacube presented in Sect.~\ref{sect:spectra-cat}, which we briefly describe in this section. In contrast to B24, we use our own measurement of the cluster member light profile to define extraction regions of the spectra, where we use dPIE instead of S\'ersic profiles.

Galaxy spectra are extracted in elliptical apertures of circularised radius, $R_e$ (i.e. half-light radius). Ellipticity and $R_e$ estimations are based on dPIE fits performed in Sect.~\ref{sect:galaxy-light} using the ${\rm F160W}$ image. We used the Python package \textsc{MPDAF} \citep{Bacon2016} to extract spectra, which uses an optimal extraction algorithm for CCD spectroscopy \citep{horne1986}. We use the estimation of the local background from \textsc{MPDAF}, which we subtract from the galaxy spectra. These spectra are restricted to $485$~nm-$716$~nm in the observer frame, similarly to \citep{Bergamini2019}. We exclude the reddest part of the MUSE wavelength range as it has lower S/N due to the MUSE sensitivity curve, and is affected by stronger sky subtraction residuals.

The LOSVD measurements are done with the python package \textsc{pPXF} \citep{Cappellari2017} combined with the nested sampling engine \textit{pyMultiNest} \citep{multinest,pymultinest} as the non-linear optimiser method. To model spectra, we use spectral templates from the Indo-US Library of Coudé Feed Stellar Spectra \citep{valdes2004}, which have a full width half maximum (FWHM) of $1.35$~$\mathring{\rm A}$, and a pixel-scale of $0.44$~$\mathring{\rm A}$ pixel$^{-1}$. We parametrize the galaxy kinematics with the velocity, $V$, the velocity dispersion, $\sigma_e$, and the first two Hermite moments, ($h_3,\,h_4$). We add a multiplicative Legendre polynomial of degree $3$ to improve the continuum. The reliability of this \textsc{pPXF} configuration has been tested in B24 with $10000$ mock spectra with a stellar population and observational noise tuned to represent AS1063 observations.

The optimisation is done in two steps. We start by a fit with a Trust Region Reflective algorithm implemented in \textsc{Scipy} \citep{2020SciPy-NMeth} to detect outlier points automatically. We mask these points during the nested sampling runs, providing us with the final estimation of $\sigma_e$. We apply the polynomial correction defined in Appendix~B from B24 with the updated coefficients presented in \citet{Beauchesne2024-erratum}. We use these fits to constrain the cluster members scaling relations, for which we add a scatter to represent the inability of such a relation to reproduce exactly each galaxy properties. Hence, we do not rescale the error to include systematics, however, we include it in the overall scatter of the relation. On average, this scatter is one order of magnitude larger than the total error of $\sigma_e$.

\subsection{Stellar masses}
\label{sect:cluster-member-stellar-masses}

To be able to successfully disentangle baryons and DM contents in cluster members, we need insights on their stellar masses. Indeed, stellar kinematics is sensitive to the light distribution and total mass but not directly to the stellar-mass-to-light ratio ($\Upsilon_*=M_*/L$). To obtain such measurements, we fit the SED of cluster galaxies. However, to account for biases in the SED fitting, we use three different SED models. We use the computing service proposed by GAZPAR\footnote{\url{https://gazpar.lam.fr/home}} operated by CeSAM-LAM and IAP, to run the software \textsc{LePhare} \citep{Arnouts2013} with the SED model presented in \citet{Ilbert2015}. The SED model is based on the stellar library presented in \citet{Bruzual2003}, the \citet{Chabrier2003} Initial Mass Function (IMF), and as a star formation history (SFH), a combination of two delayed-tau models and an exponentially declining model. In addition, we use the Python package \textsc{Bagpipes} \citep{Carnall18,Carnall19} with two similar SED models with a different SFH. Both models use the stellar population model from the 2016 updated version of the \citet{Bruzual2003} models using the MILES stellar spectral library \citep{Falcon-Barros2011}. The IMF is based on \citet{Kroupa2001}, and we assume a dust attenuation based on the Calzetti law \citep{Calzetti2000}. We use the same parametrisation as \citet{Carnall18} for the nebular emission. For the SFH model, we use a delayed-tau SFH model or a double power-law parametrisation. For the \textsc{Bagpipes} models, the SED model priors are presented in Table~\ref{table:SED-param}. We note that the IMFs of \textsc{LePhare} and \textsc{Bagpipes} are slightly different, with the \citet{Kroupa2001} IMF providing stellar mass estimates $6$ per cent higher than \citet{Chabrier2003} ones \citep[e.g.][]{Speagle2014}. However, that difference is negligible in our case as it is well dominated by the measurement uncertainties, which are on average around $30$ per cent for the three SED configurations.

\begin{table*}
\caption{Priors of the SED model parameters used by \textsc{Bagpipes}. $\mathcal{N}$, $\mathcal{U}$, $\mathcal{N}_{\rm trunc}(a,b)$ refers to a normal, uniform and truncated normal distribution with $a$ and $b$ the bounds of the truncated normal. For $\mathcal{N}$ and $\mathcal{N}_{\rm trunc}(a,b)$, prior parameters correspond to the mean and standard deviation of the distribution, while for $\mathcal{U}$, it is its bounds.}             
\label{table:SED-param}      
\centering                          
\begin{tabular}{c c c}        
\hline\hline                 
Parameter  & Prior parameters & Prior type\\
\hline                        
Redshift & $(z_{\rm measured},0.001)$ & $\mathcal{N}$\\
Log stellar mass formed & $(1,15)$ $\log M_\odot$ &  $\mathcal{U}$\\
stellar metallicity & $(0,2.5)$ $Z_\odot$ &  $\mathcal{U}$\\
V-band attenuation & $(0,2.)$ mag &  $\mathcal{U}$\\
Age since formation & $(0.1,t(z_{obs}))$ $Gyr$ & $\mathcal{U}$\\

\hline 
Delayed SFH: $\tau$ & $(0.3,10)$ $Gyr$ & $\mathcal{U}$\\
\hline                                   
Double Power-law SFH: $\tau$ & $(0.,15)$ $Gyr$ & $\mathcal{U}$\\
Double Power-law SFH: $\log_{10}\alpha$ & $(-2,3)$ & $\mathcal{U}$\\
Double Power-law SFH: $\log_{10}\beta$ & $(-2,3)$ & $\mathcal{U}$ \\
\hline 
Calibration: $\log_{10}\sigma_e$ & $(2.30,2.84)$ $\log_{10}km/s$ & $\mathcal{U}$ \\
Calibration: $P_0$ & $(1,0.25)$ & $\mathcal{N}_{\rm trunc}(0.5,2.5)$\\
Calibration: $P_1$ & $(0,0.25)$ & $\mathcal{N}_{\rm trunc}(-0.5,0.5)$\\
Calibration: $P_2$ & $(0,0.25)$ & $\mathcal{N}_{\rm trunc}(-0.5,0.5)$\\
White noise: $\log_{10}$ scaling & $(0,1.70)$ & $\mathcal{U}$ \\
\hline\hline   
\end{tabular}
\end{table*}

To perform the cluster member SED fits, we rely on the aperture photometry measured by \textsc{Sextractor} in the seven \textsc{HST} filters: ${\rm F435W}$, ${\rm F606W}$, ${\rm F814W}$, ${\rm F105W}$, ${\rm F125W}$, ${\rm F140W}$ and ${\rm F160W}$. As not all bands have the same FoV, we do not have photometry for all galaxies in all bands, particularly as their distance from the centre increases.

\begin{figure*}
    
    \centering
    \includegraphics[width=\linewidth]{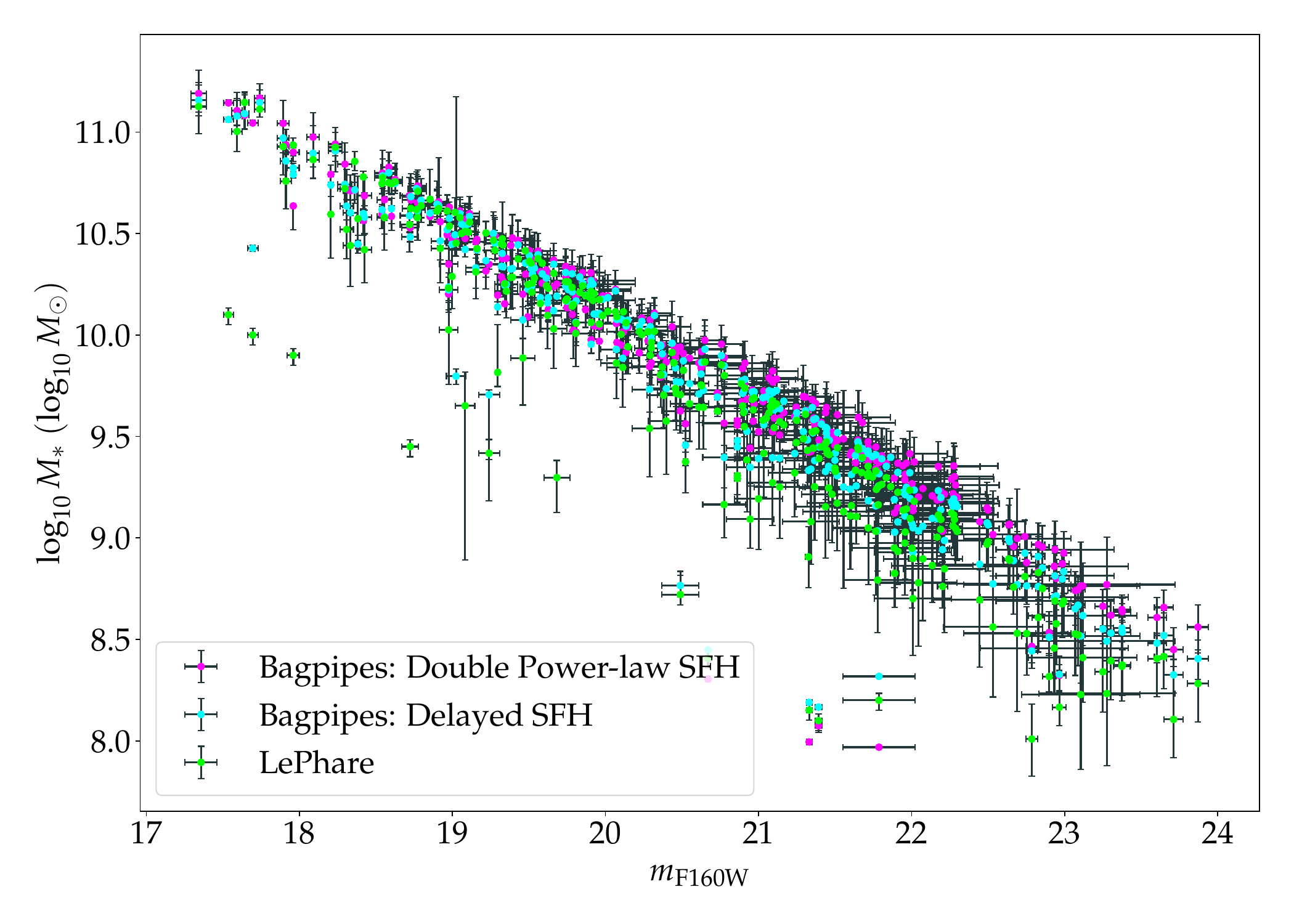}
    \caption{Cluster member stellar masses ($M_*$) as a function of their apparent magnitude in the ${\rm F160W}$ filter ($m_{\rm F160W}$). Error bars on $M_*$ present the $1\sigma\, \rm$ credible interval from the SED posterior parameters while representing the standard deviation for ${\rm F160W}$. The magenta, cyan, and lime coloured dots represent the stellar mass estimates provided by \textsc{Bagpipes} with a double power-law SFH, \textsc{Bagpipes} with a delayed SFH, and \textsc{LePhare}, respectively.}
    \label{fig:Stellar-masses-comparison-SED-model}
\end{figure*}

Fig.~\ref{fig:Stellar-masses-comparison-SED-model} presents cluster members stellar mass estimates as a function of their apparent magnitudes in the ${\rm F160W}$ filter. Apart from a few outliers, the three different SED models provide similar values within the expected systematic difference of SED fitting methods \citep[i.e. $<0.3$ dex]{Conroy2013}. One can notice systematic differences between models for the fainter end of the cluster member distribution. The SED with a double power-law SFH provides the highest stellar masses, while \textsc{LePhare} gives the lowest. Stellar masses estimated by the SED model with the delayed SFH are located between the two other models. As these differences of a few tenths of dex are expected for different SED fit methods, we keep the three estimates to understand the resulting biases on the lensing mass models. Indeed, the median of stellar mass estimates is used to define the baryonic content of each cluster member, as it is presented in B25b, section~2.2.

\section{Supplementary mass constraints 2: BCG \& ICL }
\label{sect:BCG-ICL-cons}
Similarly to cluster members, we rely on the same measurements to constrain the BCG and the ICL stellar mass through their light distribution (Sect.~\ref{sect:ICL-light}) and stellar masses (Sect.~\ref{sect:BCG_ICL_M_*}). We also recover the stellar kinematics, although in that case, it is probing the cluster total mass distribution and not only the BCG one. In particular, the light distribution will serve as a starting point for the modelling of the mass of the BCG and the ICL component, which is presented in B25b, section~2.3.

Thanks to the spatial extension and high signal-to-noise in the BCG and ICL spectra, we rely on a refined kinematic model through the \textsc{JAM} package, which requires the second moment of the velocity along the line-of-sight (LOS) axis, $V_{\rm rms}$, instead of the LOSVD. We detail how we extract the $V_{\rm rms}$ in Sect.~\ref{sect:BCG-Vrms}. This measurement allows us to define a new likelihood which complements the strong lensing observations in the very centre of the cluster. It is presented in B25b, section~3.4.

Similarly to the cluster members, we make the measurements presented in the next sections publicly available, through the repository given in Sect.~\ref{sect:data-availability}. In particular, we provide the MGE parameters across the different bands, $V_{\rm rms}$ and stellar mass estimates, as well as the extraction masks used for the kinematic measurements.

\subsection{Light distribution}
\label{sect:ICL-light}
Several works have extracted the ICL profile from imaging data, including in AS1063 \citep[e.g.][]{Montes2018}. Different methodologies have been utilised, such as composite model including multiple S\'ersic profiles \citep{Janowiecki2010}, "wavelet-like" decomposition \citep{DaRocha2005,Jimenez-Teja2016,Ellien2021} or methods taking advantage of 2D fitting algorithms used to measure galaxy light profiles \citep{Giallongo2014,Morishita2017}. In the present case, we also want to constrain the stellar kinematics. Hence, we follow the composite model approach with a MGE model. Such parametrisation has been used extensively to model galaxy stellar kinematics, assuming an axisymmetric light and mass distribution. In particular, we want to use the Jeans Anisotropic Modelling \citep[JAM;][]{Cappellari2008} which is a computationally efficient way to compute the stellar kinematics for models that do not assume spherically symmetric distributions.

\begin{figure*}
    \centering
    \includegraphics[width=.8\linewidth]{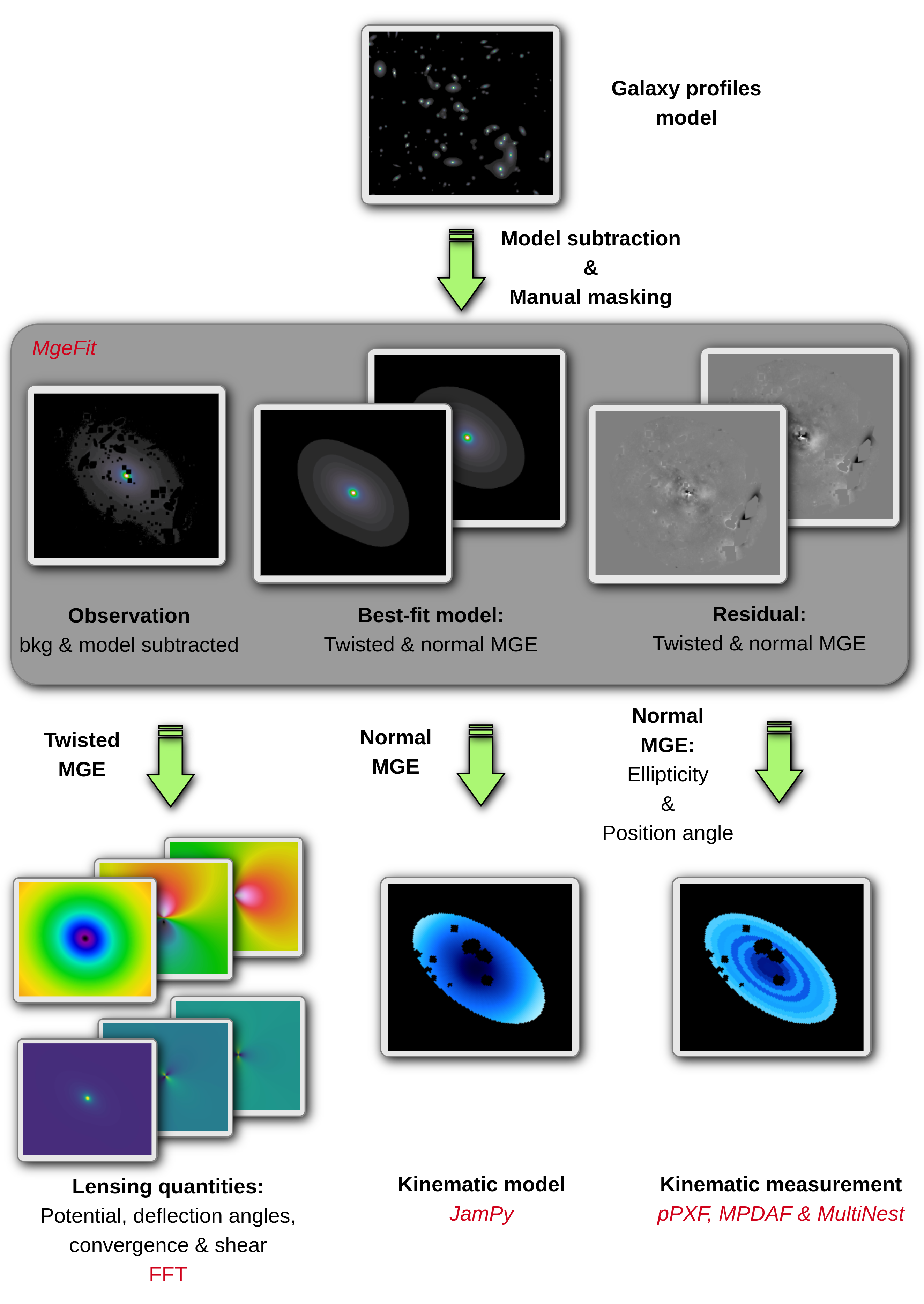}
    \caption{Diagram of the workflow to extract the BCG and the ICL kinematics, and their light distributions. The light distribution is estimated in the data image, where light profiles of galaxies have been removed. We made this estimation using MGE fits, with two parametrisations: an axisymmetric and a concentric one. We define the extraction regions for the kinematic fit and estimate lensing quantities from the MGE light profiles.}
    \label{fig:ICL-workflow}
\end{figure*}

To recover the BCG and ICL light distributions, we follow the procedure presented in Fig.\ref{fig:ICL-workflow}. We use the galaxy light profiles obtained in Sect.~\ref{sect:cluster-member-cons} to remove galaxies in the cluster field, and isolate the BCG and the ICL light emission in the ${\rm F160W}$ filter. As a single elliptical profile does not perfectly fit the galaxies in the core, we subtract the fitted model and then mask the residuals (e.g., spiral arms), rather than masking the entire galaxy. This approach allows us to use smaller masks while still effectively removing contamination. We also mask non-fitted background objects such as giant arcs. We then restrict the fitting regions to the same ellipsoidal aperture used in Sect.~\ref{sect:galaxy-light:region-fit} for the preliminary BCG and ICL models. It is a fine-tuned version of the preliminary fit, as we are taking into account the flux of field galaxies instead of simply masking them to improve the robustness of the MGE fit.

We are interested in a good, flexible model for the mass distribution (i.e. Twisted MGE) and an axisymmetric solution to be able to use JAM for the BCG and the ICL kinematics (i.e. Normal MGE). For the former, we use a MGE solution where each Gaussian is allowed to have a different position angle and is fitted on twisted isophotes, therefore a twisted MGE. Such parametrisation is well-suited for the BCG and ICL light distributions as it is expected that ellipticity and position angle will vary between the BCG and the ICL \citep{Kluge2020}. The axisymmetric MGE is fitted on ellipsoidal isophotes, and each Gaussian shares the same centre and position angle. This MGE will only be used to weight the mass in the kinematic computation, so it is less important to have the best 2D model. We enforce the weights of the Gaussian to be positive in both cases, even though negative weights can yield a better fit to the data. In particular, it enables us to split the MGE model into multiple MGE models, allowing for a varying stellar mass-to-light ratio, as presented in B25b, in section~2.3. To estimate the fit uncertainties, we rely on the \textsc{JAX} package and its automatic gradient computation to recover the Hessian of the mean squared error. We invert that Hessian and use the closest positive semi-definite matrix as the covariance matrix, following the property of the Fisher information matrix at the maximum likelihood. As we perform spectral energy distribution (SED) fits based on this MGE light model in Sect.~\ref{sect:BCG_ICL_M_*}, we are adding a systematic uncertainty of $5$ per cent to homogenise the photometry among the different depths of each \textit{HST} band.

\begin{figure}
    \centering
    \includegraphics[width=\linewidth]{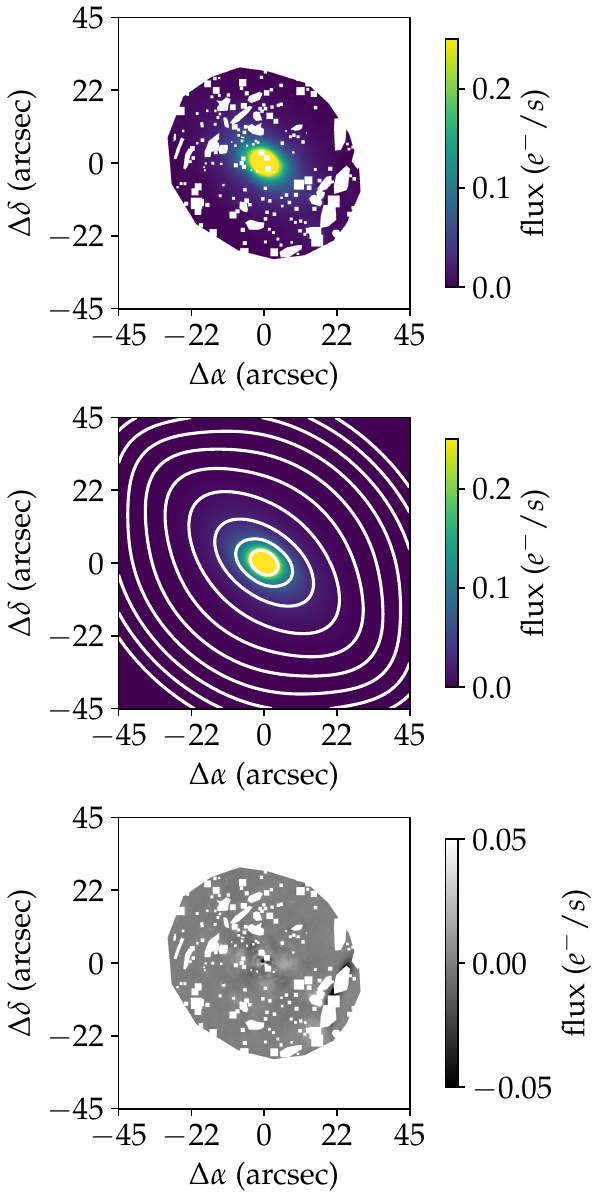}
    \caption{\textit{From top to bottom:} Masked \textit{HST} image in the ${\rm F160W}$ filter where all galaxy models have been subtracted, best-fitting MGE model on twisted isophotes and associated residuals}
    \label{fig:ICL-fit-twist}
\end{figure}

\begin{figure}
    \centering
    \includegraphics[width=\linewidth]{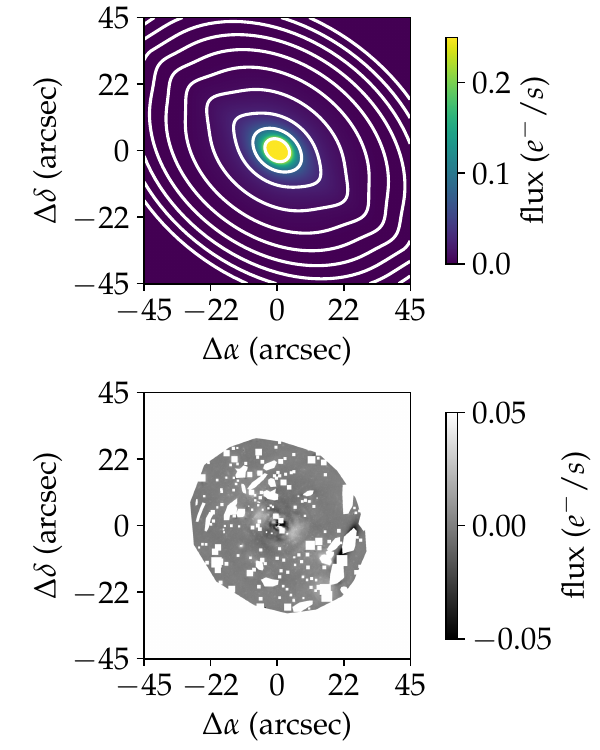}
    \caption{Best-fitting MGE model with an axisymmetric parametrisation (top) and associated residuals (bottom).}
    \label{fig:ICL-no-twist}
\end{figure}

The fitting results are presented in Fig.~\ref{fig:ICL-fit-twist} and \ref{fig:ICL-no-twist}. Fig.~\ref{fig:ICL-fit-twist} presents the masked image, the model fitted on twisted isophotes and the associated residuals. Fig.~\ref{fig:ICL-no-twist} presents the model and residuals for the normal MGE. Both profiles recover the global shape of the BCG and the ICL, though the axisymmetric fit exhibits a slightly worse orientation pattern. In particular, the twisted MGE presents a varying position angle and ellipticity with radius, as expected from this component. The ellipticity of the isophotes increases with the radius. For the normal MGE fit, we also observe the same ellipticity behaviour to a lesser extent.

Regarding the amplitude of the residuals, we gather the distribution of residuals weighted by the data errors. The twisted MGE fit presents a median and a scaled MAD of $0.037$ and $1.111$ in the fitting area, respectively. On the other hand, the axisymmetric MGE has $-0.007$ and $1.319$ for the same statistical indicator. Hence, both of these fits present a bias similar to or smaller than the global fit (i.e. $-0.071$), which is expected as the areas of bad fit have been mostly masked. As anticipated, the twisted MGE provides a better fit to the data with a MAD estimation $20$ per cent smaller than the normal MGE parametrisation. However, both fits represent the data well, as their scaled MAD estimates are close to $1$.

Finally, we convert the light distribution to a mass distribution with the mass-to-light ratio of the Sun ($M_{\rm R,\,AB}=4.60$) in the R-band from \citet{Willmer2018}. We use this mass distribution to compute the associated lensing potential, $\psi$, by fast Fourier transform using the following relations between the normalised surface mass density, also called convergence, $\kappa$, and $\psi$:
\begin{equation}
    \psi(\vec{\theta})=\frac{1}{\pi}\int\int {\rm d}^2\kappa(\vec{\theta'}) \ln|\vec{\theta}-\vec{\theta'}|
\end{equation}
based on the estimation of $\psi$, we obtain the other lensing quantities by deriving them once for the deflection and twice for the magnification matrix. These quantities can then be rescaled with the stellar mass-to-light ratio of the BCG and the ICL in units of the Sun mass-to-light ratio to match the correct mass. As \citet{Montes2014} and \citet{Montes2018} found a gradient of colour and metallicity with radius in HFF clusters, implying a varying mass-to-light ratio, we split the MGE parametrisation into multiple MGE models. Each of these MGE models has a distinct mass-to-light ratio, allowing variation in the mass-to-light ratio of the whole component, with the flexibility depending on the number of MGE models considered. The case of maximal flexibility, with one MGE model per Gaussian function in the overall representation, follows the approach presented by \citet{Collett2018}.

\subsection{$V_{\rm rms}$}
\label{sect:BCG-Vrms}
\begin{figure}
    \centering
    \includegraphics[width=.9\linewidth]{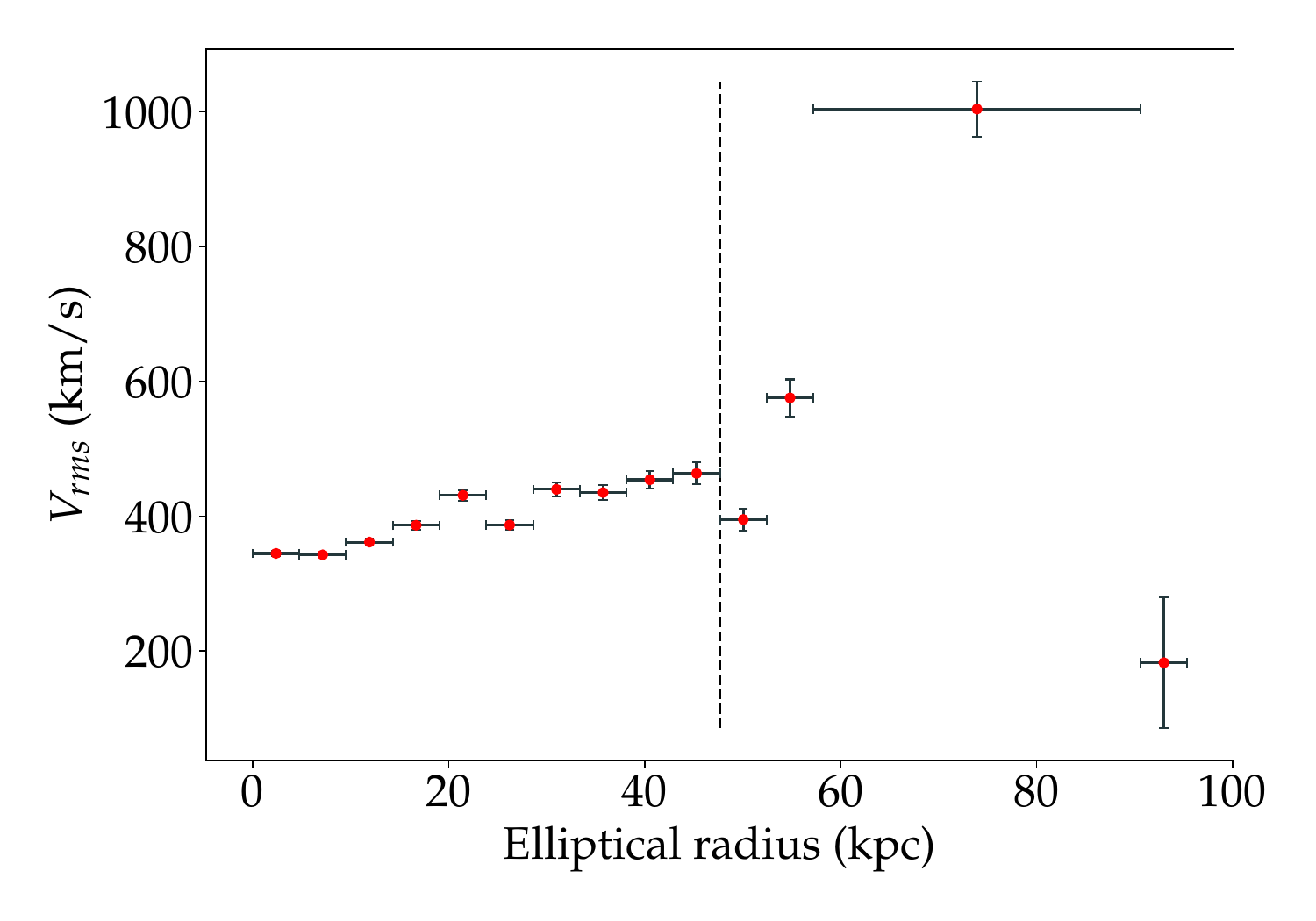}
    \includegraphics[width=\linewidth]{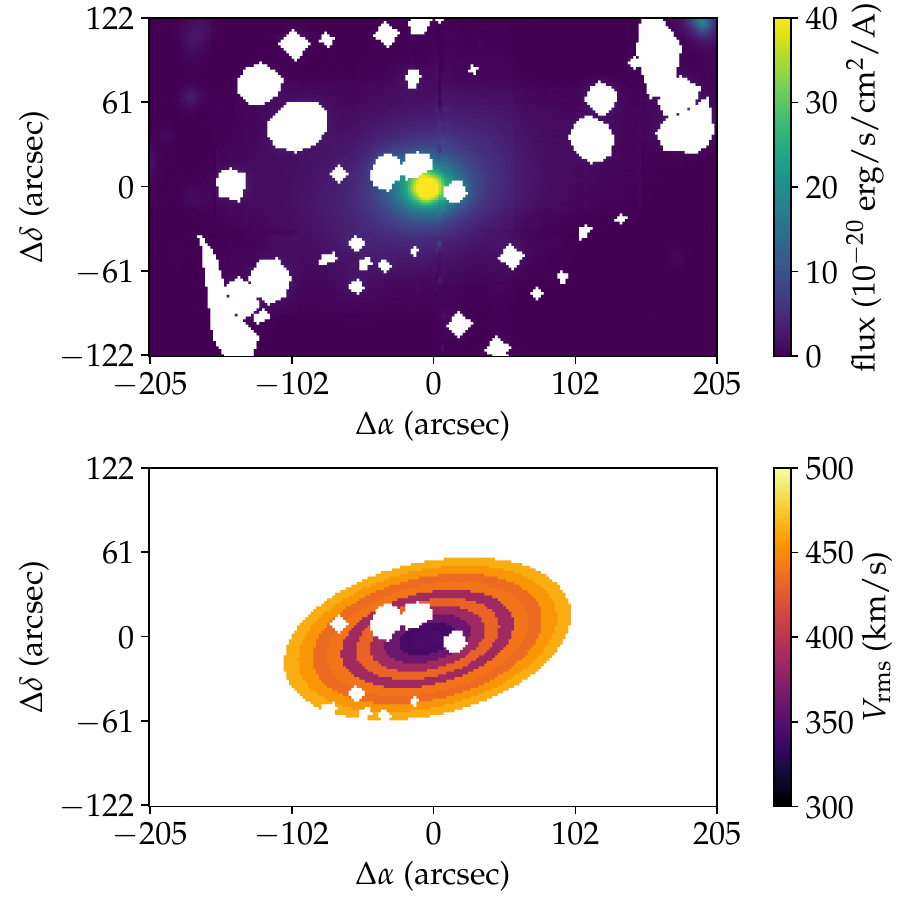}
    \caption{Kinematic estimation of the BCG \& ICL. \textit{Top panel:} $V_{\rm rms}$ estimate as a function of the elliptical radius used to define the elliptical bins. \textit{Middle panel:} Masked white-light image of the MUSE datacube. \textit{Bottom panel:} $V_{\rm rms}$ estimate in each elliptical extraction region without the rejected data points.}
    \label{fig:ICL-kinematics}
\end{figure}

To measure the evolution of $V_{\rm rms}$ as a function of the radius, we extract the BCG and the ICL MUSE spectra in increasing elliptical annuli. We fix the extraction centre coordinates, position angle and ellipticity to the same ones as the elliptical isophotes defined to fit the associated light distribution in Sect.~\ref{sect:ICL-light}. For consistency with the kinematic constraints and their use with the \textsc{JAM} library, we use the normal MGE fit (i.e. axisymmetric) as the basis for the extraction region. We use the MUSE white-light image and the \textsc{HST} images to mask galaxies present in the extraction area. The widths of the annuli are defined iteratively by increasing their radii by the size of the MUSE PSF until the associated spectrum has a median signal-to-noise, $S/N$, above a threshold of $5$. As the MUSE data were taken without adaptive optics correction, we assume that the MUSE PSF width is approximately the seeing of the observation (i.e. $\sim 1$~${\rm arcsec}$). This estimation is consistent with the PSF FWHM measured during the data reduction. We limit the radius of the aperture to $100$~${\rm kpc}$, which leads to $14$ regions. We use the same procedure for the ICL as the one for the galaxies, except that we do not remove the local background. The estimate of this background by \textsc{MPDAF} misidentifies the ICL as the background sky, which tends to modify the shapes of the spectra. The fit in the last two regions failed because the spectra were too noisy, leaving us with $12$ data points. Of those $12$ points, the last two do not follow the same pattern as the others, with a high degree of variability between them. Indeed, the last three points yield $\sigma_e$ values of $461$, $394$, and $574$~${\rm km/s}$; thus, we reject the last two data points as they appear to be dominated by systematic error, reducing our final dataset to $10$ points. The whole $V_{\rm rms}$ measurement set is presented in the top panel of Fig.~\ref{fig:ICL-kinematics}, including the last four rejected data points.

As presented in Sect.~\ref{sect:Cluster-member-LOSVD}, our procedure allows us to measure $V$ and $\sigma_e$, which are linked to $V_{\rm rms}$ such that $V_{\rm rms}=\sqrt{V^2+ \sigma^2_e}$. Hence, we use the same fitting procedure here, including the polynomial correction. As we are directly fitting these data, we rescale the error by the underestimation factor found in \citet{Beauchesne2024-erratum}, which is $1.33$. This underestimation bias was estimated using this fitting procedure on $10000$ mock MUSE spectra. The mocks were designed to mimic the stellar populations and observational uncertainties of the cluster members in AS1063. In Fig.~\ref{fig:ICL-kinematics}, we present the masked MUSE white-light image and the extraction regions with their best-fitting $V_{\rm rms}$. As shown in this plot, the $V_{\rm rms}$ increases as we move from the kinematics of the BCG (i.e. the two most inner extraction regions) to the ICL, as seen in local clusters \citep[e.g.][]{Longobardi18}. The $V_{\rm rms}$ values range from $\approx341$~${\rm km/s}$ in the BCG to $465$~${\rm km/s}$ in the elliptical bin with maximum radii.

\subsection{Stellar masses}
\label{sect:BCG_ICL_M_*}

As we use the MUSE spectroscopic data to derive the stellar mass of the BCG and ICL, we extract their photometry in each elliptical annulus where the stellar kinematics are measured. We measure the photometry by reproducing the procedure described in Sect.~\ref{sect:ICL-light} for ${\rm F435W}$, ${\rm F606W}$, ${\rm F814W}$, ${\rm F105W}$, ${\rm F125W}$ and ${\rm F140W}$ filters, which gives us MGE light models of the BCG and the ICL components. When considering \textsc{LePhare}, only the broadband filter photometry is used to constrain the SED model. In contrast, the two \textsc{Bagpipes} settings rely on combining photometric and spectroscopic data. To account for any miscalibration between photometry and spectroscopy, we follow the approach of \citet{Carnall19}, and add second-order Chebyshev polynomials as a multiplicative function of the spectra and a Gaussian white noise model. Priors used for the polynomials and white noise are listed in Table~\ref{table:SED-param}. 

\begin{figure}
    
    \centering
    \includegraphics[width=\linewidth]{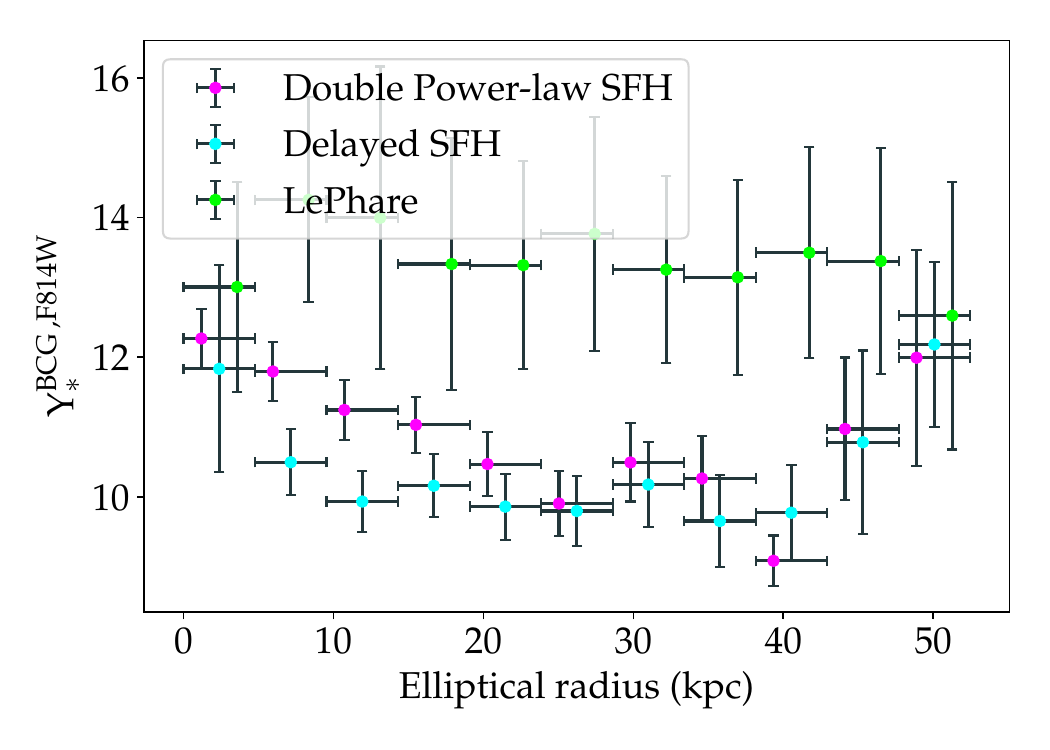}
    \includegraphics[width=\linewidth]{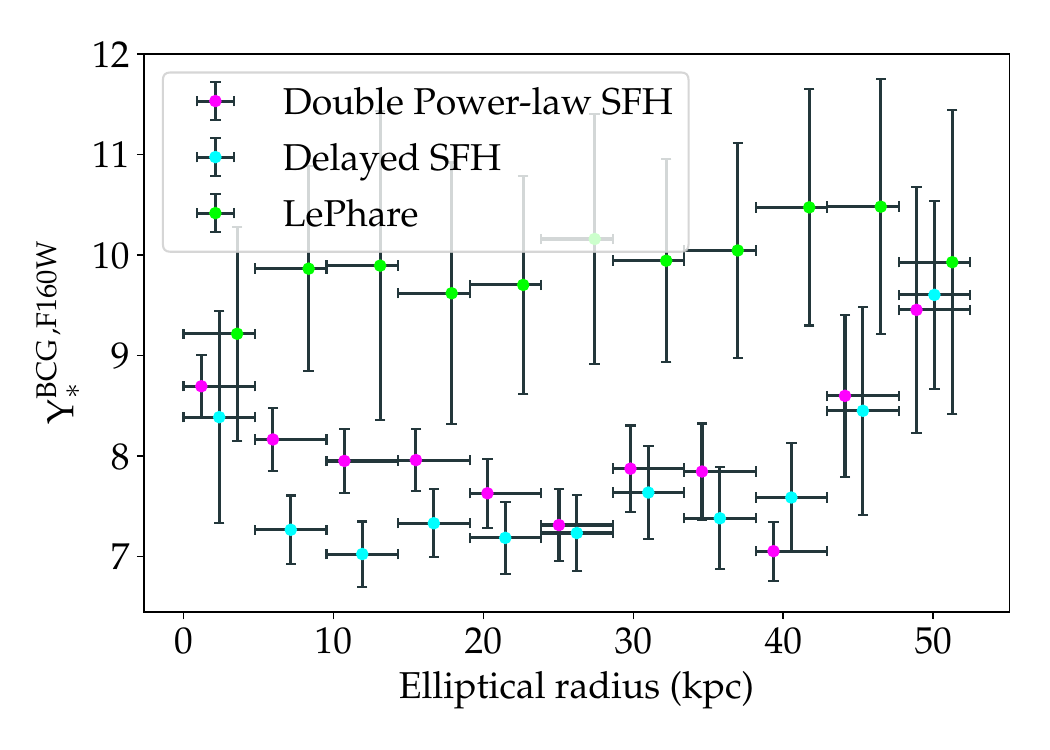}
    
    \caption{\textit{Top panel:} Stellar-mass-to-light-ratio $\Upsilon^{\rm BCG}_{*}$ derived for the combined BCG and ICL component in the ${\rm F814W}$ filter. The luminosity has been normalised by the solar luminosity in the R-band from \citet{Willmer2018}. Error bars represent standard deviations on $\Upsilon^{\rm BCG}_{*}$ made by propagating the error on the stellar mass and photometry, while the radius error represents the width of the extraction regions. To improve the readability of the plot, the results of each SED code have been slightly shifted horizontally. The magenta and cyan dots correspond to the \textsc{Bagpipes} SED models with double power-law and delayed SFHs, respectively. The lime-coloured dots represent the results obtained with \textsc{LePhare}. \textit{Bottom panel:} Same as the top panel for the ${\rm F160W}$ filter.}
    \label{fig:Stellar-masses-estimation}
\end{figure}

Fig.~\ref{fig:Stellar-masses-estimation} presents the stellar-mass-to-light ratio $\Upsilon^{\rm BCG}_{*}$ derived for the combined BCG and ICL component in the ${\rm F814W}$ (top panel) and ${\rm F160W}$ bands (bottom panel) filters. For both filters, we normalise the luminosity by the sun luminosity ($M_{\rm R,\,AB}=4.60$) in the R-band from \citet{Willmer2018}. Both \textsc{Bagpipes} SED models provide similar stellar mass estimates, with all data points agreeing within one standard deviation except the second and third ones. In contrast, \textsc{LePhare} prefers a slightly larger estimate of the stellar mass but with larger uncertainty such that it agrees with the \textsc{Bagpipes} models within two standard deviations on all data points. It is opposite to the pattern seen in Fig.~\ref{fig:Stellar-masses-comparison-SED-model}, as cluster member stellar masses estimated by \textsc{LePhare} are lower. This difference is likely due to the inclusion of MUSE spectra in the SED fits for the BCG and ICL. When \textsc{Bagpipes} was run with photometry only in preliminary tests, its stellar mass estimates were much closer to the output of \textsc{LePhare}. Stellar mass estimates obtained by combining spectroscopic and photometric data are within the $3\sigma\, \rm CI$ of the photometry-only runs. Hence, all SED models provide similar estimates of the combined BCG and ICL stellar mass. 

From these SED fits, we obtain an estimate of the stellar mass of the BCG and the ICL component. By combining all elliptical bins considered in Fig.~\ref{fig:Stellar-masses-estimation}, we obtain $6.83\pm 0.17$, $7.20\pm 0.11$ and $8.98\pm 0.34\times 10^{11}\,M_\odot$ for the fits with a delayed SFH, a double power-law SFH and the \textsc{LePhare} SED models, respectively. Assuming that all bins within the half-light radius of the BCG are a good representation of its stellar mass (i.e. the four most central bins), we obtain for the fits with a delayed SFH, a double power-law SFH and the \textsc{LePhare} SED, a stellar mass of $3.15\pm 0.21$, $3.45\pm 0.12$ and $4.05\pm 0.51\times 10^{11}\,M_\odot$, respectively. Hence, the BCG would account for roughly half of the stellar mass of the BCG and the ICL component.

Regarding a possible gradient in $\Upsilon^{\rm BCG}_{*}$, the radial variation is limited in both ${\rm F814W}$ and ${\rm F160W}$ filters. The variation between the data points is mostly within the statistical uncertainties. The \textsc{Bagpipes} SED models present a decrease of $\Upsilon^{\rm BCG}_{*}$ with the radius, followed by an increase over the last two data points. However, according to the statistical uncertainties, most of these points agree with a constant $\Upsilon^{\rm BCG}_{*}$ within two standard deviations. This pattern disagrees with \textsc{LePhare}, which does not show significant variations with the distance from the cluster centre. Variations of $\Upsilon^{\rm BCG}_{*}$ are smaller in the ${\rm F160W}$ filter. When all SED models and data points are combined, we obtain $\Upsilon^{\rm BCG,\,F160W}_{*}=8.57\pm1.42$~$M_{\odot}/L_{\odot, R}$ in comparison to $\Upsilon^{\rm BCG,\,F814W}_{*}=11.57\pm1.90$~$M_{\odot}/L_{\odot, R}$. The ${\rm F814W}$ filter thus exhibits $\approx30$ per cent more dispersion than ${\rm F160W}$, favouring the use of the ${\rm F160W}$ light distribution as a tracer of the stellar mass for modelling the baryonic content of the BCG and ICL.

As reported by \citet{Xie2024} from radio observations, there is likely an active galactic nucleus (AGN) in the BCG. To test possible bias in the estimated stellar masses, we try to add an AGN component to the \textsc{Bagpipes} SED models, similar to the one used by \citet{Carnall2023}. We fit the most central bin with this extended model, which leads to a variation of less than $1$ per cent of the median of the stellar mass posterior. Hence, we consider that the bias due to the possible presence of an AGN is negligible in our case, although the considered wavelength range is mostly dominated by stellar emission, which may bias this check. Indeed, for the SED best-fitting solution with an AGN component, the AGN emission significantly dominates the overall emission at wavelengths above $\approx3.15\,{\rm \mu m}$.

\section{Conclusion}

In the first paper of this series, we focus on the required analysis of the observational dataset to extract constraints for a comprehensive mass modelling of AS1063. In particular, we present the existing mass constraints from strong lensing and X-ray observations, which provide no knowledge of the stellar mass contained in the cluster members, the BCG, and the ICL. We then detail specific methodologies, based on photometric and spectroscopic data, to recover the light distribution, stellar kinematics, and stellar mass estimates for those components.

For cluster members, we begin by describing our procedure for recovering their light profiles, while accounting for neighbouring galaxies, in Sect.~\ref{sect:galaxy-light}. Our approach consists of three main steps: detection (Sect.~\ref{sect:galaxy-light:detect}), region fitting (Sect.~\ref{sect:galaxy-light:region-fit}), and single-object fitting (Sect.~\ref{sect:galaxy-light:bayes}). We use a standard detection procedure with cold and hot \textit{SExtractor} passes. Based on these results, we identify contiguous regions where all objects are fitted simultaneously to obtain an estimate of the light profile maximum likelihood. We then finish our procedure by fitting each object individually with a nested sampling algorithm, where the neighbouring objects are subtracted based on the previous region fit. This method allows us to fit thousands of objects in a crowded field with a moderate computing cost. We then estimate their stellar kinematics through their LOSVD from the MUSE datacube (Sect.~\ref{sect:Cluster-member-LOSVD}) and their stellar mass based on \textit{SExtractor} aperture photometry using the available \textsc{HST} bands (Sect.~\ref{sect:cluster-member-stellar-masses}).

For the BCG and the ICL component, we fit the light distribution using an MGE profile, which benefits from the cluster-member light-profile models, which are subtracted from the original datasets. We reproduce that measurement in each \textsc{HST} band to obtain the colour information required for the SED fitting. The light profile extraction is detailed in Sect.~\ref{sect:ICL-light}. Similarly to cluster members, we obtain the stellar kinematics through their $V_{\rm rms}$ within elliptical bins (Sect.~\ref{sect:BCG-Vrms}) and their stellar mass by fitting their SED through their \textsc{HST} model photometry and MUSE spectra (Sect.~\ref{sect:BCG_ICL_M_*}).

The primary use of these analyses is to parameterise the mass model presented in B25b. In particular, we are using the light profile as a tracer of the stellar mass. We present our modelling assumptions in B25b, section~2, for all components. Thanks to the stellar kinematics measurements, we are able to add two new likelihoods to the approach of B24, which are described in B25b, section~3. The first likelihood uses a \textsc{JAM} kinematic modelling combined with the $V_{\rm rms}$ measurements to probe the cluster total mass down to inside the BCG. The second one constrains the cluster member total mass based on their LOSVD, which is computed with the light profile measured in this analysis, as well as their mass distribution.

\section*{Acknowledgements}
The authors are grateful to the referee for carefully reading the manuscript and for valuable suggestions and comments, which helped improve and clarify it. The authors thank Massimo Meneghetti, Raphaël Gavazzi, Michaela Hirschmann, and Mireia Montes for their helpful discussions at various stages of the project.
The computations were performed at the University of Geneva using Yggdrasil and Boabab HPC services.
BB acknowledges the Swiss National Science Foundation (SNSF) for supporting this work.
ML acknowledges the Centre National de la Recherche Scientifique (CNRS) and the Centre National des Etudes Spatiale (CNES) for support.
MJ is supported by the United Kingdom Research and Innovation (UKRI) Future Leaders Fellowship `Using Cosmic Beasts to uncover the Nature of Dark Matter' (grant number MR/X006069/1).
\section*{Data Availability}
\label{sect:data-availability}
The measurements obtained in that work are publicly released with the paper, and are available at the Mikulski Archive for Space Telescopes as High Level Science Products via \href{https://doi.org/10.17909/t9-w6tj-wp63}{https://doi.org/10.17909/t9-w6tj-wp63}. In particular, we release a cluster member catalogue, with their light profile measurement, velocity dispersion and stellar masses. We also provide the empirical PSF and the masked F160W mosaic used in the light profile fitting procedure. For the BCG, we provide each MGE fit as well as the stellar kinematics and stellar masses. As the last two measurements have been obtained in elliptical bins from the MUSE datacube, we provide the mask of each bin. 


\bibliographystyle{mnras}
\bibliography{References} 

@ARTICLE{Niemiec2026,
       author = {{Niemiec}, A. and {Acebron}, A. and {Beauchesne}, B. and {Jauzac}, M. and {Diego}, J.~M. and {Eckert}, D. and {Harvey}, D. and {Koekemoer}, A.~M. and {Lagattuta}, D.~J. and {Limousin}, M. and {Mahler}, G. and {Patel}, N. and {Tam}, S. and {Allingham}, J.~F.~V. and {Cen}, R. and {Faisst}, A. and {Perera}, D. and {Sereno}, M.},
        title = "{Non-spherical BUFFALOs: a weak lensing view of the Frontier Field clusters and associated systematics}",
      journal = {arXiv e-prints},
         year = 2026,
        month = feb,
          eid = {arXiv:2602.06904},
        pages = {arXiv:2602.06904},
archivePrefix = {arXiv},
       eprint = {2602.06904},
 primaryClass = {astro-ph.CO},
       adsurl = {https://ui.adsabs.harvard.edu/abs/2026arXiv260206904N}
}

@ARTICLE{Dutton2011,
       author = {{Dutton}, Aaron A. and {Brewer}, Brendon J. and {Marshall}, Philip J. and {Auger}, Matthew W. and {Treu}, Tommaso and {Koo}, David C. and {Bolton}, Adam S. and {Holden}, Bradford P. and {Koopmans}, Leon V.~E.},
        title = "{The SWELLS survey - II. Breaking the disc-halo degeneracy in the spiral galaxy gravitational lens SDSS J2141-0001}",
      journal = {\mnras},
         year = 2011,
        month = nov,
       volume = {417},
       number = {3},
        pages = {1621-1642},
          doi = {10.1111/j.1365-2966.2011.18706.x},
archivePrefix = {arXiv},
       eprint = {1101.1622},
 primaryClass = {astro-ph.CO},
       adsurl = {https://ui.adsabs.harvard.edu/abs/2011MNRAS.417.1621D}
}

@ARTICLE{Beauchesne2025b,
       author = {{Beauchesne}, Benjamin and {Cl{\'e}ment}, Benjamin and {Limousin}, Marceau and {Niemiec}, Anna and {Jauzac}, Mathilde and {Alcalde Pampliega}, Bel{\'e}n and {Richard}, Johan and {Mahler}, Guillaume and {Diego}, Jose M. and {Hibon}, Pascale and {Koekemoer}, Anton M. and {Connor}, Thomas and {Kneib}, Jean-Paul and {Faisst}, Andreas L.},
        title = "{A comprehensive separation of dark matter and baryonic mass components in galaxy clusters II: an overview of the mass distribution in Abell S1063}",
      journal = {arXiv e-prints},
         year = 2025,
        month = sep,
          eid = {arXiv:2509.07777},
        pages = {arXiv:2509.07777},
          doi = {10.48550/arXiv.2509.07777},
archivePrefix = {arXiv},
       eprint = {2509.07777},
 primaryClass = {astro-ph.CO},
       adsurl = {https://ui.adsabs.harvard.edu/abs/2025arXiv250907777B}
}

@ARTICLE{Cerny2025,
       author = {{Cerny}, Catherine and {Jauzac}, Mathilde and {Lagattuta}, David and {Niemiec}, Anna and {Mahler}, Guillaume and {Edge}, Alastair and {Massey}, Richard},
        title = "{The Kaleidoscope Survey: Strong Gravitational Lensing in Galaxy Clusters with Radial Arcs}",
      journal = {arXiv e-prints},
         year = 2025,
        month = jun,
          eid = {arXiv:2506.21531},
        pages = {arXiv:2506.21531},
          doi = {10.48550/arXiv.2506.21531},
archivePrefix = {arXiv},
       eprint = {2506.21531},
 primaryClass = {astro-ph.CO},
       adsurl = {https://ui.adsabs.harvard.edu/abs/2025arXiv250621531C}
}

@ARTICLE{Koekemoer2011,
       author = {{Koekemoer}, Anton M. and {Faber}, S.~M. and {Ferguson}, Henry C. and {Grogin}, Norman A. and {Kocevski}, Dale D. and {Koo}, David C. and {Lai}, Kamson and {Lotz}, Jennifer M. and {Lucas}, Ray A. and {McGrath}, Elizabeth J. and {Ogaz}, Sara and {Rajan}, Abhijith and {Riess}, Adam G. and {Rodney}, Steve A. and {Strolger}, Louis and {Casertano}, Stefano and {Castellano}, Marco and {Dahlen}, Tomas and {Dickinson}, Mark and {Dolch}, Timothy and {Fontana}, Adriano and {Giavalisco}, Mauro and {Grazian}, Andrea and {Guo}, Yicheng and {Hathi}, Nimish P. and {Huang}, Kuang-Han and {van der Wel}, Arjen and {Yan}, Hao-Jing and {Acquaviva}, Viviana and {Alexander}, David M. and {Almaini}, Omar and {Ashby}, Matthew L.~N. and {Barden}, Marco and {Bell}, Eric F. and {Bournaud}, Fr{\'e}d{\'e}ric and {Brown}, Thomas M. and {Caputi}, Karina I. and {Cassata}, Paolo and {Challis}, Peter J. and {Chary}, Ranga-Ram and {Cheung}, Edmond and {Cirasuolo}, Michele and {Conselice}, Christopher J. and {Roshan Cooray}, Asantha and {Croton}, Darren J. and {Daddi}, Emanuele and {Dav{\'e}}, Romeel and {de Mello}, Duilia F. and {de Ravel}, Loic and {Dekel}, Avishai and {Donley}, Jennifer L. and {Dunlop}, James S. and {Dutton}, Aaron A. and {Elbaz}, David and {Fazio}, Giovanni G. and {Filippenko}, Alexei V. and {Finkelstein}, Steven L. and {Frazer}, Chris and {Gardner}, Jonathan P. and {Garnavich}, Peter M. and {Gawiser}, Eric and {Gruetzbauch}, Ruth and {Hartley}, Will G. and {H{\"a}ussler}, Boris and {Herrington}, Jessica and {Hopkins}, Philip F. and {Huang}, Jia-Sheng and {Jha}, Saurabh W. and {Johnson}, Andrew and {Kartaltepe}, Jeyhan S. and {Khostovan}, Ali A. and {Kirshner}, Robert P. and {Lani}, Caterina and {Lee}, Kyoung-Soo and {Li}, Weidong and {Madau}, Piero and {McCarthy}, Patrick J. and {McIntosh}, Daniel H. and {McLure}, Ross J. and {McPartland}, Conor and {Mobasher}, Bahram and {Moreira}, Heidi and {Mortlock}, Alice and {Moustakas}, Leonidas A. and {Mozena}, Mark and {Nandra}, Kirpal and {Newman}, Jeffrey A. and {Nielsen}, Jennifer L. and {Niemi}, Sami and {Noeske}, Kai G. and {Papovich}, Casey J. and {Pentericci}, Laura and {Pope}, Alexandra and {Primack}, Joel R. and {Ravindranath}, Swara and {Reddy}, Naveen A. and {Renzini}, Alvio and {Rix}, Hans-Walter and {Robaina}, Aday R. and {Rosario}, David J. and {Rosati}, Piero and {Salimbeni}, Sara and {Scarlata}, Claudia and {Siana}, Brian and {Simard}, Luc and {Smidt}, Joseph and {Snyder}, Diana and {Somerville}, Rachel S. and {Spinrad}, Hyron and {Straughn}, Amber N. and {Telford}, Olivia and {Teplitz}, Harry I. and {Trump}, Jonathan R. and {Vargas}, Carlos and {Villforth}, Carolin and {Wagner}, Cory R. and {Wandro}, Pat and {Wechsler}, Risa H. and {Weiner}, Benjamin J. and {Wiklind}, Tommy and {Wild}, Vivienne and {Wilson}, Grant and {Wuyts}, Stijn and {Yun}, Min S.},
        title = "{CANDELS: The Cosmic Assembly Near-infrared Deep Extragalactic Legacy Survey{\textemdash}The Hubble Space Telescope Observations, Imaging Data Products, and Mosaics}",
      journal = {\apjs},
         year = 2011,
        month = dec,
       volume = {197},
       number = {2},
          eid = {36},
        pages = {36},
          doi = {10.1088/0067-0049/197/2/36},
archivePrefix = {arXiv},
       eprint = {1105.3754},
 primaryClass = {astro-ph.CO},
       adsurl = {https://ui.adsabs.harvard.edu/abs/2011ApJS..197...36K}
}

@ARTICLE{Morishita2017,
       author = {{Morishita}, Takahiro and {Abramson}, Louis E. and {Treu}, Tommaso and {Schmidt}, Kasper B. and {Vulcani}, Benedetta and {Wang}, Xin},
        title = "{Characterizing Intracluster Light in the Hubble Frontier Fields}",
      journal = {\apj},
         year = 2017,
        month = sep,
       volume = {846},
       number = {2},
          eid = {139},
        pages = {139},
          doi = {10.3847/1538-4357/aa8403},
archivePrefix = {arXiv},
       eprint = {1610.08503},
 primaryClass = {astro-ph.GA},
       adsurl = {https://ui.adsabs.harvard.edu/abs/2017ApJ...846..139M}
}

@ARTICLE{Giallongo2014,
       author = {{Giallongo}, E. and {Menci}, N. and {Grazian}, A. and {Gallozzi}, S. and {Castellano}, M. and {Fiore}, F. and {Fontana}, A. and {Pentericci}, L. and {Boutsia}, K. and {Paris}, D. and {Speziali}, R. and {Testa}, V.},
        title = "{Diffuse Optical Intracluster Light as a Measure of Stellar Tidal Stripping: The Cluster CL0024+17 at z \raisebox{-0.5ex}\textasciitilde 0.4 Observed at the Large Binocular Telescope}",
      journal = {\apj},
         year = 2014,
        month = jan,
       volume = {781},
       number = {1},
          eid = {24},
        pages = {24},
          doi = {10.1088/0004-637X/781/1/24},
archivePrefix = {arXiv},
       eprint = {1311.1921},
 primaryClass = {astro-ph.CO},
       adsurl = {https://ui.adsabs.harvard.edu/abs/2014ApJ...781...24G}
}

@ARTICLE{Ellien2021,
       author = {{Ellien}, A. and {Slezak}, E. and {Martinet}, N. and {Durret}, F. and {Adami}, C. and {Gavazzi}, R. and {Raba{\c{c}}a}, C.~R. and {Da Rocha}, C. and {Epit{\'a}cio Pereira}, D.~N.},
        title = "{DAWIS: a detection algorithm with wavelets for intracluster light studies}",
      journal = {\aap},
         year = 2021,
        month = may,
       volume = {649},
          eid = {A38},
        pages = {A38},
          doi = {10.1051/0004-6361/202038419},
archivePrefix = {arXiv},
       eprint = {2101.03835},
 primaryClass = {astro-ph.GA},
       adsurl = {https://ui.adsabs.harvard.edu/abs/2021A&A...649A..38E}
}

@ARTICLE{Jimenez-Teja2016,
       author = {{Jim{\'e}nez-Teja}, Y. and {Dupke}, R.},
        title = "{Disentangling the ICL with the CHEFs: Abell 2744 as a Case Study}",
      journal = {\apj},
         year = 2016,
        month = mar,
       volume = {820},
       number = {1},
          eid = {49},
        pages = {49},
          doi = {10.3847/0004-637X/820/1/49},
archivePrefix = {arXiv},
       eprint = {1602.07306},
 primaryClass = {astro-ph.GA},
       adsurl = {https://ui.adsabs.harvard.edu/abs/2016ApJ...820...49J}
}

@ARTICLE{DaRocha2005,
       author = {{Da Rocha}, C. and {Mendes de Oliveira}, C.},
        title = "{Intragroup diffuse light in compact groups of galaxies: HCG 79, 88 and 95}",
      journal = {\mnras},
         year = 2005,
        month = dec,
       volume = {364},
       number = {3},
        pages = {1069-1081},
          doi = {10.1111/j.1365-2966.2005.09641.x},
archivePrefix = {arXiv},
       eprint = {astro-ph/0509908},
 primaryClass = {astro-ph},
       adsurl = {https://ui.adsabs.harvard.edu/abs/2005MNRAS.364.1069D}
}

@ARTICLE{Janowiecki2010,
       author = {{Janowiecki}, Steven and {Mihos}, J. Christopher and {Harding}, Paul and {Feldmeier}, John J. and {Rudick}, Craig and {Morrison}, Heather},
        title = "{Diffuse Tidal Structures in the Halos of Virgo Ellipticals}",
      journal = {\apj},
         year = 2010,
        month = jun,
       volume = {715},
       number = {2},
        pages = {972-985},
          doi = {10.1088/0004-637X/715/2/972},
archivePrefix = {arXiv},
       eprint = {1004.1473},
 primaryClass = {astro-ph.CO},
       adsurl = {https://ui.adsabs.harvard.edu/abs/2010ApJ...715..972J}
}

@ARTICLE{Caminha2016,
       author = {{Caminha}, G.~B. and {Grillo}, C. and {Rosati}, P. and {Balestra}, I. and {Karman}, W. and {Lombardi}, M. and {Mercurio}, A. and {Nonino}, M. and {Tozzi}, P. and {Zitrin}, A. and {Biviano}, A. and {Girardi}, M. and {Koekemoer}, A.~M. and {Melchior}, P. and {Meneghetti}, M. and {Munari}, E. and {Suyu}, S.~H. and {Umetsu}, K. and {Annunziatella}, M. and {Borgani}, S. and {Broadhurst}, T. and {Caputi}, K.~I. and {Coe}, D. and {Delgado-Correal}, C. and {Ettori}, S. and {Fritz}, A. and {Frye}, B. and {Gobat}, R. and {Maier}, C. and {Monna}, A. and {Postman}, M. and {Sartoris}, B. and {Seitz}, S. and {Vanzella}, E. and {Ziegler}, B.},
        title = "{CLASH-VLT: A highly precise strong lensing model of the galaxy cluster RXC J2248.7-4431 (Abell S1063) and prospects for cosmography}",
      journal = {\aap},
         year = 2016,
        month = mar,
       volume = {587},
          eid = {A80},
        pages = {A80},
          doi = {10.1051/0004-6361/201527670},
archivePrefix = {arXiv},
       eprint = {1512.04555},
 primaryClass = {astro-ph.CO},
       adsurl = {https://ui.adsabs.harvard.edu/abs/2016A&A...587A..80C}
}

@ARTICLE{Diego2016,
       author = {{Diego}, Jose M. and {Broadhurst}, Tom and {Wong}, Jess and {Silk}, Joseph and {Lim}, Jeremy and {Zheng}, Wei and {Lam}, Daniel and {Ford}, Holland},
        title = "{A free-form mass model of the Hubble Frontier Fields cluster AS1063 (RXC J2248.7-4431) with over one hundred constraints}",
      journal = {\mnras},
         year = 2016,
        month = jul,
       volume = {459},
       number = {4},
        pages = {3447-3459},
          doi = {10.1093/mnras/stw865},
archivePrefix = {arXiv},
       eprint = {1512.07916},
 primaryClass = {astro-ph.CO},
       adsurl = {https://ui.adsabs.harvard.edu/abs/2016MNRAS.459.3447D}
}

@ARTICLE{Monna2014,
       author = {{Monna}, A. and {Seitz}, S. and {Greisel}, N. and {Eichner}, T. and {Drory}, N. and {Postman}, M. and {Zitrin}, A. and {Coe}, D. and {Halkola}, A. and {Suyu}, S.~H. and {Grillo}, C. and {Rosati}, P. and {Lemze}, D. and {Balestra}, I. and {Snigula}, J. and {Bradley}, L. and {Umetsu}, K. and {Koekemoer}, A. and {Kuchner}, U. and {Moustakas}, L. and {Bartelmann}, M. and {Ben{\'\i}tez}, N. and {Bouwens}, R. and {Broadhurst}, T. and {Donahue}, M. and {Ford}, H. and {Host}, O. and {Infante}, L. and {Jimenez-Teja}, Y. and {Jouvel}, S. and {Kelson}, D. and {Lahav}, O. and {Medezinski}, E. and {Melchior}, P. and {Meneghetti}, M. and {Merten}, J. and {Molino}, A. and {Moustakas}, J. and {Nonino}, M. and {Zheng}, W.},
        title = "{CLASH: z {\ensuremath{\sim}} 6 young galaxy candidate quintuply lensed by the frontier field cluster RXC J2248.7-4431}",
      journal = {\mnras},
         year = 2014,
        month = feb,
       volume = {438},
       number = {2},
        pages = {1417-1434},
          doi = {10.1093/mnras/stt2284},
archivePrefix = {arXiv},
       eprint = {1308.6280},
 primaryClass = {astro-ph.CO},
       adsurl = {https://ui.adsabs.harvard.edu/abs/2014MNRAS.438.1417M}
}

@ARTICLE{Johnson2014,
       author = {{Johnson}, Traci L. and {Sharon}, Keren and {Bayliss}, Matthew B. and {Gladders}, Michael D. and {Coe}, Dan and {Ebeling}, Harald},
        title = "{Lens Models and Magnification Maps of the Six Hubble Frontier Fields Clusters}",
      journal = {\apj},
         year = 2014,
        month = dec,
       volume = {797},
       number = {1},
          eid = {48},
        pages = {48},
          doi = {10.1088/0004-637X/797/1/48},
archivePrefix = {arXiv},
       eprint = {1405.0222},
 primaryClass = {astro-ph.CO},
       adsurl = {https://ui.adsabs.harvard.edu/abs/2014ApJ...797...48J}
}

@ARTICLE{Richard2014,
       author = {{Richard}, Johan and {Jauzac}, Mathilde and {Limousin}, Marceau and {Jullo}, Eric and {Cl{\'e}ment}, Benjamin and {Ebeling}, Harald and {Kneib}, Jean-Paul and {Atek}, Hakim and {Natarajan}, Priya and {Egami}, Eiichi and {Livermore}, Rachael and {Bower}, Richard},
        title = "{Mass and magnification maps for the Hubble Space Telescope Frontier Fields clusters: implications for high-redshift studies}",
      journal = {\mnras},
         year = 2014,
        month = oct,
       volume = {444},
       number = {1},
        pages = {268-289},
          doi = {10.1093/mnras/stu1395},
archivePrefix = {arXiv},
       eprint = {1405.3303},
 primaryClass = {astro-ph.CO},
       adsurl = {https://ui.adsabs.harvard.edu/abs/2014MNRAS.444..268R}
}

@ARTICLE{Zitrin2015,
       author = {{Zitrin}, Adi and {Fabris}, Agnese and {Merten}, Julian and {Melchior}, Peter and {Meneghetti}, Massimo and {Koekemoer}, Anton and {Coe}, Dan and {Maturi}, Matteo and {Bartelmann}, Matthias and {Postman}, Marc and {Umetsu}, Keiichi and {Seidel}, Gregor and {Sendra}, Irene and {Broadhurst}, Tom and {Balestra}, Italo and {Biviano}, Andrea and {Grillo}, Claudio and {Mercurio}, Amata and {Nonino}, Mario and {Rosati}, Piero and {Bradley}, Larry and {Carrasco}, Mauricio and {Donahue}, Megan and {Ford}, Holland and {Frye}, Brenda L. and {Moustakas}, John},
        title = "{Hubble Space Telescope Combined Strong and Weak Lensing Analysis of the CLASH Sample: Mass and Magnification Models and Systematic Uncertainties}",
      journal = {\apj},
         year = 2015,
        month = mar,
       volume = {801},
       number = {1},
          eid = {44},
        pages = {44},
          doi = {10.1088/0004-637X/801/1/44},
archivePrefix = {arXiv},
       eprint = {1411.1414},
 primaryClass = {astro-ph.CO},
       adsurl = {https://ui.adsabs.harvard.edu/abs/2015ApJ...801...44Z}
}

@ARTICLE{Sereno2010,
       author = {{Sereno}, M. and {Lubini}, M. and {Jetzer}, Ph.},
        title = "{A multiwavelength strong lensing analysis of baryons and dark matter in the dynamically active cluster AC 114}",
      journal = {\aap},
         year = 2010,
        month = jul,
       volume = {518},
          eid = {A55},
        pages = {A55},
          doi = {10.1051/0004-6361/200913843},
archivePrefix = {arXiv},
       eprint = {0904.0018},
 primaryClass = {astro-ph.CO},
       adsurl = {https://ui.adsabs.harvard.edu/abs/2010A&A...518A..55S}
}

@ARTICLE{Carnall2023,
       author = {{Carnall}, Adam C. and {McLure}, Ross J. and {Dunlop}, James S. and {McLeod}, Derek J. and {Wild}, Vivienne and {Cullen}, Fergus and {Magee}, Dan and {Begley}, Ryan and {Cimatti}, Andrea and {Donnan}, Callum T. and {Hamadouche}, Massissilia L. and {Jewell}, Sophie M. and {Walker}, Sam},
        title = "{A massive quiescent galaxy at redshift 4.658}",
      journal = {\nat},
         year = 2023,
        month = jul,
       volume = {619},
       number = {7971},
        pages = {716-719},
          doi = {10.1038/s41586-023-06158-6},
archivePrefix = {arXiv},
       eprint = {2301.11413},
 primaryClass = {astro-ph.GA},
       adsurl = {https://ui.adsabs.harvard.edu/abs/2023Natur.619..716C}
}

@ARTICLE{Beauchesne2024-erratum,
       author = {{Beauchesne}, Benjamin and {Cl{\'e}ment}, Benjamin and {Hibon}, Pascale and {Limousin}, Marceau and {Eckert}, Dominique and {Kneib}, Jean-Paul and {Richard}, Johan and {Natarajan}, Priyamvada and {Jauzac}, Mathilde and {Montes}, Mireia and {Mahler}, Guillaume and {Claeyssens}, Ad{\'e}la{\"\i}de and {Jeanneau}, Alexandre and {Koekemoer}, Anton M. and {Lagattuta}, David and {Pagul}, Amanda and {S{\'a}nchez}, Javier},
        title = "{Correction to: A new step forward in realistic cluster lens mass modelling: analysis of Hubble Frontier Field Cluster Abell S1063 from joint lensing, X-ray, and galaxy kinematics data}",
      journal = {\mnras},
         year = 2025,
        month = jan,
       volume = {536},
       number = {3},
        pages = {2086-2088},
          doi = {10.1093/mnras/stae2695},
       adsurl = {https://ui.adsabs.harvard.edu/abs/2025MNRAS.536.2086B}
}

@ARTICLE{Willmer2018,
       author = {{Willmer}, Christopher N.~A.},
        title = "{The Absolute Magnitude of the Sun in Several Filters}",
      journal = {\apjs},
         year = 2018,
        month = jun,
       volume = {236},
       number = {2},
          eid = {47},
        pages = {47},
          doi = {10.3847/1538-4365/aabfdf},
archivePrefix = {arXiv},
       eprint = {1804.07788},
 primaryClass = {astro-ph.SR},
       adsurl = {https://ui.adsabs.harvard.edu/abs/2018ApJS..236...47W}
}

@ARTICLE{Chabrier2003,
       author = {{Chabrier}, Gilles},
        title = "{Galactic Stellar and Substellar Initial Mass Function}",
      journal = {\pasp},
         year = 2003,
        month = jul,
       volume = {115},
       number = {809},
        pages = {763-795},
          doi = {10.1086/376392},
archivePrefix = {arXiv},
       eprint = {astro-ph/0304382},
 primaryClass = {astro-ph},
       adsurl = {https://ui.adsabs.harvard.edu/abs/2003PASP..115..763C}
}

@ARTICLE{Arnouts2013,
       author = {{Arnouts}, S. and {Le Floc'h}, E. and {Chevallard}, J. and {Johnson}, B.~D. and {Ilbert}, O. and {Treyer}, M. and {Aussel}, H. and {Capak}, P. and {Sanders}, D.~B. and {Scoville}, N. and {McCracken}, H.~J. and {Milliard}, B. and {Pozzetti}, L. and {Salvato}, M.},
        title = "{Encoding of the infrared excess in the NUVrK color diagram for star-forming galaxies}",
      journal = {\aap},
         year = 2013,
        month = oct,
       volume = {558},
          eid = {A67},
        pages = {A67},
          doi = {10.1051/0004-6361/201321768},
archivePrefix = {arXiv},
       eprint = {1309.0008},
 primaryClass = {astro-ph.CO},
       adsurl = {https://ui.adsabs.harvard.edu/abs/2013A&A...558A..67A}
}

@ARTICLE{Ilbert2015,
       author = {{Ilbert}, O. and {Arnouts}, S. and {Le Floc'h}, E. and {Aussel}, H. and {Bethermin}, M. and {Capak}, P. and {Hsieh}, B. -C. and {Kajisawa}, M. and {Karim}, A. and {Le F{\`e}vre}, O. and {Lee}, N. and {Lilly}, S. and {McCracken}, H.~J. and {Michel-Dansac}, L. and {Moutard}, T. and {Renzini}, M.~A. and {Salvato}, M. and {Sanders}, D.~B. and {Scoville}, N. and {Sheth}, K. and {Silverman}, J.~D. and {Smol{\v{c}}i{\'c}}, V. and {Taniguchi}, Y. and {Tresse}, L.},
        title = "{Evolution of the specific star formation rate function at z< 1.4 Dissecting the mass-SFR plane in COSMOS and GOODS}",
      journal = {\aap},
         year = 2015,
        month = jul,
       volume = {579},
          eid = {A2},
        pages = {A2},
          doi = {10.1051/0004-6361/201425176},
archivePrefix = {arXiv},
       eprint = {1410.4875},
 primaryClass = {astro-ph.GA},
       adsurl = {https://ui.adsabs.harvard.edu/abs/2015A&A...579A...2I}
}

@article{Abell1989,
 adsurl = {https://ui.adsabs.harvard.edu/abs/1989ApJS...70....1A},
 author = {{Abell}, George O. and {Corwin}, Harold G., Jr. and {Olowin}, Ronald P.},
 doi = {10.1086/191333},
 journal = {\apjs},
 month = {May},
 pages = {1},
 title = {{A Catalog of Rich Clusters of Galaxies}},
 volume = {70},
 year = {1989}
}

@ARTICLE{Meneghetti2020,
       author = {{Meneghetti}, Massimo and {Davoli}, Guido and {Bergamini}, Pietro and {Rosati}, Piero and {Natarajan}, Priyamvada and {Giocoli}, Carlo and {Caminha}, Gabriel B. and {Metcalf}, R. Benton and {Rasia}, Elena and {Borgani}, Stefano and {Calura}, Francesco and {Grillo}, Claudio and {Mercurio}, Amata and {Vanzella}, Eros},
        title = "{An excess of small-scale gravitational lenses observed in galaxy clusters}",
      journal = {Science},
         year = 2020,
        month = sep,
       volume = {369},
       number = {6509},
        pages = {1347-1351},
          doi = {10.1126/science.aax5164},
archivePrefix = {arXiv},
       eprint = {2009.04471},
 primaryClass = {astro-ph.GA},
       adsurl = {https://ui.adsabs.harvard.edu/abs/2020Sci...369.1347M}
}

@article{Harvey2015,
 adsurl = {https://ui.adsabs.harvard.edu/abs/2015Sci...347.1462H},
 archiveprefix = {arXiv},
 author = {{Harvey}, David and {Massey}, Richard and {Kitching}, Thomas and {Taylor}, Andy and {Tittley}, Eric},
 doi = {10.1126/science.1261381},
 eprint = {1503.07675},
 journal = {Science},
 month = {March},
 number = {6229},
 pages = {1462-1465},
 primaryclass = {astro-ph.CO},
 title = {{The nongravitational interactions of dark matter in colliding galaxy clusters}},
 volume = {347},
 year = {2015}
}

@article{Clowe2004,
 adsurl = {https://ui.adsabs.harvard.edu/abs/2004ApJ...604..596C},
 archiveprefix = {arXiv},
 author = {{Clowe}, Douglas and {Gonzalez}, Anthony and {Markevitch}, Maxim},
 doi = {10.1086/381970},
 eprint = {astro-ph/0312273},
 journal = {\apj},
 month = {April},
 number = {2},
 pages = {596-603},
 primaryclass = {astro-ph},
 title = {{Weak-Lensing Mass Reconstruction of the Interacting Cluster 1E 0657-558: Direct Evidence for the Existence of Dark Matter}},
 volume = {604},
 year = {2004}
}

@ARTICLE{Spergel2000,
       author = {{Spergel}, David N. and {Steinhardt}, Paul J.},
        title = "{Observational Evidence for Self-Interacting Cold Dark Matter}",
      journal = {\prl},
         year = 2000,
        month = apr,
       volume = {84},
       number = {17},
        pages = {3760-3763},
          doi = {10.1103/PhysRevLett.84.3760},
archivePrefix = {arXiv},
       eprint = {astro-ph/9909386},
 primaryClass = {astro-ph},
       adsurl = {https://ui.adsabs.harvard.edu/abs/2000PhRvL..84.3760S}
}

@ARTICLE{Dai2008,
       author = {{Dai}, De-Chang and {Matsuo}, Reijiro and {Starkman}, Glenn},
        title = "{Gravitational lenses in generalized Einstein-aether theory: The bullet cluster}",
      journal = {\prd},
         year = 2008,
        month = nov,
       volume = {78},
       number = {10},
          eid = {104004},
        pages = {104004},
          doi = {10.1103/PhysRevD.78.104004},
archivePrefix = {arXiv},
       eprint = {0806.4319},
 primaryClass = {gr-qc},
       adsurl = {https://ui.adsabs.harvard.edu/abs/2008PhRvD..78j4004D}
}

@ARTICLE{Eckert2022b,
       author = {{Eckert}, D. and {Ettori}, S. and {Robertson}, A. and {Massey}, R. and {Pointecouteau}, E. and {Harvey}, D. and {McCarthy}, I.~G.},
        title = "{Constraints on dark matter self-interaction from the internal density profiles of X-COP galaxy clusters}",
      journal = {\aap},
         year = 2022,
        month = oct,
       volume = {666},
          eid = {A41},
        pages = {A41},
          doi = {10.1051/0004-6361/202243205},
archivePrefix = {arXiv},
       eprint = {2205.01123},
 primaryClass = {astro-ph.CO},
       adsurl = {https://ui.adsabs.harvard.edu/abs/2022A&A...666A..41E}
}

@ARTICLE{Andrade2022,
       author = {{Andrade}, Kevin E. and {Fuson}, Jackson and {Gad-Nasr}, Sophia and {Kong}, Demao and {Minor}, Quinn and {Roberts}, M. Grant and {Kaplinghat}, Manoj},
        title = "{A stringent upper limit on dark matter self-interaction cross-section from cluster strong lensing}",
      journal = {\mnras},
         year = 2022,
        month = feb,
       volume = {510},
       number = {1},
        pages = {54-81},
          doi = {10.1093/mnras/stab3241},
archivePrefix = {arXiv},
       eprint = {2012.06611},
 primaryClass = {astro-ph.CO},
       adsurl = {https://ui.adsabs.harvard.edu/abs/2022MNRAS.510...54A}
}

@ARTICLE{Natarajan2024,
       author = {{Natarajan}, P. and {Williams}, L.~L.~R. and {Brada{\v{c}}}, M. and {Grillo}, C. and {Ghosh}, A. and {Sharon}, K. and {Wagner}, J.},
        title = "{Strong Lensing by Galaxy Clusters}",
      journal = {\ssr},
         year = 2024,
        month = feb,
       volume = {220},
       number = {2},
          eid = {19},
        pages = {19},
          doi = {10.1007/s11214-024-01051-8},
archivePrefix = {arXiv},
       eprint = {2403.06245},
 primaryClass = {astro-ph.CO},
       adsurl = {https://ui.adsabs.harvard.edu/abs/2024SSRv..220...19N}
}

@article{Mamon2013,
   title={MAMPOSSt: Modelling Anisotropy and Mass Profiles of Observed Spherical Systems – I. Gaussian 3D velocities},
   volume={429},
   ISSN={0035-8711},
   url={http://dx.doi.org/10.1093/mnras/sts565},
   DOI={10.1093/mnras/sts565},
   number={4},
   journal={Monthly Notices of the Royal Astronomical Society},
   publisher={Oxford University Press (OUP)},
   author={Mamon, Gary A. and Biviano, Andrea and Boué, Gwenaël},
   year={2013},
   month=jan, pages={3079–3098} }

@software{Mamon2022,
       author = {{Mamon}, Gary A.},
        title = "{MAMPOSSt: Mass/orbit modeling of spherical systems}",
 howpublished = {Astrophysics Source Code Library, record ascl:2203.020},
         year = 2022,
        month = mar,
          eid = {ascl:2203.020},
       adsurl = {https://ui.adsabs.harvard.edu/abs/2022ascl.soft03020M}
}

@string{june = {June}}

@ARTICLE{Buchner2014,
       author = {{Buchner}, J. and {Georgakakis}, A. and {Nandra}, K. and {Hsu}, L. and {Rangel}, C. and {Brightman}, M. and {Merloni}, A. and {Salvato}, M. and {Donley}, J. and {Kocevski}, D.},
        title = "{X-ray spectral modelling of the AGN obscuring region in the CDFS: Bayesian model selection and catalogue}",
      journal = {\aap},
         year = 2014,
        month = apr,
       volume = {564},
          eid = {A125},
        pages = {A125},
          doi = {10.1051/0004-6361/201322971},
archivePrefix = {arXiv},
       eprint = {1402.0004},
 primaryClass = {astro-ph.HE},
       adsurl = {https://ui.adsabs.harvard.edu/abs/2014A&A...564A.125B}
}

@article{scikit-learn,
  title={Scikit-learn: Machine Learning in {P}ython},
  author={Pedregosa, F. and Varoquaux, G. and Gramfort, A. and Michel, V.
          and Thirion, B. and Grisel, O. and Blondel, M. and Prettenhofer, P.
          and Weiss, R. and Dubourg, V. and Vanderplas, J. and Passos, A. and
          Cournapeau, D. and Brucher, M. and Perrot, M. and Duchesnay, E.},
  journal={Journal of Machine Learning Research},
  volume={12},
  pages={2825--2830},
  year={2011}
}

@ARTICLE{Rix2004,
       author = {{Rix}, Hans-Walter and {Barden}, Marco and {Beckwith}, Steven V.~W. and {Bell}, Eric F. and {Borch}, Andrea and {Caldwell}, John A.~R. and {H{\"a}ussler}, Boris and {Jahnke}, Knud and {Jogee}, Shardha and {McIntosh}, Daniel H. and {Meisenheimer}, Klaus and {Peng}, Chien Y. and {Sanchez}, Sebastian F. and {Somerville}, Rachel S. and {Wisotzki}, Lutz and {Wolf}, Christian},
        title = "{GEMS: Galaxy Evolution from Morphologies and SEDs}",
      journal = {\apjs},
         year = 2004,
        month = jun,
       volume = {152},
       number = {2},
        pages = {163-173},
          doi = {10.1086/420885},
archivePrefix = {arXiv},
       eprint = {astro-ph/0401427},
 primaryClass = {astro-ph},
       adsurl = {https://ui.adsabs.harvard.edu/abs/2004ApJS..152..163R}
}

@ARTICLE{Evans2010,
       author = {{Evans}, Ian N. and {Primini}, Francis A. and {Glotfelty}, Kenny J. and {Anderson}, Craig S. and {Bonaventura}, Nina R. and {Chen}, Judy C. and {Davis}, John E. and {Doe}, Stephen M. and {Evans}, Janet D. and {Fabbiano}, Giuseppina and {Galle}, Elizabeth C. and {Gibbs}, Danny G., II and {Grier}, John D. and {Hain}, Roger M. and {Hall}, Diane M. and {Harbo}, Peter N. and {He}, Xiangqun Helen and {Houck}, John C. and {Karovska}, Margarita and {Kashyap}, Vinay L. and {Lauer}, Jennifer and {McCollough}, Michael L. and {McDowell}, Jonathan C. and {Miller}, Joseph B. and {Mitschang}, Arik W. and {Morgan}, Douglas L. and {Mossman}, Amy E. and {Nichols}, Joy S. and {Nowak}, Michael A. and {Plummer}, David A. and {Refsdal}, Brian L. and {Rots}, Arnold H. and {Siemiginowska}, Aneta and {Sundheim}, Beth A. and {Tibbetts}, Michael S. and {Van Stone}, David W. and {Winkelman}, Sherry L. and {Zografou}, Panagoula},
        title = "{The Chandra Source Catalog}",
      journal = {\apjs},
         year = 2010,
        month = jul,
       volume = {189},
       number = {1},
        pages = {37-82},
          doi = {10.1088/0067-0049/189/1/37},
archivePrefix = {arXiv},
       eprint = {1005.4665},
 primaryClass = {astro-ph.HE},
       adsurl = {https://ui.adsabs.harvard.edu/abs/2010ApJS..189...37E}
}

@ARTICLE{Xie2024,
       author = {{Xie}, Yushan and {Shan}, Huanyuan and {Li}, Nan and {Li}, Ran and {Jullo}, Eric and {Su}, Chen and {Cao}, Xiaoyue and {Kneib}, Jean-Paul and {Acebron}, Ana and {He}, Mengfan and {Yao}, Ji and {Wang}, Chunxiang and {Li}, Jiadong and {Li}, Yin},
        title = "{CURLING - I. The influence of point-like image approximation on the outcomes of cluster strong lens modelling}",
      journal = {\mnras},
         year = 2024,
        month = jun,
       volume = {531},
       number = {1},
        pages = {1179-1190},
          doi = {10.1093/mnras/stae1221},
archivePrefix = {arXiv},
       eprint = {2405.03135},
 primaryClass = {astro-ph.CO},
       adsurl = {https://ui.adsabs.harvard.edu/abs/2024MNRAS.531.1179X}
}

@article{2020SciPy-NMeth,
 adsurl = {https://ui.adsabs.harvard.edu/abs/2020NatMe..17..261V},
 archiveprefix = {arXiv},
 author = {{Virtanen}, Pauli and {Gommers}, Ralf and {Oliphant}, Travis E. and {Haberland}, Matt and {Reddy}, Tyler and {Cournapeau}, David and {Burovski}, Evgeni and {Peterson}, Pearu and {Weckesser}, Warren and {Bright}, Jonathan and {van der Walt}, St{\'e}fan J. and {Brett}, Matthew and {Wilson}, Joshua and {Millman}, K. Jarrod and {Mayorov}, Nikolay and {Nelson}, Andrew R.~J. and {Jones}, Eric and {Kern}, Robert and {Larson}, Eric and {Carey}, C.~J. and {Polat}, {\.I}lhan and {Feng}, Yu and {Moore}, Eric W. and {VanderPlas}, Jake and {Laxalde}, Denis and {Perktold}, Josef and {Cimrman}, Robert and {Henriksen}, Ian and {Quintero}, E.~A. and {Harris}, Charles R. and {Archibald}, Anne M. and {Ribeiro}, Ant{\^o}nio H. and {Pedregosa}, Fabian and {van Mulbregt}, Paul and {SciPy 1. 0 Contributors}},
 doi = {10.1038/s41592-019-0686-2},
 eprint = {1907.10121},
 journal = {Nature Methods},
 month = {February},
 pages = {261-272},
 primaryclass = {cs.MS},
 title = {{SciPy 1.0: fundamental algorithms for scientific computing in Python}},
 volume = {17},
 year = {2020}
}

@article{Amara2012,
 adsurl = {https://ui.adsabs.harvard.edu/abs/2012MNRAS.427..948A},
 archiveprefix = {arXiv},
 author = {{Amara}, Adam and {Quanz}, Sascha P.},
 doi = {10.1111/j.1365-2966.2012.21918.x},
 eprint = {1207.6637},
 journal = {\mnras},
 month = {December},
 number = {2},
 pages = {948-955},
 primaryclass = {astro-ph.IM},
 title = {{PYNPOINT: an image processing package for finding exoplanets}},
 volume = {427},
 year = {2012}
}

@INPROCEEDINGS{Bacon2010,
       author = {{Bacon}, R. and {Accardo}, M. and {Adjali}, L. and {Anwand}, H. and {Bauer}, S. and {Biswas}, I. and {Blaizot}, J. and {Boudon}, D. and {Brau-Nogue}, S. and {Brinchmann}, J. and {Caillier}, P. and {Capoani}, L. and {Carollo}, C.~M. and {Contini}, T. and {Couderc}, P. and {Daguis{\'e}}, E. and {Deiries}, S. and {Delabre}, B. and {Dreizler}, S. and {Dubois}, J. and {Dupieux}, M. and {Dupuy}, C. and {Emsellem}, E. and {Fechner}, T. and {Fleischmann}, A. and {Fran{\c{c}}ois}, M. and {Gallou}, G. and {Gharsa}, T. and {Glindemann}, A. and {Gojak}, D. and {Guiderdoni}, B. and {Hansali}, G. and {Hahn}, T. and {Jarno}, A. and {Kelz}, A. and {Koehler}, C. and {Kosmalski}, J. and {Laurent}, F. and {Le Floch}, M. and {Lilly}, S.~J. and {Lizon}, J. -L. and {Loupias}, M. and {Manescau}, A. and {Monstein}, C. and {Nicklas}, H. and {Olaya}, J. -C. and {Pares}, L. and {Pasquini}, L. and {P{\'e}contal-Rousset}, A. and {Pell{\'o}}, R. and {Petit}, C. and {Popow}, E. and {Reiss}, R. and {Remillieux}, A. and {Renault}, E. and {Roth}, M. and {Rupprecht}, G. and {Serre}, D. and {Schaye}, J. and {Soucail}, G. and {Steinmetz}, M. and {Streicher}, O. and {Stuik}, R. and {Valentin},, H. and {Vernet}, J. and {Weilbacher}, P. and {Wisotzki}, L. and {Yerle}, N.},
        title = "{The MUSE second-generation VLT instrument}",
    booktitle = {Ground-based and Airborne Instrumentation for Astronomy III},
         year = 2010,
       editor = {{McLean}, Ian S. and {Ramsay}, Suzanne K. and {Takami}, Hideki},
       series = {Society of Photo-Optical Instrumentation Engineers (SPIE) Conference Series},
       volume = {7735},
        month = jul,
          eid = {773508},
        pages = {773508},
          doi = {10.1117/12.856027},
archivePrefix = {arXiv},
       eprint = {2211.16795},
 primaryClass = {astro-ph.IM},
       adsurl = {https://ui.adsabs.harvard.edu/abs/2010SPIE.7735E..08B}
}

@software{Bacon2016,
 adsurl = {https://ui.adsabs.harvard.edu/abs/2016ascl.soft11003B},
 author = {{Bacon}, Roland and {Piqueras}, Laure and {Conseil}, Simon and {Richard}, Johan and {Shepherd}, Martin},
 eid = {ascl:1611.003},
 howpublished = {Astrophysics Source Code Library, record ascl:1611.003},
 month = {November},
 title = {{MPDAF: MUSE Python Data Analysis Framework}},
 year = {2016}
}

@article{Beauchesne2024,
 adsurl = {https://ui.adsabs.harvard.edu/abs/2024MNRAS.527.3246B},
 archiveprefix = {arXiv},
 author = {{Beauchesne}, Benjamin and {Cl{\'e}ment}, Benjamin and {Hibon}, Pascale and {Limousin}, Marceau and {Eckert}, Dominique and {Kneib}, Jean-Paul and {Richard}, Johan and {Natarajan}, Priyamvada and {Jauzac}, Mathilde and {Montes}, Mireia and {Mahler}, Guillaume and {Claeyssens}, Ad{\'e}la{\"\i}de and {Jeanneau}, Alexandre and {Koekemoer}, Anton M. and {Lagattuta}, David and {Pagul}, Amanda and {S{\'a}nchez}, Javier},
 doi = {10.1093/mnras/stad3308},
 eprint = {2301.10907},
 journal = {\mnras},
 month = {January},
 number = {2},
 pages = {3246-3275},
 primaryclass = {astro-ph.CO},
 title = {{A new step forward in realistic cluster lens mass modelling: analysis of Hubble Frontier Field Cluster Abell S1063 from joint lensing, X-ray, and galaxy kinematics data}},
 volume = {527},
 year = {2024}
}

@article{Bergamini2019,
 adsurl = {https://ui.adsabs.harvard.edu/abs/2019A&A...631A.130B},
 archiveprefix = {arXiv},
 author = {{Bergamini}, P. and {Rosati}, P. and {Mercurio}, A. and {Grillo}, C. and {Caminha}, G.~B. and {Meneghetti}, M. and {Agnello}, A. and {Biviano}, A. and {Calura}, F. and {Giocoli}, C. and {Lombardi}, M. and {Rodighiero}, G. and {Vanzella}, E.},
 doi = {10.1051/0004-6361/201935974},
 eid = {A130},
 eprint = {1905.13236},
 journal = {\aap},
 month = {November},
 pages = {A130},
 primaryclass = {astro-ph.GA},
 title = {{Enhanced cluster lensing models with measured galaxy kinematics}},
 volume = {631},
 year = {2019}
}

@article{Bertin1996,
 adsurl = {https://ui.adsabs.harvard.edu/abs/1996A&AS..117..393B},
 author = {{Bertin}, E. and {Arnouts}, S.},
 doi = {10.1051/aas:1996164},
 journal = {\aaps},
 month = {June},
 pages = {393-404},
 title = {{SExtractor: Software for source extraction.}},
 volume = {117},
 year = {1996}
}

@article{Bonamigo2017,
 adsurl = {https://ui.adsabs.harvard.edu/abs/2017ApJ...842..132B},
 archiveprefix = {arXiv},
 author = {{Bonamigo}, M. and {Grillo}, C. and {Ettori}, S. and {Caminha}, G.~B. and {Rosati}, P. and {Mercurio}, A. and {Annunziatella}, M. and {Balestra}, I. and {Lombardi}, M.},
 doi = {10.3847/1538-4357/aa75cc},
 eid = {132},
 eprint = {1705.10322},
 journal = {\apj},
 month = {June},
 number = {2},
 pages = {132},
 primaryclass = {astro-ph.GA},
 title = {{Joining X-Ray to Lensing: An Accurate Combined Analysis of MACS J0416.1-2403}},
 volume = {842},
 year = {2017}
}

@article{Bonamigo2018,
 adsurl = {https://ui.adsabs.harvard.edu/abs/2018ApJ...864...98B},
 archiveprefix = {arXiv},
 author = {{Bonamigo}, M. and {Grillo}, C. and {Ettori}, S. and {Caminha}, G.~B. and {Rosati}, P. and {Mercurio}, A. and {Munari}, E. and {Annunziatella}, M. and {Balestra}, I. and {Lombardi}, M.},
 doi = {10.3847/1538-4357/aad4a7},
 eid = {98},
 eprint = {1807.10286},
 journal = {\apj},
 month = {September},
 number = {1},
 pages = {98},
 primaryclass = {astro-ph.GA},
 title = {{Dissection of the Collisional and Collisionless Mass Components in a Mini Sample of CLASH and HFF Massive Galaxy Clusters at z {\ensuremath{\approx}} 0.4}},
 volume = {864},
 year = {2018}
}

@article{Bradac2008,
 adsurl = {https://ui.adsabs.harvard.edu/abs/2008ApJ...687..959B},
 archiveprefix = {arXiv},
 author = {{Brada{\v{c}}}, Maru{\v{s}}a and {Allen}, Steven W. and {Treu}, Tommaso and
{Ebeling}, Harald and {Massey}, Richard and {Morris}, R. Glenn and
{von der Linden}, Anja and {Applegate}, Douglas},
 doi = {10.1086/591246},
 eprint = {0806.2320},
 journal = {\apj},
 month = {November},
 number = {2},
 pages = {959-967},
 primaryclass = {astro-ph},
 title = {{Revealing the Properties of Dark Matter in the Merging Cluster MACS J0025.4-1222}},
 volume = {687},
 year = {2008}
}

@article{Brownstein2007,
 adsurl = {https://ui.adsabs.harvard.edu/abs/2007MNRAS.382...29B},
 archiveprefix = {arXiv},
 author = {{Brownstein}, J.~R. and {Moffat}, J.~W.},
 doi = {10.1111/j.1365-2966.2007.12275.x},
 eprint = {astro-ph/0702146},
 journal = {\mnras},
 month = {November},
 number = {1},
 pages = {29-47},
 primaryclass = {astro-ph},
 title = {{The Bullet Cluster 1E0657-558 evidence shows modified gravity in the absence of dark matter}},
 volume = {382},
 year = {2007}
}

@article{Bruzual2003,
 adsurl = {https://ui.adsabs.harvard.edu/abs/2003MNRAS.344.1000B},
 archiveprefix = {arXiv},
 author = {{Bruzual}, G. and {Charlot}, S.},
 doi = {10.1046/j.1365-8711.2003.06897.x},
 eprint = {astro-ph/0309134},
 journal = {\mnras},
 month = {October},
 number = {4},
 pages = {1000-1028},
 primaryclass = {astro-ph},
 title = {{Stellar population synthesis at the resolution of 2003}},
 volume = {344},
 year = {2003}
}

@article{Buchner2019,
 adsurl = {https://ui.adsabs.harvard.edu/abs/2019PASP..131j8005B},
 archiveprefix = {arXiv},
 author = {{Buchner}, Johannes},
 doi = {10.1088/1538-3873/aae7fc},
 eprint = {1707.04476},
 journal = {\pasp},
 month = {October},
 number = {1004},
 pages = {108005},
 primaryclass = {stat.CO},
 title = {{Collaborative Nested Sampling: Big Data versus Complex Physical Models}},
 volume = {131},
 year = {2019}
}

@article{Buchner2021,
 adsurl = {https://ui.adsabs.harvard.edu/abs/2021JOSS....6.3001B},
 archiveprefix = {arXiv},
 author = {{Buchner}, Johannes},
 doi = {10.21105/joss.03001},
 eid = {3001},
 eprint = {2101.09604},
 journal = {The Journal of Open Source Software},
 month = {April},
 number = {60},
 pages = {3001},
 primaryclass = {stat.CO},
 title = {{UltraNest - a robust, general purpose Bayesian inference engine}},
 volume = {6},
 year = {2021}
}

@article{Calzetti2000,
 adsurl = {https://ui.adsabs.harvard.edu/abs/2000ApJ...533..682C},
 archiveprefix = {arXiv},
 author = {{Calzetti}, Daniela and {Armus}, Lee and {Bohlin}, Ralph C. and {Kinney}, Anne L. and {Koornneef}, Jan and {Storchi-Bergmann}, Thaisa},
 doi = {10.1086/308692},
 eprint = {astro-ph/9911459},
 journal = {\apj},
 month = {April},
 number = {2},
 pages = {682-695},
 primaryclass = {astro-ph},
 title = {{The Dust Content and Opacity of Actively Star-forming Galaxies}},
 volume = {533},
 year = {2000}
}

@article{Cappellari2002,
 adsurl = {https://ui.adsabs.harvard.edu/abs/2002MNRAS.333..400C},
 archiveprefix = {arXiv},
 author = {{Cappellari}, Michele},
 doi = {10.1046/j.1365-8711.2002.05412.x},
 eprint = {astro-ph/0201430},
 journal = {\mnras},
 month = {June},
 number = {2},
 pages = {400-410},
 primaryclass = {astro-ph},
 title = {{Efficient multi-Gaussian expansion of galaxies}},
 volume = {333},
 year = {2002}
}

@article{Cappellari2003,
 adsurl = {https://ui.adsabs.harvard.edu/abs/2003MNRAS.342..345C},
 archiveprefix = {arXiv},
 author = {{Cappellari}, Michele and {Copin}, Yannick},
 doi = {10.1046/j.1365-8711.2003.06541.x},
 eprint = {astro-ph/0302262},
 journal = {\mnras},
 month = {June},
 number = {2},
 pages = {345-354},
 primaryclass = {astro-ph},
 title = {{Adaptive spatial binning of integral-field spectroscopic data using Voronoi tessellations}},
 volume = {342},
 year = {2003}
}

@article{Cappellari2008,
 adsurl = {https://ui.adsabs.harvard.edu/abs/2008MNRAS.390...71C},
 archiveprefix = {arXiv},
 author = {{Cappellari}, Michele},
 doi = {10.1111/j.1365-2966.2008.13754.x},
 eprint = {0806.0042},
 journal = {\mnras},
 month = {October},
 number = {1},
 pages = {71-86},
 primaryclass = {astro-ph},
 title = {{Measuring the inclination and mass-to-light ratio of axisymmetric galaxies via anisotropic Jeans models of stellar kinematics}},
 volume = {390},
 year = {2008}
}

@article{Cappellari2017,
 adsurl = {https://ui.adsabs.harvard.edu/abs/2017MNRAS.466..798C},
 archiveprefix = {arXiv},
 author = {{Cappellari}, Michele},
 doi = {10.1093/mnras/stw3020},
 eprint = {1607.08538},
 journal = {\mnras},
 month = {April},
 number = {1},
 pages = {798-811},
 primaryclass = {astro-ph.GA},
 title = {{Improving the full spectrum fitting method: accurate convolution with Gauss-Hermite functions}},
 volume = {466},
 year = {2017}
}

@article{Carnall18,
 adsurl = {https://ui.adsabs.harvard.edu/abs/2018MNRAS.480.4379C},
 archiveprefix = {arXiv},
 author = {{Carnall}, A.~C. and {McLure}, R.~J. and {Dunlop}, J.~S. and {Dav{\'e}}, R.},
 doi = {10.1093/mnras/sty2169},
 eprint = {1712.04452},
 journal = {\mnras},
 month = {November},
 number = {4},
 pages = {4379-4401},
 primaryclass = {astro-ph.GA},
 title = {{Inferring the star formation histories of massive quiescent galaxies with BAGPIPES: evidence for multiple quenching mechanisms}},
 volume = {480},
 year = {2018}
}

@article{Carnall19,
 adsurl = {https://ui.adsabs.harvard.edu/abs/2019ApJ...873...44C},
 archiveprefix = {arXiv},
 author = {{Carnall}, Adam C. and {Leja}, Joel and {Johnson}, Benjamin D. and {McLure}, Ross J. and {Dunlop}, James S. and {Conroy}, Charlie},
 doi = {10.3847/1538-4357/ab04a2},
 eid = {44},
 eprint = {1811.03635},
 journal = {\apj},
 month = {March},
 number = {1},
 pages = {44},
 primaryclass = {astro-ph.GA},
 title = {{How to Measure Galaxy Star Formation Histories. I. Parametric Models}},
 volume = {873},
 year = {2019}
}

@article{Clowe2006,
 adsurl = {https://ui.adsabs.harvard.edu/abs/2006ApJ...648L.109C},
 archiveprefix = {arXiv},
 author = {{Clowe}, Douglas and {Brada{\v{c}}}, Maru{\v{s}}a and
{Gonzalez}, Anthony H. and {Markevitch}, Maxim and {Randall}, Scott W. and
{Jones}, Christine and {Zaritsky}, Dennis},
 doi = {10.1086/508162},
 eprint = {astro-ph/0608407},
 journal = {\apjl},
 month = {September},
 number = {2},
 pages = {L109-L113},
 primaryclass = {astro-ph},
 title = {{A Direct Empirical Proof of the Existence of Dark Matter}},
 volume = {648},
 year = {2006}
}

@article{Collett2018,
 adsurl = {https://ui.adsabs.harvard.edu/abs/2018Sci...360.1342C},
 archiveprefix = {arXiv},
 author = {{Collett}, Thomas E. and {Oldham}, Lindsay J. and {Smith}, Russell J. and {Auger}, Matthew W. and {Westfall}, Kyle B. and {Bacon}, David and {Nichol}, Robert C. and {Masters}, Karen L. and {Koyama}, Kazuya and {van den Bosch}, Remco},
 doi = {10.1126/science.aao2469},
 eprint = {1806.08300},
 journal = {Science},
 month = {June},
 number = {6395},
 pages = {1342-1346},
 primaryclass = {astro-ph.CO},
 title = {{A precise extragalactic test of General Relativity}},
 volume = {360},
 year = {2018}
}

@article{Conroy2013,
 adsurl = {https://ui.adsabs.harvard.edu/abs/2013ARA&A..51..393C},
 archiveprefix = {arXiv},
 author = {{Conroy}, Charlie},
 doi = {10.1146/annurev-astro-082812-141017},
 eprint = {1301.7095},
 journal = {\araa},
 month = {August},
 number = {1},
 pages = {393-455},
 primaryclass = {astro-ph.CO},
 title = {{Modeling the Panchromatic Spectral Energy Distributions of Galaxies}},
 volume = {51},
 year = {2013}
}

@article{Ettori2013,
 adsurl = {https://ui.adsabs.harvard.edu/abs/2013SSRv..177..119E},
 archiveprefix = {arXiv},
 author = {{Ettori}, Stefano and {Donnarumma}, Annamaria and {Pointecouteau}, Etienne and {Reiprich}, Thomas H. and {Giodini}, Stefania and {Lovisari}, Lorenzo and {Schmidt}, Robert W.},
 doi = {10.1007/s11214-013-9976-7},
 eprint = {1303.3530},
 journal = {\ssr},
 month = {August},
 number = {1-4},
 pages = {119-154},
 primaryclass = {astro-ph.CO},
 title = {{Mass Profiles of Galaxy Clusters from X-ray Analysis}},
 volume = {177},
 year = {2013}
}

@article{Falcon-Barros2011,
 adsurl = {https://ui.adsabs.harvard.edu/abs/2011A&A...532A..95F},
 archiveprefix = {arXiv},
 author = {{Falc{\'o}n-Barroso}, J. and {S{\'a}nchez-Bl{\'a}zquez}, P. and {Vazdekis}, A. and {Ricciardelli}, E. and {Cardiel}, N. and {Cenarro}, A.~J. and {Gorgas}, J. and {Peletier}, R.~F.},
 doi = {10.1051/0004-6361/201116842},
 eid = {A95},
 eprint = {1107.2303},
 journal = {\aap},
 month = {August},
 pages = {A95},
 primaryclass = {astro-ph.CO},
 title = {{An updated MILES stellar library and stellar population models}},
 volume = {532},
 year = {2011}
}

@article{Granata2022,
 adsurl = {https://ui.adsabs.harvard.edu/abs/2022A&A...659A..24G},
 archiveprefix = {arXiv},
 author = {{Granata}, G. and {Mercurio}, A. and {Grillo}, C. and {Tortorelli}, L. and {Bergamini}, P. and {Meneghetti}, M. and {Rosati}, P. and {Caminha}, G.~B. and {Nonino}, M.},
 doi = {10.1051/0004-6361/202141817},
 eid = {A24},
 eprint = {2107.09079},
 journal = {\aap},
 month = {March},
 pages = {A24},
 primaryclass = {astro-ph.GA},
 title = {{Improved strong lensing modelling of galaxy clusters using the Fundamental Plane: Detailed mapping of the baryonic and dark matter mass distribution of Abell S1063}},
 volume = {659},
 year = {2022}
}

@article{Herbel2018,
 adsurl = {https://ui.adsabs.harvard.edu/abs/2018JCAP...07..054H},
 archiveprefix = {arXiv},
 author = {{Herbel}, J{\"o}rg and {Kacprzak}, Tomasz and {Amara}, Adam and {Refregier}, Alexandre and {Lucchi}, Aurelien},
 doi = {10.1088/1475-7516/2018/07/054},
 eid = {054},
 eprint = {1801.07615},
 journal = {\jcap},
 month = {July},
 number = {7},
 pages = {054},
 primaryclass = {astro-ph.IM},
 title = {{Fast point spread function modeling with deep learning}},
 volume = {2018},
 year = {2018}
}

@article{horne1986,
 adsurl = {https://ui.adsabs.harvard.edu/abs/1986PASP...98..609H},
 author = {{Horne}, K.},
 doi = {10.1086/131801},
 journal = {\pasp},
 month = {June},
 pages = {609-617},
 title = {{An optimal extraction algorithm for CCD spectroscopy.}},
 volume = {98},
 year = {1986}
}

@software{JAX,
 adsurl = {https://ui.adsabs.harvard.edu/abs/2021ascl.soft11002B},
 author = {{Bradbury}, James and {Frostig}, Roy and {Hawkins}, Peter and {Johnson}, Matthew James and {Leary}, Chris and {Maclaurin}, Dougal and {Necula}, George and {Paszke}, Adam and {VanderPlas}, Jake and {Wanderman-Milne}, Skye and {Zhang}, Qiao},
 eid = {ascl:2111.002},
 howpublished = {Astrophysics Source Code Library, record ascl:2111.002},
 month = {November},
 title = {{JAX: Autograd and XLA}},
 year = {2021}
}

@article{Kassiola1993,
 adsurl = {https://ui.adsabs.harvard.edu/abs/1993ApJ...417..450K},
 author = {{Kassiola}, Aggeliki and {Kovner}, Israel},
 doi = {10.1086/173325},
 journal = {\apj},
 month = {November},
 pages = {450},
 title = {{Elliptic Mass Distributions versus Elliptic Potentials in Gravitational Lenses}},
 volume = {417},
 year = {1993}
}

@article{Kluge2020,
 adsurl = {https://ui.adsabs.harvard.edu/abs/2020ApJS..247...43K},
 archiveprefix = {arXiv},
 author = {{Kluge}, M. and {Neureiter}, B. and {Riffeser}, A. and {Bender}, R. and {Goessl}, C. and {Hopp}, U. and {Schmidt}, M. and {Ries}, C. and {Brosch}, N.},
 doi = {10.3847/1538-4365/ab733b},
 eid = {43},
 eprint = {1908.08544},
 journal = {\apjs},
 month = {April},
 number = {2},
 pages = {43},
 primaryclass = {astro-ph.GA},
 title = {{Structure of Brightest Cluster Galaxies and Intracluster Light}},
 volume = {247},
 year = {2020}
}

@article{Kneib2012,
 adsurl = {https://ui.adsabs.harvard.edu/abs/2011A&ARv..19...47K},
 archiveprefix = {arXiv},
 author = {{Kneib}, Jean-Paul and {Natarajan}, Priyamvada},
 doi = {10.1007/s00159-011-0047-3},
 eid = {47},
 eprint = {1202.0185},
 journal = {\aapr},
 month = {November},
 pages = {47},
 primaryclass = {astro-ph.CO},
 title = {{Cluster lenses}},
 volume = {19},
 year = {2011}
}

@article{Kroupa2001,
 adsurl = {https://ui.adsabs.harvard.edu/abs/2001MNRAS.322..231K},
 archiveprefix = {arXiv},
 author = {{Kroupa}, Pavel},
 doi = {10.1046/j.1365-8711.2001.04022.x},
 eprint = {astro-ph/0009005},
 journal = {\mnras},
 month = {April},
 number = {2},
 pages = {231-246},
 primaryclass = {astro-ph},
 title = {{On the variation of the initial mass function}},
 volume = {322},
 year = {2001}
}

@article{Limousin2005,
 adsurl = {https://ui.adsabs.harvard.edu/abs/2005MNRAS.356..309L},
 archiveprefix = {arXiv},
 author = {{Limousin}, Marceau and {Kneib}, Jean-Paul and {Natarajan}, Priyamvada},
 doi = {10.1111/j.1365-2966.2004.08449.x},
 eprint = {astro-ph/0405607},
 journal = {\mnras},
 month = {January},
 number = {1},
 pages = {309-322},
 primaryclass = {astro-ph},
 title = {{Constraining the mass distribution of galaxies using galaxy-galaxy lensing in clusters and in the field}},
 volume = {356},
 year = {2005}
}

@article{Limousin2022,
 adsurl = {https://ui.adsabs.harvard.edu/abs/2022A&A...664A..90L},
 archiveprefix = {arXiv},
 author = {{Limousin}, Marceau and {Beauchesne}, Benjamin and {Jullo}, Eric},
 doi = {10.1051/0004-6361/202243278},
 eid = {A90},
 eprint = {2202.02992},
 journal = {\aap},
 month = {August},
 pages = {A90},
 primaryclass = {astro-ph.CO},
 title = {{Dark matter in galaxy clusters: Parametric strong-lensing approach}},
 volume = {664},
 year = {2022}
}

@article{Longobardi18,
 adsurl = {https://ui.adsabs.harvard.edu/abs/2018A&A...620A.111L},
 archiveprefix = {arXiv},
 author = {{Longobardi}, A. and {Arnaboldi}, M. and {Gerhard}, O. and {Pulsoni}, C. and {S{\"o}ldner-Rembold}, I.},
 doi = {10.1051/0004-6361/201832729},
 eid = {A111},
 eprint = {1809.10708},
 journal = {\aap},
 month = {December},
 pages = {A111},
 primaryclass = {astro-ph.GA},
 title = {{Kinematics of the outer halo of M 87 as mapped by planetary nebulae{\ensuremath{\star}}}},
 volume = {620},
 year = {2018}
}

@article{lotz2017,
 adsurl = {https://ui.adsabs.harvard.edu/abs/2017ApJ...837...97L},
 archiveprefix = {arXiv},
 author = {{Lotz}, J.~M. and {Koekemoer}, A. and {Coe}, D. and {Grogin}, N. and
{Capak}, P. and {Mack}, J. and {Anderson}, J. and {Avila}, R. and
{Barker}, E.~A. and {Borncamp}, D. and {Brammer}, G. and {Durbin}, M. and
{Gunning}, H. and {Hilbert}, B. and {Jenkner}, H. and {Khandrika}, H. and
{Levay}, Z. and {Lucas}, R.~A. and {MacKenty}, J. and {Ogaz}, S. and
{Porterfield}, B. and {Reid}, N. and {Robberto}, M. and {Royle}, P. and
{Smith}, L.~J. and {Storrie-Lombardi}, L.~J. and {Sunnquist}, B. and
{Surace}, J. and {Taylor}, D.~C. and {Williams}, R. and {Bullock}, J. and
{Dickinson}, M. and {Finkelstein}, S. and {Natarajan}, P. and
{Richard}, J. and {Robertson}, B. and {Tumlinson}, J. and {Zitrin}, A. and
{Flanagan}, K. and {Sembach}, K. and {Soifer}, B.~T. and {Mountain}, M.},
 doi = {10.3847/1538-4357/837/1/97},
 eid = {97},
 eprint = {1605.06567},
 journal = {\apj},
 month = {March},
 number = {1},
 pages = {97},
 primaryclass = {astro-ph.GA},
 title = {{The Frontier Fields: Survey Design and Initial Results}},
 volume = {837},
 year = {2017}
}

@article{Mercurio2021,
 adsurl = {https://ui.adsabs.harvard.edu/abs/2021A&A...656A.147M},
 archiveprefix = {arXiv},
 author = {{Mercurio}, A. and {Rosati}, P. and {Biviano}, A. and {Annunziatella}, M. and {Girardi}, M. and {Sartoris}, B. and {Nonino}, M. and {Brescia}, M. and {Riccio}, G. and {Grillo}, C. and {Balestra}, I. and {Caminha}, G.~B. and {De Lucia}, G. and {Gobat}, R. and {Seitz}, S. and {Tozzi}, P. and {Scodeggio}, M. and {Vanzella}, E. and {Angora}, G. and {Bergamini}, P. and {Borgani}, S. and {Demarco}, R. and {Meneghetti}, M. and {Strazzullo}, V. and {Tortorelli}, L. and {Umetsu}, K. and {Fritz}, A. and {Gruen}, D. and {Kelson}, D. and {Lombardi}, M. and {Maier}, C. and {Postman}, M. and {Rodighiero}, G. and {Ziegler}, B.},
 doi = {10.1051/0004-6361/202142168},
 eid = {A147},
 eprint = {2109.03305},
 journal = {\aap},
 month = {December},
 pages = {A147},
 primaryclass = {astro-ph.GA},
 title = {{CLASH-VLT: Abell S1063. Cluster assembly history and spectroscopic catalogue}},
 volume = {656},
 year = {2021}
}

@ARTICLE{Speagle2014,
       author = {{Speagle}, J.~S. and {Steinhardt}, C.~L. and {Capak}, P.~L. and {Silverman}, J.~D.},
        title = "{A Highly Consistent Framework for the Evolution of the Star-Forming ``Main Sequence'' from z \raisebox{-0.5ex}\textasciitilde 0-6}",
      journal = {\apjs},
         year = 2014,
        month = oct,
       volume = {214},
       number = {2},
          eid = {15},
        pages = {15},
          doi = {10.1088/0067-0049/214/2/15},
archivePrefix = {arXiv},
       eprint = {1405.2041},
 primaryClass = {astro-ph.GA},
       adsurl = {https://ui.adsabs.harvard.edu/abs/2014ApJS..214...15S}
}

@ARTICLE{Montes2014,
       author = {{Montes}, Mireia and {Trujillo}, Ignacio},
        title = "{Intracluster Light at the Frontier: A2744}",
      journal = {\apj},
         year = 2014,
        month = oct,
       volume = {794},
       number = {2},
          eid = {137},
        pages = {137},
          doi = {10.1088/0004-637X/794/2/137},
archivePrefix = {arXiv},
       eprint = {1405.2070},
 primaryClass = {astro-ph.CO},
       adsurl = {https://ui.adsabs.harvard.edu/abs/2014ApJ...794..137M}
}

@article{Montes2018,
 adsurl = {https://ui.adsabs.harvard.edu/abs/2018MNRAS.474..917M},
 archiveprefix = {arXiv},
 author = {{Montes}, Mireia and {Trujillo}, Ignacio},
 doi = {10.1093/mnras/stx2847},
 eprint = {1710.03240},
 journal = {\mnras},
 month = {February},
 number = {1},
 pages = {917-932},
 primaryclass = {astro-ph.CO},
 title = {{Intracluster light at the Frontier - II. The Frontier Fields Clusters}},
 volume = {474},
 year = {2018}
}

@article{Montes2022,
 adsurl = {https://ui.adsabs.harvard.edu/abs/2022NatAs...6..308M},
 archiveprefix = {arXiv},
 author = {{Montes}, Mireia},
 doi = {10.1038/s41550-022-01616-z},
 eprint = {2203.06199},
 journal = {Nature Astronomy},
 month = {March},
 pages = {308-316},
 primaryclass = {astro-ph.GA},
 title = {{The faint light in groups and clusters of galaxies}},
 volume = {6},
 year = {2022}
}

@article{multinest,
 adsurl = {https://ui.adsabs.harvard.edu/abs/2019OJAp....2E..10F},
 archiveprefix = {arXiv},
 author = {{Feroz}, Farhan and {Hobson}, Michael P. and {Cameron}, Ewan and {Pettitt}, Anthony N.},
 doi = {10.21105/astro.1306.2144},
 eid = {10},
 eprint = {1306.2144},
 journal = {The Open Journal of Astrophysics},
 month = {November},
 number = {1},
 pages = {10},
 primaryclass = {astro-ph.IM},
 title = {{Importance Nested Sampling and the MultiNest Algorithm}},
 volume = {2},
 year = {2019}
}

@article{Natarajan2017,
 adsurl = {https://ui.adsabs.harvard.edu/abs/2017MNRAS.468.1962N},
 archiveprefix = {arXiv},
 author = {{Natarajan}, Priyamvada and {Chadayammuri}, Urmila and
{Jauzac}, Mathilde and {Richard}, Johan and {Kneib}, Jean-Paul and
{Ebeling}, Harald and {Jiang}, Fangzhou and {van den Bosch}, Frank and
{Limousin}, Marceau and {Jullo}, Eric and {Atek}, Hakim and
{Pillepich}, Annalisa and {Popa}, Cristina and {Marinacci}, Federico and
{Hernquist}, Lars and {Meneghetti}, Massimo and {Vogelsberger}, Mark},
 doi = {10.1093/mnras/stw3385},
 eprint = {1702.04348},
 journal = {\mnras},
 month = {June},
 number = {2},
 pages = {1962-1980},
 primaryclass = {astro-ph.GA},
 title = {{Mapping substructure in the HST Frontier Fields cluster lenses and in cosmological simulations}},
 volume = {468},
 year = {2017}
}

@article{Newman2013b,
 adsurl = {https://ui.adsabs.harvard.edu/abs/2013ApJ...765...25N},
 archiveprefix = {arXiv},
 author = {{Newman}, Andrew B. and {Treu}, Tommaso and {Ellis}, Richard S. and {Sand}, David J.},
 doi = {10.1088/0004-637X/765/1/25},
 eid = {25},
 eprint = {1209.1392},
 journal = {\apj},
 month = {March},
 number = {1},
 pages = {25},
 primaryclass = {astro-ph.CO},
 title = {{The Density Profiles of Massive, Relaxed Galaxy Clusters. II. Separating Luminous and Dark Matter in Cluster Cores}},
 volume = {765},
 year = {2013}
}

@article{Postman2012,
 adsurl = {https://ui.adsabs.harvard.edu/abs/2012ApJS..199...25P},
 archiveprefix = {arXiv},
 author = {{Postman}, Marc and {Coe}, Dan and {Ben{\'\i}tez}, Narciso and
{Bradley}, Larry and {Broadhurst}, Tom and {Donahue}, Megan and
{Ford}, Holland and {Graur}, Or and {Graves}, Genevieve and
{Jouvel}, Stephanie and {Koekemoer}, Anton and {Lemze}, Doron and
{Medezinski}, Elinor and {Molino}, Alberto and {Moustakas}, Leonidas and
{Ogaz}, Sara and {Riess}, Adam and {Rodney}, Steve and {Rosati}, Piero and
{Umetsu}, Keiichi and {Zheng}, Wei and {Zitrin}, Adi and
{Bartelmann}, Matthias and {Bouwens}, Rychard and {Czakon}, Nicole and
{Golwala}, Sunil and {Host}, Ole and {Infante}, Leopoldo and
{Jha}, Saurabh and {Jimenez-Teja}, Yolanda and {Kelson}, Daniel and
{Lahav}, Ofer and {Lazkoz}, Ruth and {Maoz}, Dani and
{McCully}, Curtis and {Melchior}, Peter and {Meneghetti}, Massimo and
{Merten}, Julian and {Moustakas}, John and {Nonino}, Mario and
{Patel}, Brandon and {Reg{\"o}s}, Enik{\"o} and {Sayers}, Jack and
{Seitz}, Stella and {Van der Wel}, Arjen},
 doi = {10.1088/0067-0049/199/2/25},
 eid = {25},
 eprint = {1106.3328},
 journal = {\apjs},
 month = {April},
 number = {2},
 pages = {25},
 primaryclass = {astro-ph.CO},
 title = {{The Cluster Lensing and Supernova Survey with Hubble: An Overview}},
 volume = {199},
 year = {2012}
}

@article{pymultinest,
 adsurl = {https://ui.adsabs.harvard.edu/abs/2014A&A...564A.125B},
 archiveprefix = {arXiv},
 author = {{Buchner}, J. and {Georgakakis}, A. and {Nandra}, K. and {Hsu}, L. and {Rangel}, C. and {Brightman}, M. and {Merloni}, A. and {Salvato}, M. and {Donley}, J. and {Kocevski}, D.},
 doi = {10.1051/0004-6361/201322971},
 eid = {A125},
 eprint = {1402.0004},
 journal = {\aap},
 month = {April},
 pages = {A125},
 primaryclass = {astro-ph.HE},
 title = {{X-ray spectral modelling of the AGN obscuring region in the CDFS: Bayesian model selection and catalogue}},
 volume = {564},
 year = {2014}
}

@article{Robertson2021,
 adsurl = {https://ui.adsabs.harvard.edu/abs/2021MNRAS.501.4610R},
 archiveprefix = {arXiv},
 author = {{Robertson}, Andrew and {Massey}, Richard and {Eke}, Vincent and {Schaye}, Joop and {Theuns}, Tom},
 doi = {10.1093/mnras/staa3954},
 eprint = {2009.07844},
 journal = {\mnras},
 month = {March},
 number = {3},
 pages = {4610-4634},
 primaryclass = {astro-ph.CO},
 title = {{The surprising accuracy of isothermal Jeans modelling of self-interacting dark matter density profiles}},
 volume = {501},
 year = {2021}
}

@article{Sagunski2020,
 adsurl = {https://ui.adsabs.harvard.edu/abs/2021JCAP...01..024S},
 archiveprefix = {arXiv},
 author = {{Sagunski}, Laura and {Gad-Nasr}, Sophia and {Colquhoun}, Brian and {Robertson}, Andrew and {Tulin}, Sean},
 doi = {10.1088/1475-7516/2021/01/024},
 eid = {024},
 eprint = {2006.12515},
 journal = {\jcap},
 month = {January},
 number = {1},
 pages = {024},
 primaryclass = {astro-ph.CO},
 title = {{Velocity-dependent self-interacting dark matter from groups and clusters of galaxies}},
 volume = {2021},
 year = {2021}
}

@article{sersic1963,
 adsurl = {https://ui.adsabs.harvard.edu/abs/1963BAAA....6...41S},
 author = {{S{\'e}rsic}, J.~L.},
 journal = {Boletin de la Asociacion Argentina de Astronomia La Plata Argentina},
 month = {February},
 pages = {41-43},
 title = {{Influence of the atmospheric and instrumental dispersion on the brightness distribution in a galaxy}},
 volume = {6},
 year = {1963}
}

@article{steinhardt2020,
 adsurl = {https://ui.adsabs.harvard.edu/abs/2020ApJS..247...64S},
 archiveprefix = {arXiv},
 author = {{Steinhardt}, Charles L. and {Jauzac}, Mathilde and {Acebron}, Ana and {Atek}, Hakim and {Capak}, Peter and {Davidzon}, Iary and {Eckert}, Dominique and {Harvey}, David and {Koekemoer}, Anton M. and {Lagos}, Claudia D.~P. and {Mahler}, Guillaume and {Montes}, Mireia and {Niemiec}, Anna and {Nonino}, Mario and {Oesch}, P.~A. and {Richard}, Johan and {Rodney}, Steven A. and {Schaller}, Matthieu and {Sharon}, Keren and {Strolger}, Louis-Gregory and {Allingham}, Joseph and {Amara}, Adam and {Bah{\'e}}, Yannick and {B{\oe}hm}, C{\'e}line and {Bose}, Sownak and {Bouwens}, Rychard J. and {Bradley}, Larry D. and {Brammer}, Gabriel and {Broadhurst}, Tom and {Ca{\~n}as}, Rodrigo and {Cen}, Renyue and {Cl{\'e}ment}, Benjamin and {Clowe}, Douglas and {Coe}, Dan and {Connor}, Thomas and {Darvish}, Behnam and {Diego}, Jose M. and {Ebeling}, Harald and {Edge}, A.~C. and {Egami}, Eiichi and {Ettori}, Stefano and {Faisst}, Andreas L. and {Frye}, Brenda and {Furtak}, Lukas J. and {G{\'o}mez-Guijarro}, C. and {Remolina Gonz{\'a}lez}, J.~D. and {Gonzalez}, Anthony and {Graur}, Or and {Gruen}, Daniel and {Harvey}, David and {Hensley}, Hagan and {Hovis-Afflerbach}, Beryl and {Jablonka}, Pascale and {Jha}, Saurabh W. and {Jullo}, Eric and {Kneib}, Jean-Paul and {Kokorev}, Vasily and {Lagattuta}, David J. and {Limousin}, Marceau and {von der Linden}, Anja and {Linzer}, Nora B. and {Lopez}, Adrian and {Magdis}, Georgios E. and {Massey}, Richard and {Masters}, Daniel C. and {Maturi}, Matteo and {McCully}, Curtis and {McGee}, Sean L. and {Meneghetti}, Massimo and {Mobasher}, Bahram and {Moustakas}, Leonidas A. and {Murphy}, Eric J. and {Natarajan}, Priyamvada and {Neyrinck}, Mark and {O'Connor}, Kyle and {Oguri}, Masamune and {Pagul}, Amanda and {Rhodes}, Jason and {Rich}, R. Michael and {Robertson}, Andrew and {Sereno}, Mauro and {Shan}, Huanyuan and {Smith}, Graham P. and {Sneppen}, Albert and {Squires}, Gordon K. and {Tam}, Sut-Ieng and {Tchernin}, C{\'e}line and {Toft}, Sune and {Umetsu}, Keiichi and {Weaver}, John R. and {van Weeren}, R.~J. and {Williams}, Liliya L.~R. and {Wilson}, Tom J. and {Yan}, Lin and {Zitrin}, Adi},
 doi = {10.3847/1538-4365/ab75ed},
 eid = {64},
 eprint = {2001.09999},
 journal = {\apjs},
 month = {April},
 number = {2},
 pages = {64},
 primaryclass = {astro-ph.GA},
 title = {{The BUFFALO HST Survey}},
 volume = {247},
 year = {2020}
}

@article{tortorelli2018,
 adsurl = {https://ui.adsabs.harvard.edu/abs/2018MNRAS.477..648T},
 archiveprefix = {arXiv},
 author = {{Tortorelli}, Luca and {Mercurio}, Amata and {Paolillo}, Maurizio and {Rosati}, Piero and {Gargiulo}, Adriana and {Gobat}, Raphael and {Balestra}, Italo and {Caminha}, G.~B. and {Annunziatella}, Marianna and {Grillo}, Claudio and {Lombardi}, Marco and {Nonino}, Mario and {Rettura}, Alessandro and {Sartoris}, Barbara and {Strazzullo}, Veronica},
 doi = {10.1093/mnras/sty617},
 eprint = {1803.02375},
 journal = {\mnras},
 month = {June},
 number = {1},
 pages = {648-668},
 primaryclass = {astro-ph.GA},
 title = {{The Kormendy relation of galaxies in the Frontier Fields clusters: Abell S1063 and MACS J1149.5+2223}},
 volume = {477},
 year = {2018}
}

@article{tortorelli2023,
 adsurl = {https://ui.adsabs.harvard.edu/abs/2023FrASS..1089443T},
 archiveprefix = {arXiv},
 author = {{Tortorelli}, Luca and {Mercurio}, Amata},
 doi = {10.3389/fspas.2023.989443},
 eid = {51},
 eprint = {2302.07890},
 journal = {Frontiers in Astronomy and Space Sciences},
 month = {March},
 pages = {51},
 primaryclass = {astro-ph.GA},
 title = {{MORPHOFIT: An automated galaxy structural parameters fitting package}},
 volume = {10},
 year = {2023}
}

@article{valdes2004,
 adsurl = {https://ui.adsabs.harvard.edu/abs/2004ApJS..152..251V},
 archiveprefix = {arXiv},
 author = {{Valdes}, Francisco and {Gupta}, Ranjan and {Rose}, James A. and {Singh}, Harinder P. and {Bell}, David J.},
 doi = {10.1086/386343},
 eprint = {astro-ph/0402435},
 journal = {\apjs},
 month = {June},
 number = {2},
 pages = {251-259},
 primaryclass = {astro-ph},
 title = {{The Indo-US Library of Coud{\'e} Feed Stellar Spectra}},
 volume = {152},
 year = {2004}
}




\appendix

\section{\textsc{Sextractor} parameter}
\label{app:sextractor_param}
The \textsc{Sextractor} parameters used in the cold and hot pass are listed in Table~\ref{Tab:sextractor_param}

\begin{table}
\centering
\begin{tabular}{ccc}
    \hline
    Parameters & Cold & Hot\\
    \hline
    \textsc{DETECT\_MINAREA} &$5$ &$5$\\
    \textsc{DETECT\_THRESH} &$5$&$1.8$\\
    \textsc{ANALYSIS\_THRESH} &$6$ &$2$\\
    \textsc{FILTER} &\textsc{Y} &\textsc{Y}\\
    \textsc{FILTER\_NAME} &\textsc{default.conv} &\textsc{default.conv}\\
    \textsc{DEBLEND\_NTHRESH} &$12$ &$64$\\
    \textsc{DEBLEND\_MINCONT} &$0.001$ &$0.0001$\\
    \textsc{BACK\_SIZE} &$32$ &$32$\\
    \textsc{BACK\_FILTERSIZE} &$3$ &$3$\\
    \textsc{BACKPHOTO\_TYPE} &\textsc{GLOBAL} &\textsc{GLOBAL}\\

    \hline
\end{tabular}
\caption{\textsc{Sextractor} parameters used in the cold and hot pass to create a catalogue for AS1063.}
\label{Tab:sextractor_param}
\end{table}
\section{Light model}
\label{app:light_model}
As we automatically fit a large number of galaxies, we modify the analytical expression of the considered profiles so that there are no undefined points in our parameter bounds. This can obfuscate the meaning of the parameters, but it allows the code to avoid some crashes. The parameters are converted to the usual definitions after the fit.

The usual definitions of the S\'ersic or dPIE profiles projected on the plane of the sky are the following:
\begin{align}
    I_{\rm S\'ersic}=I_0\exp \left( -b_n\left[ \left(\frac{R}{R_e}\right)^{1/n} -1\right] \right)\\
    I_{\rm dPIE}=\frac{\Sigma_0 r_{\rm cut}}{r_{\rm cut}-r_{\rm core}}\left(\frac{1}{\sqrt{r^2_{\rm core}+R^2}}-\frac{1}{\sqrt{r^2_{\rm cut}+R^2}} \right)
\end{align}
where $I_0$, $R_e$ and $n$ are the normalisation, half-light radius and the S\'ersic index, respectively. They are the free parameters of the profile, and $b_n$ can be computed as the solution of $\gamma(2n,b_n)=\Gamma(2n)/2$. For the dPIE profile, $\Sigma_0$, $r_{\rm cut}$ and  $r_{\rm core}$ are the normalisation, cut, and core radii, respectively.

To ensure a better parameter space and improve the computation speed for the S\'ersic profile, we are not computing $b_n$ as we replaced $R_e$ with another parameter, $h$, that encompasses its values with $R_e$. We also replace $n$ by $\beta=n^{-1}$ and $I_0$ with $I_{0,n}=I_0\exp b_n$ such that we obtain:
\begin{equation}
    I_{\rm S\'ersic}=I_{0,n}\exp \left( -h R^{\beta} \right)
\end{equation}
From this expression, we can then deduce the expression of $h$ as $h=b_n R_e^{-1/n}$.

For the dPIE profile, we simplify the normalisation and switch $r_{\rm core}$ by $a_{\rm core}$ such that:
\begin{equation}
    I_{\rm dPIE}=\Sigma^*_0 \left(\frac{1}{\sqrt{a^2_{\rm core}r^2_{\rm cut}+R^2}}-\frac{1}{\sqrt{r^2_{\rm cut}+R^2}} \right)
\end{equation}
With the following expression for $a_{\rm core}$ and $\Sigma'_0$:
\begin{align}
    a_{\rm core}=r_{\rm core}/r_{\rm cut}\\
    \Sigma^*_0=\frac{\Sigma_0 r_{\rm cut}}{r_{\rm cut}-r_{\rm core}}\\
\end{align}
Thanks to our redefinition of the dPIE parameter, we can ensure that $r_{\rm cut}>r_{\rm core}$ by letting $a_{\rm core}$ in $]0,1[$.

\section{Fit on regions stamp: Initial parameter estimation}
\label{app:init_param_LM}
In section~\ref{sect:galaxy-light:region-fit}, we detail our procedure to fit multiple objects light distributions in a postage stamp image. The optimisation procedure is based on a non-linear least-squares algorithm, which requires a guess for light model parameters as a starting point. For the coordinates centre, position angle and ellipticity, we are using \textsc{Sextractor} measurements of \textsc{XWIN\_IMAGE}, \textsc{YWIN\_IMAGE},\textsc{THETAWIN\_IMAGE}, \textsc{AWIN\_IMAGE} and \textsc{BWIN\_IMAGE}. For the S\'ersic profile, we start with an index $n=2.5$, \textsc{FLUX\_RADIUS} measurements for $R_e$, and we estimate $I_0$ based on the \textsc{MAG\_BEST} values. These parameters are then converted to our effective parameters defined in Sect.~\ref{app:light_model}. To adopt similar starting points for dPIE, we fit dPIE profiles in 1D on a S\'ersic profile with the previous parameter set. We use the fit output as a starting point for the least-squares fit.

Regarding the parameter bounds, we choose to use large bounds to avoid hitting them, ensuring some safety against invalid parameter sets with our redefinition of the light profile. We make an exception for the coordinate centre, which we constrain around the \textsc{Sextractor} estimation. The bounds are defined as follows:
\begin{align}
    &\text{axis ratio}:\,\mathcal{U}(1,20)\\
    &x_c:\,\mathcal{U}(x_{c, \rm init}-1,x_{c, \rm init}+1)\\
    &y_c:\,\mathcal{U}(y_{c, \rm init}-1,y_{c, \rm init}+1)\\
    &\theta:\,\mathcal{U}(-\pi/2,\pi)\\
    &h:\,\mathcal{U}(0,100 h_{\rm init})\\
    &\beta:\,\mathcal{U}(0.1,5)\\
    &I_{0,n}:\,\mathcal{U}(0,I^{\rm init}_{0,n})\\
    &a_{\rm core}:\,\mathcal{U}(0,1)\\
    &r_{\rm cut}:\,\mathcal{U}(0,100 r_{\rm cut,init})\\
    &\Sigma^*_0:\,\mathcal{U}(0,100\Sigma^*_{0, \rm init})
\end{align}

We note that in some cases, these bounds were too large and prevented convergence for faint objects or in particularly crowded regions. The main parameters in question are the $\text{axis ratio}$ and $h$ or $r_{\rm cut}$, which tend to converge to overestimated values. 

\section{Light profile fit: $\sinh^{-1}$ likelihood}
\label{app:Likelihood_checks}
\subsection{Improved robustness}
\begin{table}
\centering
\begin{tabular}{cc}
    \hline
    Parameters & Distribution\\
    \hline
    $\log I_0$& $\mathcal{U}(-1,1.5)$\\
    $R_e$& $\mathcal{U}(1,30)$\\
    $\log I^{\rm dPIE}_0$& $\mathcal{U}(-4,4)$\\
    $\log r_{\rm cut}$& $\mathcal{U}(-4,5)$\\
    $r_{\rm core}/r_{\rm cut}$& $\mathcal{U}(0,1)$\\
    \hline
\end{tabular}
\caption{Distributions for the parameters used to create the mock data and to select initial parameters for the least-square fit.}
\label{Tab:dist-fit}
\end{table}

To show the benefit of the $\sinh^{-1}$ transformation, we did least-squares fits of a dPIE on mock 1D light profiles. We created mock profiles by simulating S\'ersic profiles with an index $n=4$ with a random half-light radius $R_e$ and central intensity $I_0$. The input distributions of $R_e$ and $I_0$ are presented in Table~\ref{Tab:dist-fit}. We select the parameter such that most mock data can be fitted reasonably well by a dPIE. We added $1$ per cent of Gaussian noise to each mock. Overall, we created $500$ distinct mocks. We sample their light profiles with $50$ points from $0.2$ arcsec to $2\times R_e$ linearly spaced, similarly to the pixels of an image.

For each mock, we performed $500$ a non-linear least-squares fits with a regular Gaussian likelihood and with the $\sinh^{-1}$. The optimiser is the same as the one used for galaxy light profile fit presented in sect.~\ref{sect:galaxy-light}. Each fit is done with a different starting point, which we randomly select. We deliberately took large parameter ranges to make the fit more prone to failure. The fit for each method is done with the same starting points. Based on those fits, we selected a good fit, as every fit for which the median relative error is below $15$ per cent.

In total, we performed $250000$ fits, for which we obtain $79108$ good fit from either the transformed or untransformed likelihood. Among those successful fits, $87.6$ per cent (i.e. $69290$ good fits) were obtained with the $\sinh^{-1}$ transformation, while we record $73.5$ per cent (i.e. $58132$ good fits) of success without. Hence, the $\sinh^{-1}$ transformation allowed us to find a good model in $15.1$ per cent more cases. This improvement can be expected by the normalisation of the gradient induced by the transformation.

\subsection{Residual differences}

\begin{figure*}
    \begin{minipage}{0.48\linewidth}
    \centering
    \includegraphics[width=\linewidth]{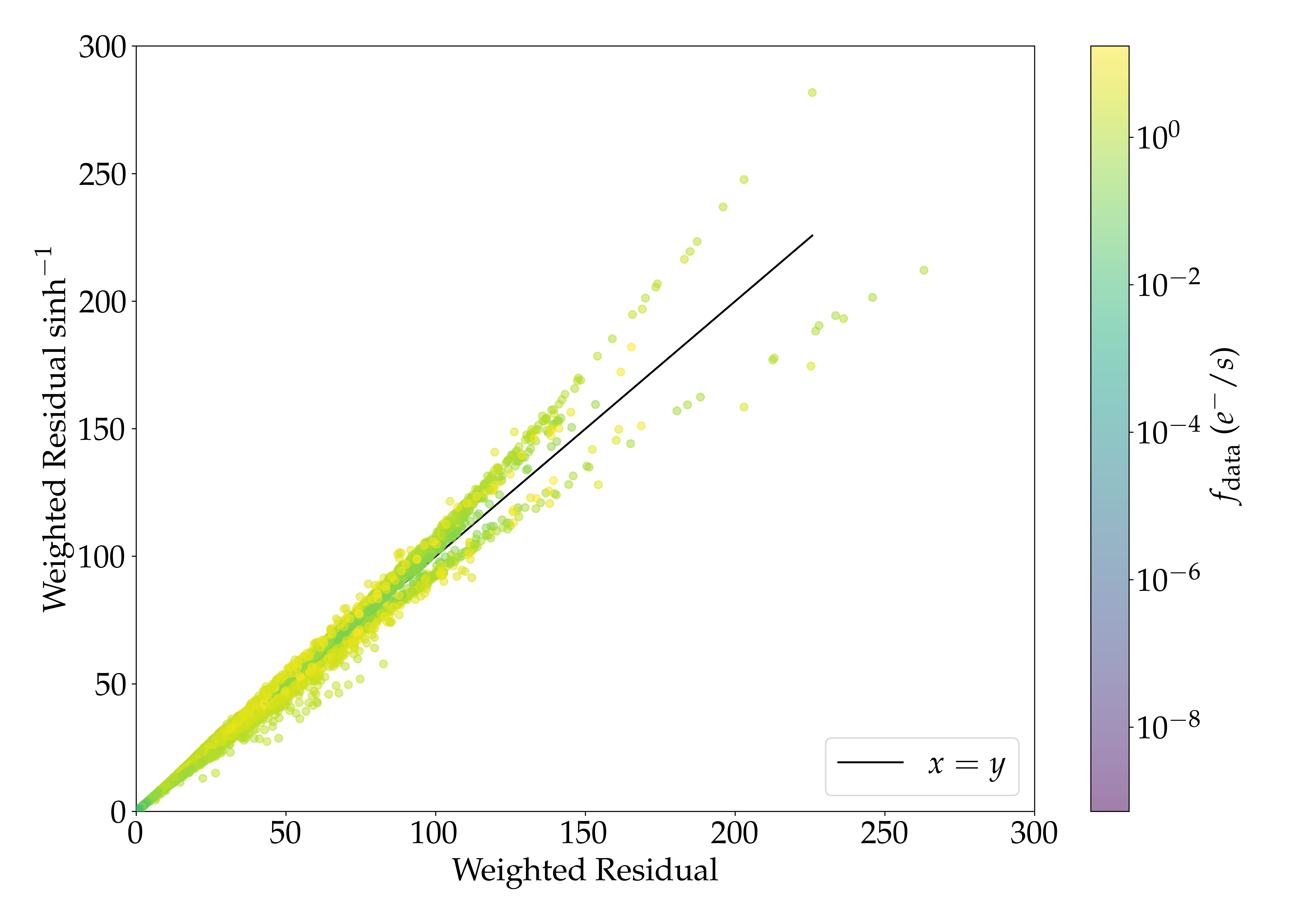} 
    \end{minipage}
    \begin{minipage}{0.48\linewidth}
    \centering
    \includegraphics[width=\linewidth]{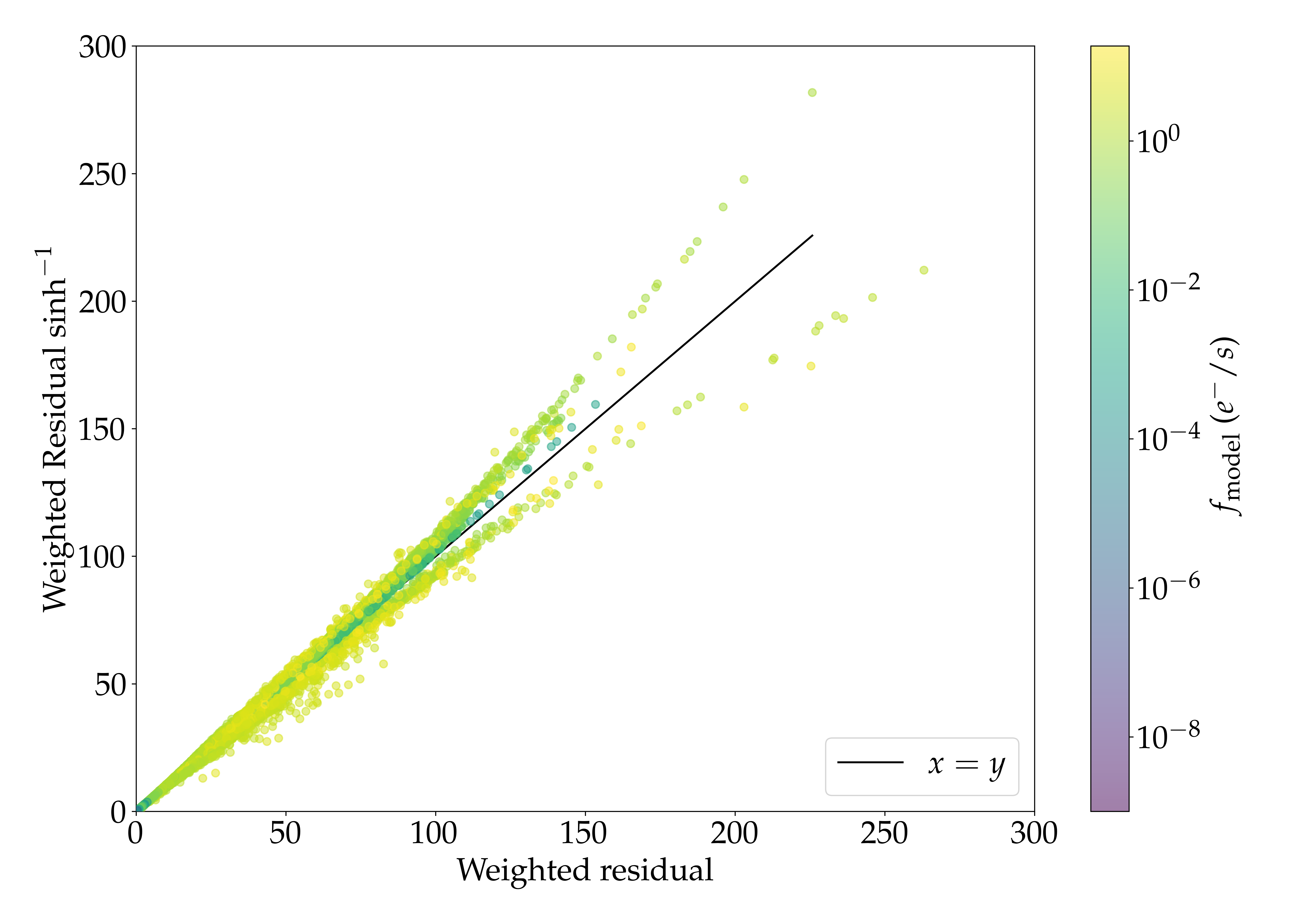}
    \end{minipage}
    
    \caption{Absolute residuals weighted by the associated errors with the $\sinh^{-1}$ transformation in function of its untransformed counterpart. The datapoints are colour coded in function of their observed flux $f_{\rm obs}$ (\textit{left panel}) or their modelled flux $f_{\rm model}$ (\textit{right panel}).}
    \label{fig:Likelihood_checks}
\end{figure*}

Figure~\ref{fig:Likelihood_checks} presents another view of the residuals from the galaxy fitting process presented in sect.~\ref{sect:galaxy-light} at Fig.~\ref{fig:galaxy-fitting-res-laplace}. Both panels show the absolute residuals, weighted by the associated errors, for the $\sinh^{-1}$ transformation as a function of its untransformed counterpart. In the left panel, the points are colour-coded corresponding to their observed flux, while in the right panel, it is based on their best-fit model flux. We retrieve the results discussed in sect.~\ref{sect:galaxy-light}, which show that when residuals are close to zero, both the transformed and untransformed schemes are almost identical. The difference appears for larger residuals as the log behaviour of the $\sinh^{-1}$ comes into effect. This transform influence is materialised by an under- and over-penalisation of high-value residual, which is the manifestation of the asymmetry of the likelihood function. Thanks to the colour coding of the data points, we can see that there is an under-penalisation when $f_{\rm obs}<f_{\rm model}$ and an over-penalisation when $f_{\rm obs}>f_{\rm model}$.

Results presented in sect.~\ref{sect:galaxy-light} could seem contradictory to the Fig.~\ref{fig:Likelihood_checks}, although the number of pixels at the bottom left corner are largely dominating the pixel distribution. Only $3$ per cent of the overall pixels present absolute normalised residuals above $10$. On the $97$ per cent remaining around $66$ per cent are background or faint pixels (i.e. $f_{\rm obs}<0.01\,e^{-}/s$), and the pixels above the background and with absolute normalised residuals inferior to $2$ represent $23$ per cent of the pixel distribution.

\section{Bayesian inference: Prior}
\label{app:init_param_Bayes}

Following the non-linear least-square fit, we perform a Bayesian inference as detailed in Sect.~\ref{sect:galaxy-light:bayes}. For the prior, we use uniform priors for all parameters. We reuse the bounds defined in the previous section for most parameters except $\Sigma^*_0$, $r_{\rm cut}$, $h$ and $I_{0,n}$. For these last parameters, we compute their statistical errors from the least-squares fit and redefine their bounds so that they are centred on the best-fitting solutions from the least-squares fit, with a width on each side equal to $5$ times their statistical errors. We ensure that these bounds are included in the least-squares bounds to avoid overestimating them when the statistical errors are large.

As stated in Sect.~\ref{app:init_param_LM}, some of the bounds are too large, which prevents convergence from reaching a good solution. Hence, for some objects, we manually change the bounds associated with the $\text{axis ratio}$ and $h$ or $r_{\rm cut}$. As all fits are independent, they do not affect neighbouring objects.


\bsp	
\label{lastpage}
\end{document}